\documentclass[twocolumn,trackchanges]{aastex7}

\usepackage{siunitx}
\usepackage{booktabs}   
\usepackage{amsmath}    

\begin{document}

\title{Spatially Resolved Physical Properties of Young Star Clusters and Star-forming Clumps \\ 
in the Brightest $z>6$ Galaxy, the Strongly Lensed Cosmic Spear at $z=6.2$}

\author[0000-0002-5258-8761]{Abdurro'uf}\email{fnuabdur@iu.edu}
\affiliation{Department of Astronomy, Indiana University, 727 East Third Street, Bloomington, IN 47405, USA}
\affiliation{Department of Physics and Astronomy, The Johns Hopkins University, 3400 N Charles St. Baltimore, MD 21218, USA}
\affiliation{Space Telescope Science Institute (STScI), 3700 San Martin Drive, Baltimore, MD 21218, USA}

\author[0000-0001-7410-7669]{Dan Coe}\email{dcoe@stsci.edu}
\affiliation{Space Telescope Science Institute (STScI), 
3700 San Martin Drive, Baltimore, MD 21218, USA}
\affiliation{Association of Universities for Research in Astronomy (AURA), Inc.
for the European Space Agency (ESA)}
\affiliation{Department of Physics and Astronomy, The Johns Hopkins University, 3400 N Charles St. Baltimore, MD 21218, USA}

\author[0009-0007-0522-7326]{Tom Resseguier}\email{tresseg1@jhu.edu}
\affiliation{Department of Physics and Astronomy, The Johns Hopkins University, 3400 N Charles St. Baltimore, MD 21218, USA}
\affiliation{Space Telescope Science Institute (STScI), 3700 San Martin Drive, Baltimore, MD 21218, USA}

\author[]{Calla Murphy}\email{cmurph71@jh.edu}
\affiliation{Department of Physics and Astronomy, The Johns Hopkins University, 3400 N Charles St. Baltimore, MD 21218, USA}

\author[0000-0002-9217-7051]{Xinfeng Xu}\email{xinfeng.xu@northwestern.edu}
\affiliation{Department of Physics and Astronomy, Northwestern University, 2145 Sheridan Road, Evanston, IL, 60208, USA}

\author[0000-0002-8192-8091]{Angela Adamo}\email{angela.adamo@astro.su.se}
\affiliation{Department of Astronomy, The Oskar Klein Centre, Stockholm University, AlbaNova, SE-10691 Stockholm, Sweden}

\author[0000-0002-4430-8846]{Namrata Roy}\email{namratar@asu.edu} 
\affiliation{School of Earth and Space Exploration, Arizona State University, Tempe, AZ 85287-6004, USA}

\author[0000-0002-6586-4446]{Alaina Henry}\email{ahenry@stsci.edu}
\affiliation{Space Telescope Science Institute (STScI), 
3700 San Martin Drive, Baltimore, MD 21218, USA}

\author[0000-0002-5588-9156]{Vasily Kokorev}\email{vkokorev@utexas.edu}
\affiliation{Department of Astronomy, The University of Texas at Austin, Austin, TX 78712, USA}

\author[0000-0003-2680-005X]{Gabriel Brammer}\email{gabriel.brammer@nbi.ku.dk}
\affiliation{Cosmic Dawn Center (DAWN), Denmark}
\affiliation{Niels Bohr Institute, University of Copenhagen, Jagtvej 128, 2200 Copenhagen N, Denmark}

\author[0000-0001-7201-5066]{Seiji Fujimoto}\email{seiji.fujimoto@utoronto.ca}
\affiliation{Dunlap Institute for Astronomy and Astrophysics, 50 St. George Street, Toronto, Ontario, M5S 3H4, Canada}

\author[0000-0001-7113-2738]{Henry C. Ferguson}\email{ferguson@stsci.edu}
\affiliation{Space Telescope Science Institute (STScI), 3700 San Martin Drive, Baltimore, MD 21218, USA}

\author[0000-0002-6015-8614]{Amanda Pagul}\email{apagul@stsci.edu}
\affiliation{Space Telescope Science Institute (STScI), 
3700 San Martin Drive, Baltimore, MD 21218, USA}

\author[0000-0001-8156-6281]{Rogier A. Windhorst}\email{Rogier.Windhorst@asu.edu}
\affiliation{School of Earth and Space Exploration, Arizona State University, Tempe, AZ 85287-6004, USA}

\author[0000-0001-6670-6370]{Timothy Heckman}\email{theckma1@jhu.edu}
\affiliation{Department of Physics and Astronomy, The Johns Hopkins University, 3400 N Charles St. Baltimore, MD 21218, USA}
\affiliation{School of Earth and Space Exploration, Arizona State University, Tempe, AZ 85287-6004, USA}

\author[0000-0001-9065-3926]{Jose M. Diego}\email{jdiego@ifca.unican.es}
\affiliation{Instituto de F\'isica de Cantabria (CSIC-UC). Avda. Los Castros s/n. 39005 Santander, Spain}


\author[0000-0003-3596-8794]{Hollis B. Akins}\email{hollis.akins@gmail.com}
\affiliation{Department of Astronomy, The University of Texas at Austin, 2515 Speedway, Stop C1400, Austin, TX 78712, USA} 

\author[0000-0003-2718-8640]{Joseph F. V. Allingham}\email{allingha@post.bgu.ac.il}
\affiliation{Physics Department, Ben-Gurion University of the Negev, P.O. Box 653, Be'er-Sheva 84105, Israel}

\author[0000-0001-5758-1000]{Ricardo O. Amor\'{i}n}\email{ricardo.amorin@userena.cl} 
\affiliation{Instituto de Astrof\'{i}sica de Andaluc\'{i}a (CSIC), Apartado 3004, 18080 Granada, Spain}

\author[0000-0002-4153-053X]{Danielle A. Berg}\email{daberg@austin.utexas.edu}
\affiliation{Department of Astronomy, The University of Texas at Austin, 2515 Speedway, Stop C1400, Austin, TX 78712, USA}
\affiliation{Cosmic Frontier Center, The University of Texas at Austin, Austin, TX 78712, USA} 

\author[0000-0001-5984-0395]{Maru\v{s}a Brada\v{c}}\email{marusa.bradac@fmf.uni-lj.si}
\affiliation{University of Ljubljana, Faculty of Mathematics and Physics, Jadranska ulica 19, SI-1000 Ljubljana, Slovenia}

\author[0000-0002-7908-9284]{Larry D. Bradley}\email{
lbradley@stsci.edu}
\affiliation{Space Telescope Science Institute (STScI), 3700 San Martin Drive, Baltimore, MD 21218, USA}

\author[0000-0003-1060-0723]{Wenlei Chen}\email{wenlei.chen@okstate.edu}
\affiliation{Department of Physics, Oklahoma State University, 145 Physical Sciences Bldg, Stillwater, OK 74078, USA}

\author[0000-0002-0302-2577]{John Chisholm}\email{chisholm@austin.utexas.edu}
\affiliation{Department of Astronomy, The University of Texas at Austin, 2515 Speedway, Stop C1400, Austin, TX 78712, USA} 
\affiliation{Cosmic Frontier Center, The University of Texas at Austin, Austin, TX 78712, USA}

\author[0000-0003-1949-7638]{Christopher J. Conselice}\email{conselice@gmail.com}
\affiliation{Jodrell Bank Centre for Astrophysics, University of Manchester, Oxford Road, Manchester M13 9PL, UK}

\author[0000-0001-8460-1564]{Pratika Dayal}\email{pdayal@cita.utoronto.ca}
\affiliation{Canadian Institute for Theoretical Astrophysics, 60 St George St, University of Toronto, Toronto, ON M5S 3H8, Canada}
\affiliation{David A. Dunlap Department of Astronomy and Astrophysics, University of Toronto, 50 St George St, Toronto ON M5S 3H8, Canada}
\affiliation{Department of Physics, 60 St George St, University of Toronto, Toronto, ON M5S 3H8, Canada}

\author[0000-0003-0348-2917]{Miroslava Dessauges-Zavadsky}\email{miroslava.dessauges@unige.ch} 
\affiliation{Department of Astronomy, University of Geneva, Chemin Pegasi 51, 1290 Sauverny, Switzerland}

 \author[0000-0002-9382-9832]{Andreas L. Faisst}\email{afaisst@caltech.edu}
\email{afaisst@caltech.edu}
\affiliation{IPAC, California Institute of Technology, 1200 E. California Blvd. Pasadena, CA 91125, USA}

\author[0000-0001-8519-1130]{Steven L. Finkelstein}\email{stevenf@astro.as.utexas.edu}
\affiliation{Department of Astronomy, The University of Texas at Austin, 2515 Speedway, Stop C1400, Austin, TX 78712, USA} 
\affiliation{Cosmic Frontier Center, The University of Texas at Austin, Austin, TX 78712, USA}

\author[0000-0001-7440-8832]{Yoshinobu Fudamoto}\email{y.fudamoto@chiba-u.jp}
\affiliation{Center for Frontier Science, Chiba University, 1-33 Yayoi-cho, Inage-ku, Chiba 263-8522, Japan}

\author[orcid=0000-0001-6278-032X]{Lukas J. Furtak}\email{furtak@utexas.edu}
\affiliation{Cosmic Frontier Center, The University of Texas at Austin, Austin, TX 78712, USA} 
\affiliation{Department of Astronomy, The University of Texas at Austin, Austin, TX 78712, USA}

\author[0000-0002-6047-430X]{Yuichi Harikane}\email{hari@icrr.u-tokyo.ac.jp}
\affiliation{Institute for Cosmic Ray Research, The University of Tokyo, 5-1-5 Kashiwanoha, Kashiwa, Chiba 277-8582, Japan}

\author[0000-0003-4512-8705]{Tiger Yu-Yang Hsiao}\email{tiger.hsiao@utexas.edu}
\affiliation{Department of Astronomy, The University of Texas at Austin, Austin, TX 78712, USA}
\affiliation{Cosmic Frontier Center, The University of Texas at Austin, Austin, TX 78712, USA}

\author[0000-0002-6090-2853]{Yolanda Jimenez-Teja}\email{yojite@iaa.csic.es}
\affiliation{Instituto de Astrofísica de Andalucía–CSIC, Glorieta de la Astronomía s/n, E-18008, Granada, Spain}

\author[0000-0002-6610-2048]{Anton M. Koekemoer}\email{koekemoer@stsci.edu}
\affiliation{Space Telescope Science Institute, 3700 San Martin Drive,
Baltimore, MD 21218, USA}

\author[0000-0003-2366-8858]{Rebecca L. Larson}\email{rlarson@stsci.edu}
\affiliation{Space Telescope Science Institute (STScI), 
3700 San Martin Drive, Baltimore, MD 21218, USA}

\author[0000-0003-1581-7825]{Ray A. Lucas}\email{lucas@stsci.edu}
\affiliation{Space Telescope Science Institute, 3700 San Martin Drive, Baltimore, MD 21218, USA}

\author[0000-0003-1427-2456]{Matteo Messa}\email{matteo.messa@inaf.it}
\affiliation{INAF -- OAS, Osservatorio di Astrofisica e Scienza dello Spazio di Bologna, via Gobetti 93/3, I-40129 Bologna, Italy}

\author[0000-0002-8530-9765]{Lamiya Mowla}\email{lmowla@wellesley.edu}
\affiliation{Department of Physics and Astronomy, Wellesley College, Wellesley, MA 02481, USA}

\author[0009-0000-1999-5472]{Minami Nakane}\email{nakanem@icrr.u-tokyo.ac.jp}
\affiliation{Institute for Cosmic Ray Research, The University of Tokyo, 5-1-5 Kashiwanoha, Kashiwa, Chiba 277-8582, Japan}
\affiliation{Department of Physics, Graduate School of Science, The University of Tokyo, 7-3-1 Hongo, Bunkyo, Tokyo 113-0033, Japan}

\author[]{Ga\"el Noirot}\email{gnoirot@stsci.edu} 
\affiliation{Space Telescope Science Institute, 3700 San Martin Drive, Baltimore, MD 21218, USA}

\author[0000-0002-9651-5716]{Richard Pan}\email{richard.pan@tufts.edu}
\affiliation{Department of Physics \& Astronomy, Tufts University, Medford, MA 02155, USA}

\author[0000-0002-2282-8795]{Massimo Pascale}\email{massimopascale@berkeley.edu}
\affiliation{Department of Physics \& Astronomy, University of California, Los Angeles, 430 Portola Plaza, Los Angeles, CA 90095, USA}

\author[0000-0001-5492-1049]{Johan Richard}\email{johan.richard@univ-lyon1.fr}
\affiliation{Univ Lyon, Univ Lyon 1, Ens de Lyon, CNRS, Centre de Recherche Astrophysique de Lyon UMR5574, F-69230, Saint-Genis-Laval, France}

\author[0000-0003-4223-7324]{Massimo Ricotti}\email{ricotti@astro.umd.edu}
\affiliation{Department of Astronomy, University of Maryland, College Park, 20742, USA}

\author[0000-0002-6265-2675]{Luke Robbins}\email{andrew.robbins@tufts.edu}
\affiliation{Department of Physics and Astronomy, Tufts University, Medford, MA 02155, USA}

\author[0000-0000-0000-0000]{Daniel Schaerer}\email{daniel.schaerer@unige.ch}
\affiliation{Observatoire de Genève, Université de Genève, Chemin Pegasi 51, 1290 Versoix, Switzerland}
\affiliation{CNRS, IRAP, 14 Avenue E. Belin, 31400 Toulouse, France}

\author[0000-0002-4622-6617]{Fengwu Sun}\email{fengwu.sun@cfa.harvard.edu}
\affiliation{Center for Astrophysics $|$ Harvard \& Smithsonian, 60 Garden St., Cambridge, MA 02138, USA}

\author[0000-0002-5057-135X]{Eros Vanzella}\email{eros.vanzella@inaf.it}
\affiliation{INAF -- OAS, Osservatorio di Astrofisica e Scienza dello Spazio di Bologna, via Gobetti 93/3, I-40129 Bologna, Italy}

\author[0000-0003-1815-0114]{Brian Welch}\email{brian.welch@issibern.ch}
\affiliation{International Space Science Institute, Hallerstrasse 6, 3012 Bern, Switzerland}

\author[0000-0002-4201-7367]{Chris Willott}\email{chris.willott@nrc.ca} 
\affiliation{NRC Herzberg, 5071 West Saanich Rd, Victoria, BC V9E 2E7, Canada}

\author[0000-0002-0350-4488]{Adi Zitrin}\email{adizitrin@gmail.com}
\affiliation{Physics Department, Ben-Gurion University of the Negev, P.O. Box 653, Be'er-Sheva 84105, Israel}





\newcommand{\LCDM}{$\Lambda$CDM}

\newcommand{\red}[1]{{\color{red} #1}}
\newcommand{\redss}[1]{{\color{red} ** #1}}
\newcommand{\redbf}[1]{{\color{red}\bf #1 \color{black}}}

\newcommand{\ny}{$\tilde {\rm n}$}
\newcommand{\about}{$\sim$}
\newcommand{\appr}{$\approx$}
\newcommand{\gt}{$>$}
\newcommand{\uJy}{$\mu$Jy}
\newcommand{\sig}{$\sigma$}
\newcommand{\Lya}{Lyman-$\alpha$}
\renewcommand{\th}{$^{\rm th}$}
\newcommand{\lam}{$\lambda$}

\newcommand{\tentothe}[1]{$10^{#1}$}
\newcommand{\tentotheminus}[1]{$10^{-#1}$}
\newcommand{\e}[1]{$\times 10^{#1}$}
\newcommand{\en}[1]{$\times 10^{-#1}$}
\newcommand{\cgsfluxunits}{erg$\,$s$^{-1}\,$cm$^{-2}$}
\newcommand{\cgsfluxdensityunits}{erg$\,$s$^{-1}\,$cm$^{-2}$\,\AA$^{-1}$}
\newcommand{\linefluxunits}{\tentotheminus{20} \cgsfluxunits}

\newcommand{\logU}{$\log(U)$}
\newcommand{\logOH}{12+log(O/H)}

\newcommand{\sinv}{s$^{-1}$}

\newcommand{\footnoteurl}[1]{\footnote{\url{#1}}}

\newcommand{\tnm}[1]{\tablenotemark{#1}}
\newcommand{\super}[1]{$^{\rm #1}$}
\newcommand{\supa}{$^{\rm a}$}
\newcommand{\supb}{$^{\rm b}$}
\newcommand{\supc}{$^{\rm c}$}
\newcommand{\supd}{$^{\rm d}$}
\newcommand{\supe}{$^{\rm e}$}
\newcommand{\supf}{$^{\rm f}$}
\newcommand{\supg}{$^{\rm g}$}
\newcommand{\suph}{$^{\rm h}$}
\newcommand{\supi}{$^{\rm i}$}
\newcommand{\supj}{$^{\rm j}$}
\newcommand{\supk}{$^{\rm k}$}
\newcommand{\supl}{$^{\rm l}$}
\newcommand{\supm}{$^{\rm m}$}
\newcommand{\supn}{$^{\rm n}$}
\newcommand{\supo}{$^{\rm o}$}

\newcommand{\sqarcmin}{arcmin\squared}

\newcommand{\supcomma}{$^{\rm ,}$}

\newcommand{\rhalf}{$r_{1/2}$}

\newcommand{\chisq}{$\chi^2$}

\newcommand{\Zgas}{$Z_{\rm gas}$}  
\newcommand{\Zstar}{$Z_*$}  

\newcommand{\inv}{\per}
\newcommand{\Mstar}{$M^*$}
\newcommand{\Lstar}{$L^*$}
\newcommand{\phistar}{$\phi^*$}

\newcommand{\sigmaMsunpc}{$M_{\odot}\,\rm{pc}^{-2}$}
\newcommand{\sigmaMsunkpc}{$M_{\odot}\,\rm{kpc}^{-2}$}
\newcommand{\mutan}{$\mu_{\rm tan}$}

\newcommand{\logM}{log($M_*$/\Msun)}

\newcommand{\LUV}{$L_{UV}$}
\newcommand{\MUV}{$M_{UV}$}

\newcommand{\Msun}{$M_\odot$}
\newcommand{\Lsun}{$L_\odot$}
\newcommand{\Zsun}{$Z_\odot$}

\newcommand{\Mvir}{$M_{vir}$}
\newcommand{\Mt}{$M_{200}$}
\newcommand{\Mf}{$M_{500}$}

\newcommand{\Ndotion}{$\dot{N}_{\rm ion}$}
\newcommand{\xiion}{$\xi_{\rm ion}$}
\newcommand{\logxiion}{log(\xiion)}
\newcommand{\fesc}{$f_{\rm esc}$}

\newcommand{\XHI}{$X_{\rm HI}$}
\newcommand{\XHII}{$X_{\rm HII}$}
\newcommand{\RHII}{$R_{\rm HII}$}

\newcommand{\Halpha}{H$\alpha$}
\newcommand{\Hbeta}{H$\beta$}
\newcommand{\Hgamma}{H$\gamma$}
\newcommand{\Hdelta}{H$\delta$}
\newcommand{\Halphaw}{\Halpha\,$\lambda$6563}
\newcommand{\Hbetaw}{\Hbeta\,$\lambda$4861}
\newcommand{\Hgammaw}{H$\gamma$\,$\lambda$4342}
\newcommand{\Hdeltaw}{H$\delta$\,$\lambda$4103}
\newcommand{\Ha}{\Halpha}
\newcommand{\Hb}{\Hbeta}

\newcommand{\I}{\,{\sc i}}
\newcommand{\II}{\,{\sc ii}}
\newcommand{\III}{\,{\sc iii}}
\newcommand{\IV}{\,{\sc iv}}
\newcommand{\VI}{\,{\sc vi}}
\newcommand{\VII}{\,{\sc vii}}
\newcommand{\VIII}{\,{\sc viii}}

\newcommand{\HI}{H\,{\sc i}}
\newcommand{\HII}{H\,{\sc ii}}
\newcommand{\HeI}{He\,{\sc i}}
\newcommand{\HeII}{He\,{\sc ii}}

\newcommand{\CII}{[C\,{\sc ii}]}
\newcommand{\CIIw}{\CII\,$\lambda$2325 (blend)}
\newcommand{\CIII}{C\,{\sc iii}]}
\newcommand{\CIIIwa}{\CIII\,$\lambda$1907}
\newcommand{\CIIIwb}{\CIII\,$\lambda$1909}
\newcommand{\CIIId}{C\,{\sc iii}]}
\newcommand{\CIIIdw}{C\,{\sc iii}]\,$\lambda\lambda$1907,1909}
\newcommand{\CIV}{C\,{\sc iv}}
\newcommand{\CIVw}{\CIV\,$\lambda$1549}
\newcommand{\OII}{[O\,{\sc ii}]}
\newcommand{\OIIwa}{\OII\,$\lambda$3726}
\newcommand{\OIIwb}{\OII\,$\lambda$3729}
\newcommand{\OIIdw}{\OII\,$\lambda\lambda$3726,3729}
\newcommand{\OIII}{[O\,{\sc iii}]}
\newcommand{\OIIIw}{\OIII\,$\lambda$5007}
\newcommand{\OIIIww}{\OIII\,$\lambda\lambda$4959,5007}
\newcommand{\OIIIwa}{\OIII\,$\lambda$4363}
\newcommand{\OIIIwc}{\OIII\,$\lambda$4959}
\newcommand{\NeIII}{[Ne\,{\sc iii}]}
\newcommand{\NeIIIw}{\NeIII\,$\lambda$3870}
\newcommand{\NeIIIwb}{\NeIII\,$\lambda$3969}
\newcommand{\HeIw}{HeI\,$\lambda$3890}
\newcommand{\HeIwa}{HeI\,$\lambda$4473}
\newcommand{\HeIIw}{HeII\,$\lambda$1640}
\newcommand{\NII}{[N\,{\sc ii}]}
\newcommand{\NIII}{N\,{\sc iii}]}
\newcommand{\NIV}{N\,{\sc iv}]}
\newcommand{\NIIIw}{\NIII\,$\lambda$1748}
\newcommand{\NIVw}{\NIV\,$\lambda$1486}
\newcommand{\MgII}{Mg\,{\sc ii}}
\newcommand{\MgIIw}{\MgII\,$\lambda$2800}

\newcommand{\SII}{[S\,{\sc ii}]}
\newcommand{\SIIwa}{\SII\,$\lambda$6716}
\newcommand{\SIIwb}{\SII\,$\lambda$6731}
\newcommand{\SIIdw}{\SII\,$\lambda\lambda$6716,6731}

\newcommand{\Lyaw}{Ly$\alpha$\,$\lambda$1216}

\newcommand{\OIIIwfivem}{\OIII\,$\lambda$52$\mu$m}
\newcommand{\OIIIweightm}{\OIII\,$\lambda$88$\mu$m}



\newcommand{\Om}{\Omega_{\rm M}}
\newcommand{\OL}{\Omega_\Lambda}

\newcommand{\etal}{et al.}

\newcommand{\citeps}{\citep}

\newcommand{\HST}{{\em HST}}
\newcommand{\SST}{{\em SST}}
\newcommand{\Hubble}{{\em Hubble}}
\newcommand{\Spitzer}{{\em Spitzer}}
\newcommand{\Chandra}{{\em Chandra}}
\newcommand{\JWST}{{\em JWST}}
\newcommand{\Planck}{{\em Planck}}

\newcommand{\Bradac}{{Brada\v{c}}}

\newcommand{\citepeg}[1]{\citep[e.g.,][]{#1}}

\newcommand{\range}[2]{\! \left[ _{#1} ^{#2} \right] \!}  

\newcommand{\grizli}{\textsc{grizli}}
\newcommand{\eazypy}{\textsc{eazypy}}
\newcommand{\msaexp}{\textsc{msaexp}}
\newcommand{\trilogy}{\textsc{trilogy}}
\newcommand{\bagpipes}{\textsc{bagpipes}}
\newcommand{\beagle}{\textsc{beagle}}
\newcommand{\photutils}{\textsc{photutils}}
\newcommand{\SEP}{\textsc{sep}}
\newcommand{\piXedfit}{\textsc{piXedfit}}
\newcommand{\pyneb}{\textsc{pyneb}}
\newcommand{\HIIC}{\textsc{hii-chi-mistry}}
\newcommand{\astropy}{\textsc{astropy}}
\newcommand{\astrodrizzle}{\textsc{astrodrizzle}}
\newcommand{\multinest}{\textsc{multinest}}
\newcommand{\cloudy}{\textsc{Cloudy}}
\newcommand{\jdaviz}{\textsc{Jdaviz}}
\newcommand{\emcee}{\textsc{emcee}}
\newcommand{\galfit}{\textsc{GALFIT}}
\newcommand{\prospector}{\textsc{prospector}}
\newcommand{\db}{\textsc{Dense Basis}}

\renewcommand{\tt}[1]{\texttt{#1}}

\newcommand{\SE}{\tt{SourceExtractor}}

\newcommand{\PD}[1]{\textcolor{blue}{[PD: #1\;]}}

\newcommand{\JD}{MACS0647$-$JD}

\newcommand{\edense}{$n_{e}$}
\newcommand{\ROII}{$R_{[\rm{OII}]}$}
\newcommand{\OIIratio}{\OII\,$\lambda$3729/$\lambda$3726}

\newcommand{\RSII}{$R_{[\rm{SII}]}$}
\newcommand{\SIIratio}{\SII\,$\lambda$6716/$\lambda$6731}
\newcommand{\CIIIratio}{C\,{\sc iii}]\,$\lambda$1909/$\lambda$1907}

\newcommand{\lya}{\hbox{Ly$\alpha$}}        
\newcommand{\nv}{\hbox{\sc N\,v}}           
\newcommand{\niv}{\hbox{\sc N\,iv]}}      
\newcommand{\ariv}{\hbox{\sc Ar\,iv]}}      
\newcommand{\civ}{\hbox{\sc C\,iv}}         
\newcommand{\heii}{\hbox{He\,{\sc ii}}}     
\newcommand{\cii}{\hbox{\sc C\,ii]}}      
\newcommand{\hei}{\hbox{He\,{\sc i}}}     
\newcommand{\oiiisemi}{\hbox{\sc O\,iii]}}  
\newcommand{\ciii}{\hbox{\sc C\,iii]}}      
\newcommand{\siiii}{\hbox{Si\,{\sc iii]}}}  
\newcommand{\mgii}{\hbox{Mg\,{\sc ii}}}     
\newcommand{\oi}{\hbox{O\,{\sc i}}}     
\newcommand{\oii}{\hbox{\sc [O\,ii]}}     
\newcommand{\hb}{\hbox{\sc H$\beta$}}       
\newcommand{\oiii}{\hbox{\sc [O\,iii]}}     
\newcommand{\ha}{\hbox{\sc H$\alpha$}}      
\newcommand{\hd}{\hbox{\sc H$\delta$}}      
\newcommand{\hg}{\hbox{\sc H$\gamma$}}      
\newcommand{\he}{\hbox{\sc H$\epsilon$}}      
\newcommand{\het}{\hbox{\sc H$\eta$}}      
\newcommand{\sii}{\hbox{[S\,{\sc ii}]}}     
\newcommand{\siii}{\hbox{[S\,{\sc iii}]}}   
\newcommand{\pab}{\hbox{Pa$\beta$}}      
\newcommand{\pag}{\hbox{Pa$\gamma$}}
\newcommand{\neiii}{\hbox{[Ne\,{\sc iii]}}}  
\newcommand{\siiiIR}{\hbox{[S\,{\sc iii]}}}  
\newcommand{\feii}{\hbox{[Fe\,{\sc ii]}}}  
\newcommand{\nev}{\hbox{[Ne\,{\sc v]}}}  
\newcommand{\oiv}{\hbox{O\,{\sc iv]}}}  
\newcommand{\siv}{\hbox{[S\,{\sc iv]}}}  

\newcommand{\kms}{km\,s$^{-1}$}

\begin{abstract}
We present spatially resolved analysis of stellar populations in the brightest $z>6$ galaxy known to date (AB mag 23), the strongly lensed MACS0308$-$zD1 (dubbed the ``Cosmic Spear'') at $z_{\rm spec}=6.2$.
New JWST NIRCam imaging and high-resolution NIRSpec IFU spectroscopy span the rest-frame ultraviolet to optical. 
The NIRCam imaging reveals bright star-forming clumps and a tail consisting of three distinct, extremely compact star clusters that are multiply-imaged by gravitational lensing.
The star clusters have delensed effective radii of $R_{\rm{eff}} \lesssim 8$ pc, stellar masses of $M_{*} \sim 10^{6}-10^{7}\,M_{\odot}$, and high stellar mass surface densities of $\Sigma_{*} \gtrsim 2\times 10^{4}\,M_{\odot}~\rm{pc}^{-2}$. While their stellar populations are very young ($\sim 6-11$ Myr), their dynamical ages exceed unity, consistent with the clusters being gravitationally bound systems. Placing the star clusters in the size vs.~stellar mass density plane, we find they occupy a region similar to other high-redshift star clusters within galaxies observed recently with JWST, being significantly more massive and denser than local star clusters. Spatially resolved analysis of the brightest clump reveals a compact, intensely star-forming core. The ionizing photon production efficiency ($\xi_{\rm{ion}}$) is slightly suppressed in this central region, potentially indicating a locally elevated Lyman continuum escape fraction facilitated by feedback-driven channels.
\end{abstract}



\section{Introduction} \label{sec:intro}

Understanding how galaxies assembled their stellar contents during the first billion years of cosmic history is an important question in extragalactic astronomy. Observations of galaxies during the epoch of reionization ($z\gtrsim 6$) have revealed a surprisingly diverse population, ranging from low-mass dwarf systems thought to dominate the ionizing photon budget \citep[e.g.,][]{Atek2024} to relatively massive dusty star-forming systems \citep[e.g.,][]{Watson2015, Fudamoto2021, Xiao2024}. Within this complex landscape, the formation of dense, compact stellar systems, namely young star clusters, offers a unique window into the physics of early star formation and the origin of ancient stellar populations such as globular clusters (GCs) observed in local galaxies.

The hierarchical nature of star formation implies that stars form in clustered environments across a wide range of spatial and mass scales, spanning approximately three orders of magnitude in size and six in mass, ranging from Giant Molecular Clouds ($R \sim 20$--$100$~\text{pc}; $M_{*} \sim 10^4$--$10^6~\text{M}_{\odot}$) down to the dense pre-stellar cores ($R \sim 0.05$~\text{pc}; $M_{*} \sim 0.5$--$5~\text{M}_{\odot}$) where individual stars are born \citep[e.g.,][]{Larson1981, Elmegreen1996, Lada2003, Grasha2017}. While this has been extensively demonstrated in local starbursts and general star-forming galaxies \citep[e.g.,][]{Calzetti2015, Adamo2020, Mehta2021}, resolving such scales at high redshift has long been hindered by the limited spatial resolution and sensitivity of pre-JWST facilities. However, recent breakthroughs enabled by the JWST \citep{Rigby2023,Gardner2023}, particularly when combined with the natural magnification provided by strong gravitational lensing of galaxy clusters, have pushed the observational frontier down to parsec scales in galaxies at $z > 4$ \citep[e.g.,][]{Vanzella2022, Vanzella2023, Adamo2024, Mowla2024, 2025Fujimoto, Messa2025a, Nakane2025}.

These studies have uncovered young, massive star clusters with stellar masses of $\sim 10^{5}-10^{7}\,M_{\odot}$, effective radii ranging from a few to tens of parsecs, and high stellar mass surface densities ($\Sigma_{*}$) of up to $\sim 10^{6}\,M_{\odot}\,\rm{pc}^{-2}$, consistent with the properties expected for proto-globular clusters \citep[PGCs; e.g.,][]{Mowla2022, Vanzella2022, Faisst2022, Claeyssens2023, Vanzella2023, Adamo2024, Mowla2024, Vanzella2025, Messa2025, 2025Bradac, Claeyssens2026, Whitaker2026}. In some cases, such clusters contribute a substantial fraction ($10–60$\%) of the host galaxy’s far-ultraviolet (FUV) light or stellar mass ($M_{*}$), pointing toward clustered star formation as a dominant mode in the early universe \citep[e.g.,][]{Ricotti2002, Katz2013, Garcia2023}. Moreover, their intense ionizing radiation and mechanical feedback may facilitate the escape of Lyman continuum photons, making them plausible contributors to reionization \citep[e.g.,][]{Rivera-Thorsen2019, He2020, Vanzella2020, Vanzella2022a}.

\begin{figure*}[ht]
\centering
\includegraphics[width=1.0\textwidth]{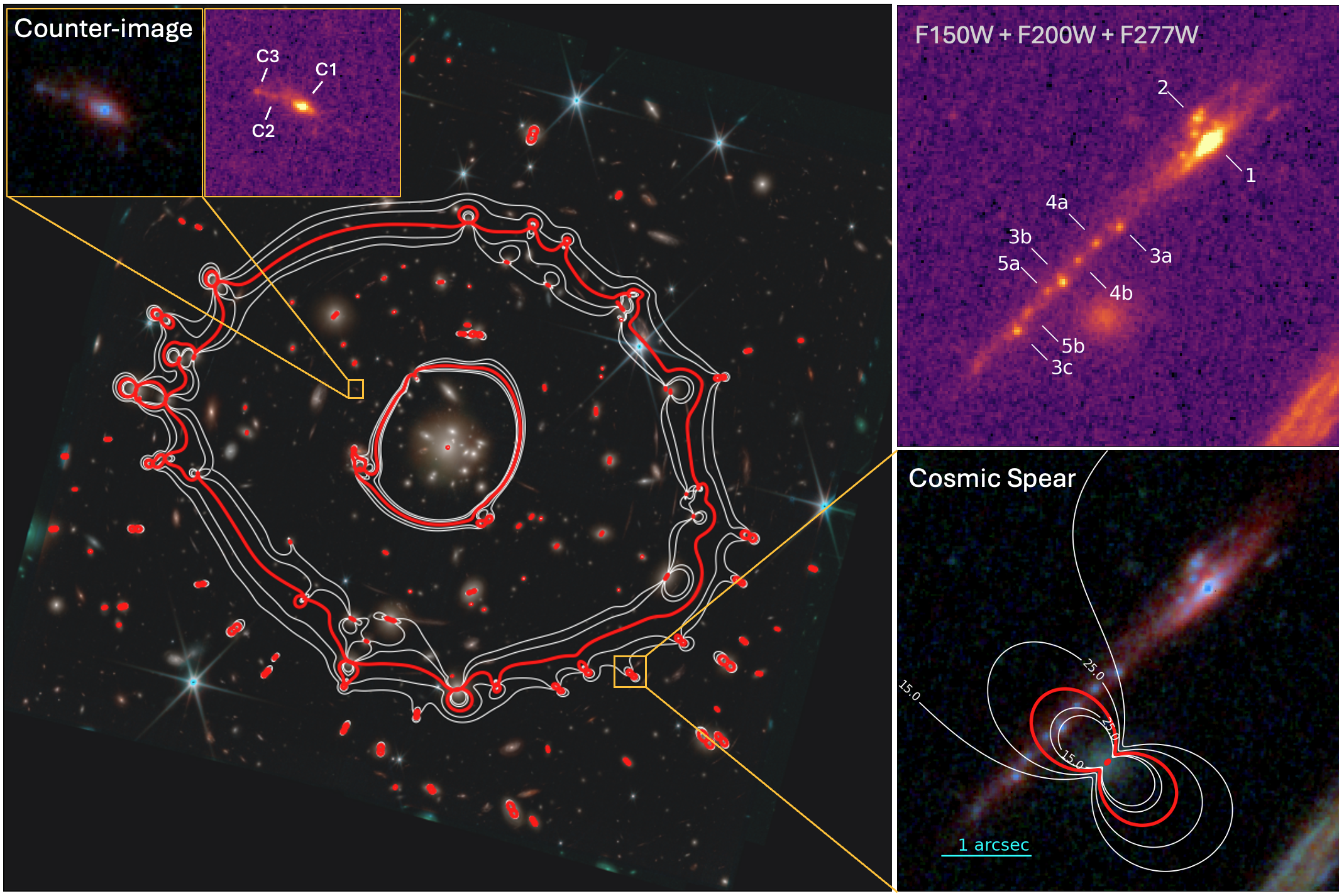}
\caption{NIRCam color image (red: F277W+F356W, green: F150W+F200W, and blue: F090W+F115W) of the Cosmic Spear, at $z=6.2$, in the background of galaxy cluster MACS J0308+2645 at $z=0.356$. The strongly-lensed galaxy (shown in the two right panels) is resolved into two main components: a pair of large, bright star-forming clumps (1 and 2), and an arc containing three distinct star clusters (3, 4, and 5) which are multiply imaged into seven sources (3a, 3b, 3c, 4a, 4b, 5a, and 5b). The top-left corner inset shows a counter-image identified from the lens model. It reveals a clumpy morphology, with three distinct substructures (lower limit). The red line represents a lensing critical curve at $z=6.2$, while the white lines show contours of constant magnifications.}
\label{fig:macs0308-zD1_img}
\end{figure*}

Despite these advances, only a handful of such systems have been identified to date \citep[e.g.,][]{Vanzella2023,Adamo2024,Mowla2024,2025Bradac,Messa2025a}, owing to the rarity of suitable lensing configurations and the need for deep, high-resolution imaging. Each new detection, therefore, provides a critical data point for constraining the formation efficiency, evolution, and survival of massive star clusters at early times.

In this work, we present JWST/NIRCam and NIRSpec Integral Field Unit (IFU) observations of a lensed arc galaxy MACS0308$-$zD1 (dubbed the ``Cosmic Spear'') at $z_{\rm spec} = 6.2$, which is spatially resolved down to a few parsecs by the gravitational lensing effect of a foreground galaxy cluster MACS J0308$+$2645 (henceforth MACS0308$+$26; \citealt{Ebeling2001}), such that individual massive star clusters within it are revealed. This galaxy was discovered by the Reionization Lensing Cluster Survey \citep[RELICS;][]{Coe2019} as the brightest ($23.4$ mag) known lensed galaxy at $z>6$ \citep{Acebron2018,Salmon2020,Welch2023}, and initially confirmed spectroscopically through ALMA observation of the \CII\ 158 $\mu$m line \citep{Fudamoto2024}.

While previous observations with the Hubble Space Telescope (HST) and Spitzer Space Telescope showed that this lensed system consists of a bright clump (knot) and a diffuse arc ($2\farcs3$ long) with no apparent substructure \citep{Salmon2020,Welch2023}, recent JWST/NIRCam imaging from GO 5293 (PI: X. Xu) and VENUS\footnote{\url{https://jwst-venus.github.io/}} (Vast Exploration for Nascent, Unexplored Sources; GO 6882; PIs: S.~Fujimoto and D.~Coe) programs have revealed three compact star clusters multiply-imaged along the arc (see Figure~\ref{fig:macs0308-zD1_img}). Moreover, the bright knot is resolved into at least two substructures. In this paper, we analyze the physical properties of these newly identified clusters alongside the two bright clumps. The star clusters exhibit signatures of young stellar populations ($\sim 6-11$ Myr), high stellar mass surface density ($\Sigma_{*} \gtrsim 2 \times 10^{4}\,M_{\odot}\,\rm{pc}^{-2}$), and dynamical ages consistent with gravitationally bound systems. 

Our analysis builds on and extends previous studies of clustered star formation in lensed systems at $z\sim 2-10$, offering new insights into the formation of star clusters during the reionization epoch. We discuss their implications for the understanding of star cluster formation and contributions to reionizing the intergalactic medium at the early cosmic epoch.

This paper is structured as follows. We describe the observations in Section~\ref{sec:data}. In Section~\ref{sec:method}, we describe our analysis that includes photometric measurement, morphological modeling, and SED fitting on both integrated (i.e.,~global) and spatially resolved scales. We present our results in Section~\ref{sec:results}. In Section~\ref{sec:discuss}, we discuss their implications in the context of clustered formation efficiency, star cluster evolution, and their contributions to the reionization. Throughout this paper, we adopt a flat $\Lambda$CDM cosmology with $\Omega_{m}=0.3$, $\Omega_{\Lambda}=0.7$, and $H_{0}=70\text{ km}\text{ s}^{-1}\text{ Mpc}^{-1}$. All magnitudes are given in the AB system.

\section{JWST Data}\label{sec:data}

The MACS J0308+2645 galaxy cluster \citep{Ebeling2001} was observed with HST as part of the RELICS survey \citep[HST GO 14096;][]{Coe2019}, using three ACS filters (F435W, F606W, and F814W) and four WFC3/IR filters (F105W, F125W, F140W, and F160W). The cluster lies at redshift $z=0.356$ and has a Sunyaev–Zel’dovich mass of $M_{500} = 10.8 \times 10^{14}\,M_{\odot}$, making it one of the most massive clusters in the Planck PSZ2 catalog \citep{Planck2016}. 

Follow-up JWST observations were first obtained in Cycle 3 through GO 5293 (PIs: X. Xu \& A. Henry). These included NIRCam imaging and NIRSpec Integral Field Unit (IFU) spectroscopy. The NIRCam observation was carried out on 4 February 2025, obtaining imaging in six filters (F115W, F150W, F200W, F250M, F300M, and F410M), covering the wavelength range of $1.0$–$4.0~\mu$m. The exposure time was 1100~s for F115W, F150W, F300M, and F410M, and 1600~s for F200W and F250M. The NIRSpec IFU observation, which was carried out on 10 August 2024, used three high-resolution gratings (G140H/F100LP, G235H/F170LP, and G395H/F290LP), spanning $1.0$–$5.3~\mu$m, with an exposure time of 6600 s. While the spectroscopy primarily targeted the bright clumps of the $z=6.2$ Cosmic Spear, the field of view (FoV) also encompassed two star clusters in the arc (see Section~\ref{sec:properties_clumps}).

In Cycle 4, additional NIRCam imaging was acquired as part of the VENUS  program (GO 6882; PIs: S. Fujimoto \& D. Coe). The observations, carried out on 25 August 2025, used 10 filters (F090W, F115W, F150W, F200W, F210M, F277W, F300M, F356W, F410M, and F444W), spanning $1.0$–$5.0~\mu$m, with exposure times of $15-25$ minutes to obtain a homogeneous point source detection limit ($5\sigma$) of $\sim 28$ mag across all of these filters. The VENUS program re-observed all of the filters used in GO 5293 except F250M. To maximize depth in our analysis, we stack the overlapping images from both programs using inverse-variance weighting.

The point spread function (PSF) full width at half maximum (FWHM) of NIRCam imaging data varies across filters, widening with wavelength from $0\farcs03$ to $0\farcs16$. Similarly, the spatial resolution of the NIRSpec IFU data increases toward longer wavelengths, ranging from $0\farcs03$ to $0\farcs22$.  

All the imaging data from the two programs are uniformly processed with the \grizli\ pipeline \citep{brammer_2022_grizli}, similarly to the public DJA (Dawn JWST Archive) product\footnote{\url{https://dawn-cph.github.io/dja/}}. It starts with JWST level-2 products from MAST and performs a photometric calibration with the Calibration Reference Data System (CRDS) context \texttt{jwst\_1456.pmap}. The \grizli\ pipeline implements several important improvements over the standard STScI pipeline, including corrections for cosmic rays, stray light, and detector artifacts \citep{Bradley2023,Rigby2023}. Moreover, it implements additional background, 1/f noise, and diffraction spike subtraction procedures, both at the amplifier level, for each filter, and then the final drizzled mosaic. This pipeline aligns the JWST images to a common world coordinate system (WCS) registered to GAIA DR3 \citep{Gaia2021} and combines the exposures using \textsc{drizzlepac}\footnote{\url{https://github.com/spacetelescope/drizzlepac}}, producing mosaics on a common pixel grid of $0\farcs03$ pixel$^{-1}$. The uncertainty (sigma) images were generated by taking the inverse square root of the weight maps produced by the \grizli\ pipeline reduction. These maps account for the primary variance components, including detector read noise and Poisson noise.

The NIRSpec IFU data are reduced through the standard three-stage JWST pipeline with customized steps adopting similar procedures as applied by the TEMPLATES team \citep{Hutchison2024, Rigby2025}. The detailed description of the NIRSpec IFU reduction will be presented in X. Xu et al. (in prep.).

\section{Methodology} \label{sec:method}

\subsection{Generating Combined Data Cubes} \label{sec:generate_datacubes}

\begin{figure*}[ht]
\centering
\includegraphics[width=1.0\textwidth]{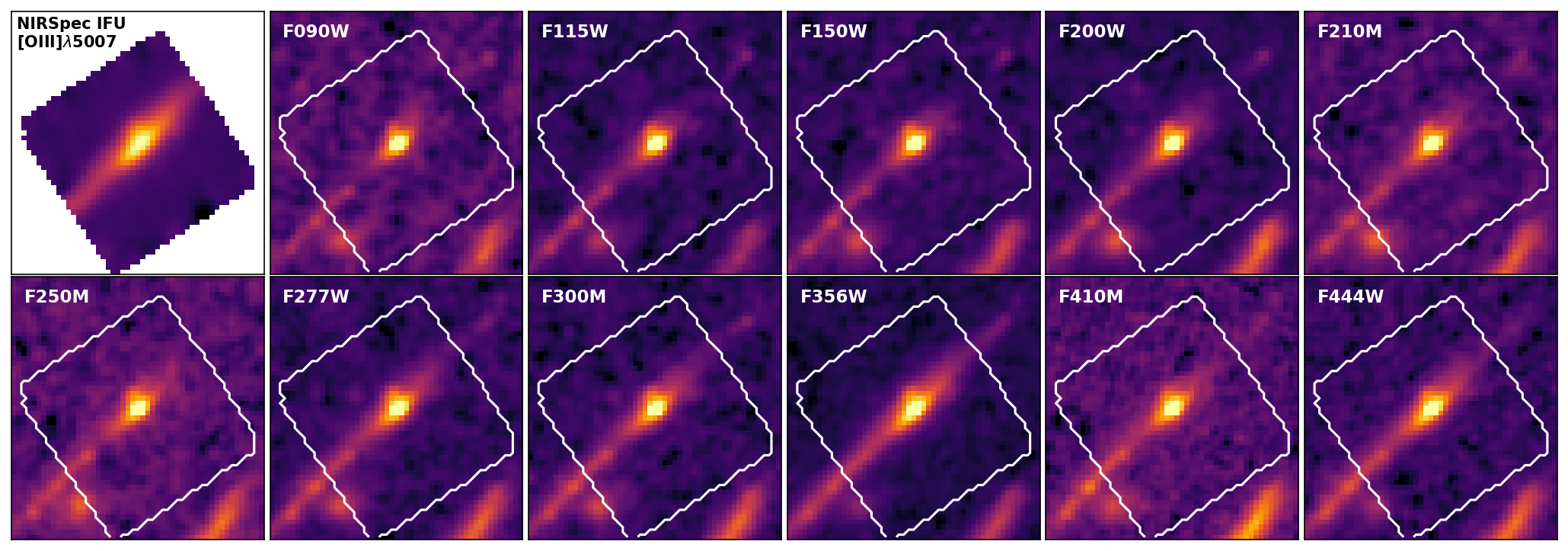}
\caption{Spectrophotometric data cube constructed by combining NIRCam imaging and NIRSpec IFU data using \piXedfit. This data cube is used only for spatially resolved SED fitting of the clumps, and not for aperture photometry and morphological modeling. To create the data cube, the datasets were spatially and spectrally aligned through PSF matching to the spatial resolution of the longest-wavelength grid of the NIRSpec G395H/F290LP data, resampling to a pixel scale of $0\farcs1$, and aligning normalization between the spectra and the photometry for consistency. The top-left panel displays the [OIII]$\lambda\,5007$ map, derived from the NIRSpec IFU data by integrating the spectra across a narrow (100 \AA) wavelength grid centered at the emission line wavelength. The remaining panels show the NIRCam images following spatial alignment and PSF matching to the NIRSpec IFU resolution. The IFU's FoV is shown in the top-left panel and outlined by the white box in the other panels.} 
\label{fig:plot_datacube_nirspec_nircam}
\end{figure*}

The unprecedented depth and spatial resolution of the NIRCam instrument, combined with the natural magnification and stretching from gravitational lensing by the foreground galaxy cluster, enable us to resolve individual star clusters in the Cosmic Spear (see Figure~\ref{fig:macs0308-zD1_img}). With the aid of lens modeling (T.~Resseguier et al., in prep.), we identify three distinct star clusters, labeled 3, 4, and 5, along the arc. These clusters are multiply imaged into seven sources (3a, 3b, 3c, 4a, 4b, 5a, and 5b). This multiplicity arises from an additional magnification caused by a cluster member galaxy (i.e.,~perturber) located southwest of the arc in the projected plane along the line of sight (see Figure~\ref{fig:macs0308-zD1_img}). The lensing critical curve at $z=6.2$ passes through the arcs between 4a and 4b and between 5a and 5b, resulting in mirror symmetry. 

Furthermore, the extended bright knot is resolved into two bright clump components, labeled as 1 and 2. We identify two additional candidates of compact sources in the vicinity of these clumps; lens modeling confirms that these are likely genuine substructures rather than artifacts (T.~Resseguier et al., in prep.). However, because they are small and only resolved in the short-wavelength filters, while remaining undetected or unresolved at longer wavelengths, we do not analyze them individually and instead treat them as part of clump 2. 

To maximize the scientific yield from the rich datasets available for this system, we combine the 11-band NIRCam imaging (including the stacked images from the VENUS and GO 5293 programs) and NIRSpec IFS data. Using \piXedfit\ \citep{Abdurrouf2021,Abdurrouf2022}, we construct two types of data cube: an imaging-only (i.e.,~photometric) data cube and a joint imaging and IFS (i.e.,~spectrophotometric) data cube. The procedures for constructing these data cubes are described in the following.

The photometric data cube is built by resampling and reprojecting the multiband imaging data, cropping the frames around the coordinates of the Cosmic Spear, and combining them into a single FITS file. This image processing pipeline intentionally bypasses PSF matching to preserve the native spatial resolution of the images, as this photometric cube is intended for subsequent aperture photometry and morphological modeling of individual clumps and star clusters, which do not require PSF matching across images.

The spectrophotometric data cube, which forms the basis for our spatially resolved SED fitting and derivation of localized physical properties, is constructed by combining the NIRCam imaging with NIRSpec IFU observations across three grating/filter channels (G140H/F100LP, G235H/F170LP, and G395H/F290LP). This processing accounts for both spatial and spectral alignment, including PSF matching, spatial resampling, reprojection, and spectral matching. PSF matching is applied consistently across the wavelength slices of the NIRSpec IFU data and across the imaging filters. Because the largest PSF belongs to the longest-wavelength grid in the G395H/F290LP data (with a FWHM of $0\farcs22$), \piXedfit\ matches all other datasets, both the imaging and the IFU slices, to this spatial resolution. For each IFU slice and each imaging band, a matching kernel is generated using the \texttt{photutils} package \citep{larry_bradley_2024_13989456}. This procedure requires input PSFs for all datasets; for the IFU data, we model these PSFs using \texttt{STPSF} \citep{2012Perrin}. The generated kernels are then used to convolve the images and IFU slices, except the longest-wavelength slice of the G395H/F290LP data, which serves as the reference. Following PSF matching, we perform spatial resampling and reprojection. Since the NIRSpec IFU data have the coarsest pixel scale ($0\farcs1$ pix$^{-1}$), all datasets are resampled and re-projected to match their spatial sampling and projection. 

To this end, the data cube has been spatially aligned. We then correct for artifacts in the spaxel-level NIRSpec IFU spectra, commonly referred to as ``wiggles'', which arise from the significantly undersampled PSF and the subsequent resampling of the 2D raw spectra into a 3D cube, producing low-frequency sinusoidal-like noise \citep[see e.g.,][]{2023Perna,2023Law,Dumont2025,2025Shajib}. These artifacts are not corrected in the JWST data pipeline, yet they can distort the spectral shape and potentially bias our resolved SED fitting analysis. In our NIRSpec IFU data, the ``wiggles'' features appear both in individual spaxels and in spectra summed over multiple spaxels within an aperture, and they are most pronounced at short wavelengths ($\lambda \lesssim 2\,\mu$m). Several public tools have been developed to remove these artifacts \citep[e.g.,][]{2023Perna,Dumont2025,2025Shajib} by forward modeling them with sinusoidal functions; however, these methods do not perform effectively for our data, likely due to the faint and noisy nature of the spectra, which makes analytical modeling difficult. We therefore adopt a different approach that avoids explicit sinusoidal modeling. First, we mask spectral wavelength grids associated with emission lines to isolate the stellar continuum, which we then fit with a fifth-order Legendre polynomial. Subtracting this fit from the continuum yields the residual ``wiggle'' component. To extract a smooth representation of this feature, we convolve the residual spectrum with a Gaussian kernel of large standard deviation ($\Delta\lambda = 200$ Å). Subtracting this smoothed component from the original spectrum produces a flat spectrum with ``wiggle'' removed. Finally, we reconstruct the full spectrum by adding back the best-fit polynomial continuum and the previously masked emission lines, resulting in a clean spectral data cube.              

The next step is aligning the normalization of the spectra and photometry at each pixel. Small mismatches are observed between the NIRSpec IFU spectra and the imaging photometry. To correct for this, we rescale the spectra to match the overall normalization and shape of the photometric SEDs. Before this spectral matching, the spectra from the three NIRSpec channels are merged to provide full wavelength coverage and then resampled to $10$ \AA\ bins using the \texttt{SpectRes} package \citep{2017Carnall}. The spectral matching procedure follows the approach described in \citet{Abdurrouf2021}, with one modification in how the correction factor is applied. Instead of using a smooth multiplicative factor derived from a Legendre polynomial (order 3) as in that work, we adopt an additive correction factor with order 5. This choice is motivated by the faint and noisy nature of the spectra, which often contain negative flux values, making a multiplicative correction unsuitable. 

Figure~\ref{fig:plot_datacube_nirspec_nircam} shows the generated spectrophotometric data cube. As a representation of the NIRSpec IFU data, the top left panel shows the [OIII]$\lambda\,5007$ map obtained by summing up fluxes in the wavelength grids around the line. The rest of the panels show processed images that have been spatially aligned to the NIRSpec IFU.

\subsection{Clump Photometry} \label{sec:clump_photometry}

We use the photometric data cube to measure the photometry of the star clusters through aperture photometry. We use circular apertures of $0\farcs1$ radius centered at the coordinates obtained from \galfit\ modeling (see Section~\ref{sec:morpho_modeling}). This aperture size is large enough to capture the bulk of the cluster emission, as the corresponding diameter ($0\farcs2$) exceeds the PSF FWHM across all filters, assuming a Gaussian profile. Concurrently, it remains small enough to prevent flux contamination or blending between adjacent clusters.

To obtain clean photometry of the star clusters, we subtract the contribution of the diffuse background light associated with the host galaxy. The diffuse background flux for each cluster is estimated by averaging the pixel fluxes within an annulus spanning radii of $0\farcs16$ to $0\farcs31$ from the cluster center, and then multiplying this average value by the cluster's aperture area. We subsequently apply aperture corrections to the fluxes across all filters using their respective PSF curves of growth. The specific apertures and background annuli used for this photometry are illustrated in Appendix~\ref{sec:aper_photo_clumps_arc} (Figure~\ref{fig:plot_arc_sed_apertures}). The aperture loss correction ranges from 1.4 to 2.1 across filters. We obtained the empirical PSFs from the VENUS data products. These PSFs are constructed by stacking cutouts of bright, unsaturated stars identified in each field, and the same set of stars are used across filters.

For clumps 1 and 2, we use elliptical apertures with a radius of $0\farcs32$ and $0\farcs2$, respectively, and elliptical geometry adopting the \galfit\ fits (Section~\ref{sec:morpho_modeling}). The radii (along the semi-major axis) are chosen to ensure that the two apertures do not overlap. Because the apertures for clumps 1 and 2 do not cover their full extents, we rescale the measured fluxes by factors of 2.5 and 1.8, respectively. These factors are derived from the average ratios (across filters) between the fluxes obtained through the aperture photometry and those from \galfit\ fits performed without flux constraints (i.e.,~priors; see Section~\ref{sec:morpho_modeling}). 

We do not adopt photometry derived directly from \galfit. Due to image noise and the faintness of both the diffuse arc component and the star clusters, the \galfit\ photometry is often unstable and yields non-smooth SEDs. This instability likely arises from the Levenberg–Marquardt least-squares algorithm in \galfit\ converging to local minima in low signal-to-noise data.

To obtain representative photometry of the three star clusters for subsequent SED fitting analysis, we stack the photometry of their multiple lensed images, after correcting each for the lensing magnification. The stacking is performed using inverse-variance weighting, as described further in Appendix~\ref{sec:photometry_stacking}. As can be seen from Figure~\ref{fig:stacking_photonetry}, the delensed SEDs of the mirror images of each star cluster are generally consistent with one another, confirming that they are plausibly multiple images of the same objects. The photometry of individual clusters are summarized in Table~\ref{tab:clump_photometry}, while the stacked photometry is given in Table~\ref{tab:clump_stacked_photometry}.

\subsection{Morphological Modeling} \label{sec:morpho_modeling}

\begin{figure*}[ht]
\centering
\includegraphics[width=0.9\textwidth]{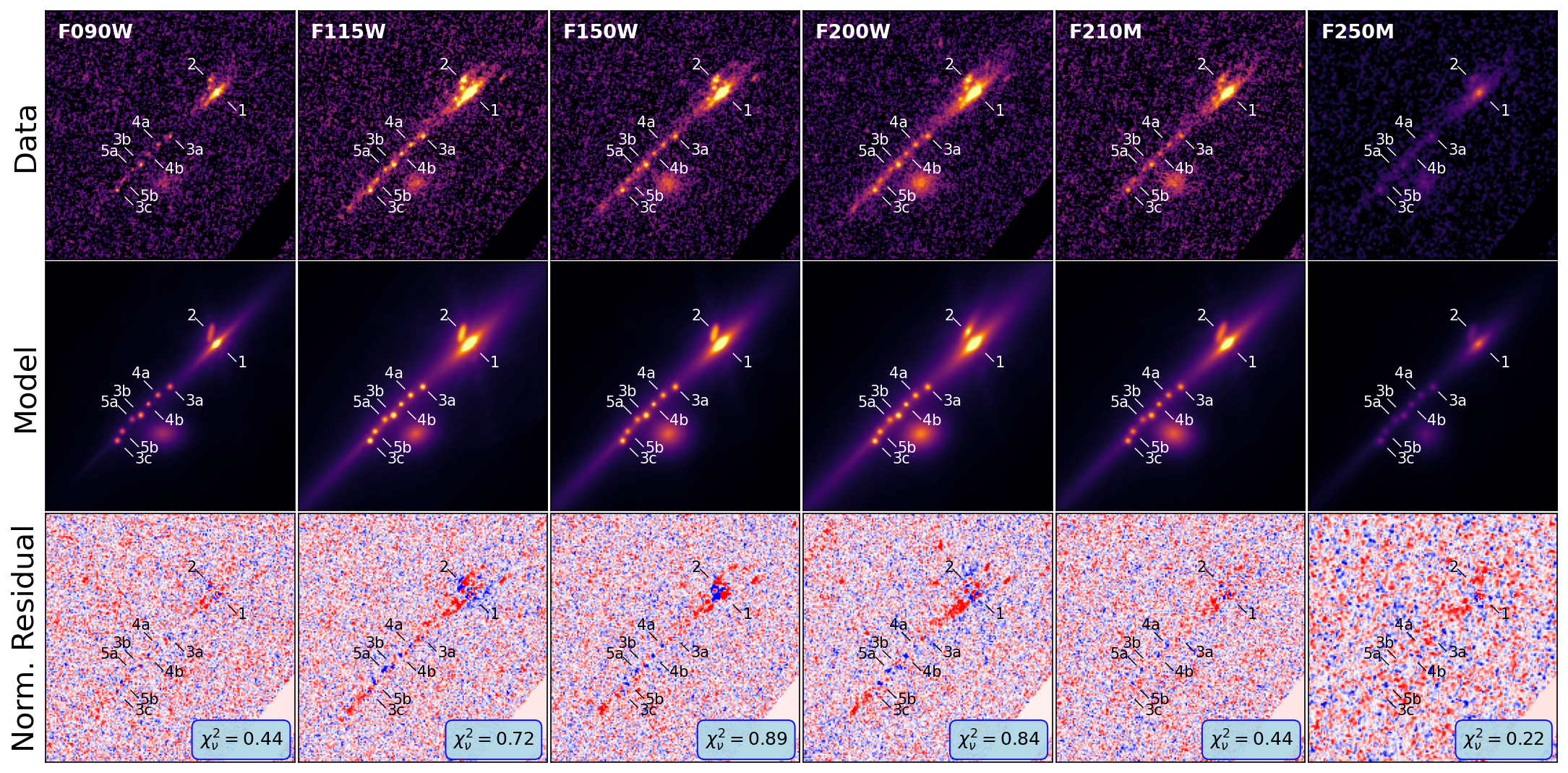}
\includegraphics[width=0.9\textwidth]{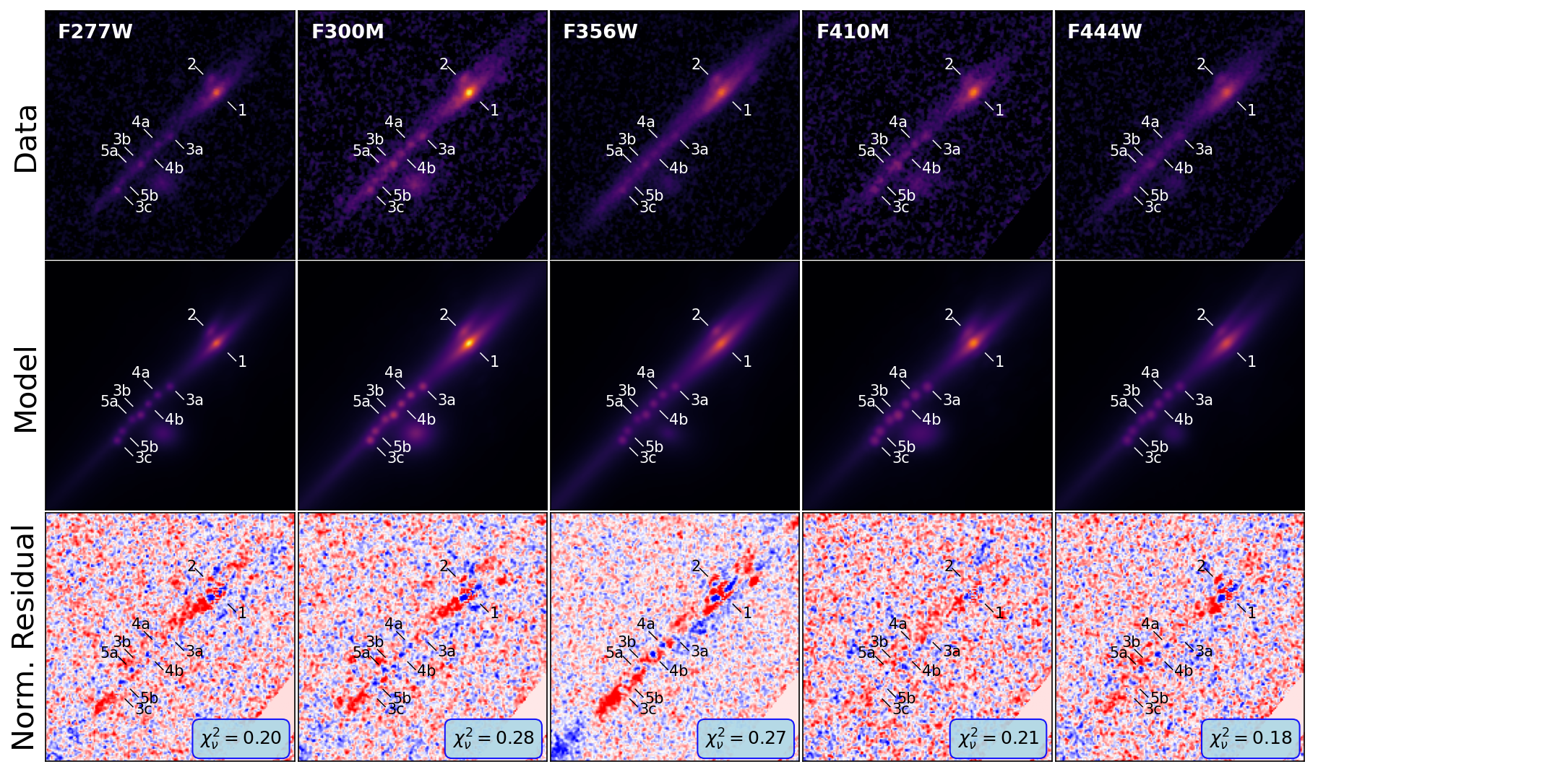}
\caption{Results of morphological modeling using \galfit\ applied to 11-band NIRCam images in their native resolutions. In each column, the rows from top to bottom show data, best-fit model, and normalized residual images. The latter is the residual divided by the error image ($\sigma$). In the modeling, the galaxy is decomposed into ten components: the diffuse light of the host galaxy underlying the arc (i.e.,~tail), two bright clumps, and seven compact sources associated with the multiple images of the star clusters. In the normalized residual images, red, white, and blue colors are for positive, zero, and negative values, respectively. The reduced chi-square ($\chi_{\nu}^{2}$) values are shown in the bottom-right corner of the residual maps.}
\label{fig:galfit_result_final1}
\end{figure*}

We model the morphology of the bright clumps and star clusters with \galfit\ \citep{Peng2002} to obtain their PSF-deconvolved effective radii, which are then used to estimate the intrinsic sizes (after correcting for lensing magnification). We run \galfit\ on the photometric data cube, beginning with the F150W band, where the clumps are brighter and more spatially resolved. In the modeling, the galaxy system is decomposed into ten components: the diffuse light of the host galaxy underlying the arc (i.e.,~tail), two bright clumps, and seven compact sources representing the star cluster images. The diffuse arc component and clumps are modeled with a S\'ersic profile \citep{Sersic1968}, as they have extended size, while the star clusters are modeled with 2D Gaussian profiles, following previous studies \citep[e.g.,][]{Claeyssens2023,Vanzella2023,Adamo2024}. To account for the contaminating light from the galaxy cluster member on the southwest side of the arc, we also model the galaxy with a S\'ersic profile. 

Each component is first modeled individually to obtain initial estimates of its morphological parameters. For this step, we isolate the light from each component by cropping its surrounding region and masking neighboring sources to minimize contamination. Elliptical apertures are used for the diffuse arc and the two bright clumps, while circular apertures of radius $0\farcs18$ are applied to the star clusters. After obtaining initial parameters from these individual fits, we perform a global fit on the full, uncropped image using a composite model of all components. In this stage, the central coordinates and position angles are fixed to the values from the individual fits, while the other parameters are allowed to vary. This global fitting provides a complete set of best-fit morphological parameters and integrated photometry for the F150W image. The resulting parameters are then used as priors for fitting the other NIRCam bands, where we again fix the coordinates and position angles while allowing the remaining parameters to vary.

To mitigate the possibility of the \galfit\ fitting converging to local minima within a potentially multimodal parameter space, we generate 700 perturbed model realizations based on the initial best-fit parameters. The magnitudes of the two bright clumps and the star clusters are perturbed using normal distributions centered on the values obtained from aperture photometry (Section~\ref{sec:clump_photometry}) with a standard deviation of $0.15$ mag. From these realizations, we select the model with the lowest chi-square value. This approach yields morphological models that not only reproduce the structural shapes of the clumps and star clusters but also provide photometry consistent with aperture measurements. We note that this process does not significantly alter the overall chi-square values compared to those obtained from the original \galfit\ fitting.

\subsection{Integrated SED Fitting of Clumps and Star Clusters} \label{sec:sedfit}

We perform SED fitting on the measured photometry of the clumps and star clusters to derive their stellar population properties, primarily using \bagpipes\ \citep{Carnall2018,Carnall2019}. To ensure the robustness of our results and assess systematics from different modeling assumptions, we complement this analysis with \piXedfit\ \citep{Abdurrouf2021,Abdurrouf2022} and \textsc{Prospector} \citep{Leja2017,Johnson2021}. These additional codes are particularly useful for the star-forming clumps and the counter image (i.e.,~host galaxy; see Section~\ref{sec:host_counter_properties}), which may exhibit complex star formation history (SFH) and stellar populations. 

\subsubsection{\bagpipes} \label{sec:bagpipes_fitting}

Our \bagpipes\ fitting uses the Binary Population and Stellar Synthesis (BPASS) v2.2.1 templates \citep{Stanway2018}, which include modeling of stellar binary evolution. The initial mass function (IMF) is assumed to be in the form of \citet{Kroupa2001} with an upper mass cutoff of $300\,M_{\odot}$. Nebular emission is included through the photoionization code CLOUDY \citep{Ferland2017}. We allow the ionization parameter ($U$) to vary bwteeen $-3$ and $-1$. To account for the effect of dust attenuation, we try two dust attenuation models: \citet{Calzetti2000} and \citet{Salim2018}. Based on the spatially resolved spectroscopic analysis which suggests a low dust attenuation level from the Balmer decrement (N.~Roy et al. in prep.), we narrow down the range for $A_{\rm V}$ to $0-0.5$ mag. We model the SFH in various forms. For clumps 1 and 2, which have extended size and expected to contain stellar population with a range of ages, we explore various extended SFH models: (1) the non-parametric SFH with continuity prior following \citet{Leja2019}, (2) delayed tau with a flat $\tau$ prior over a wide range of 1 Myr $-$ 1 Gyr, (3) exponentially declining model with similarly extended $\tau$ prior, and (4) double power-law model.

In contrast to clumps 1 and 2, we adopt short SFH model for the star clusters, as their small sizes ($\lesssim 15$ pc) suggest they host relatively uniform, coeval stellar populations. Specifically, we apply two models: exponentially declining SFH with short $\tau$ fixed to 1 Myr and a single burst model.

\subsubsection{\prospector} \label{sec:prospector_fitting}

\prospector\ is based on the Flexible Stellar Population Synthesis \citep[FSPS;][]{Conroy2009,Conroy2010}, which has flexible options for SED modeling ingredients. We apply the Mesa Isochrones and Stellar Tracks \citep[MIST;][]{Choi2016} isochrone, the MILES stellar spectral library \citep{Sanchez-Blazquez2006,Falcon-Barroso2011}, the \citet{Kroupa2001} IMF, and \citet{Calzetti2000} dust attenuation model. We run \prospector\ on the SEDs of the star-forming clumps (1 and 2) and the counter image, applying the non-parametric SFH model with continuity prior following \citet{Leja2019}.

\subsubsection{\piXedfit} \label{sec:pixedfit_fitting}

Similar to \prospector, \piXedfit\ uses the FSPS for generating model SEDs. We apply the Mesa Isochrones and Stellar Tracks (MIST) isochrones \cite{Choi2016} and MILES stellar spectral library \citep{Sanchez-Blazquez2006,Falcon-Barroso2011}. The IMF is assumed to be the form of \citet{Chabrier2003}, which has a very small mass offset ($\sim 0.03$ dex) compared to \citet{Kroupa2001} IMF \citep[e.g.,][]{Speagle2014}. For modeling dust attenuation, we assume \citet{Calzetti2000} law. For the SFH form, we assume the delayed tau for star-forming clumps as they have an extended region and are expected to contain multiple stellar populations, and a simple exponentially declining (i.e.,~tau model) with short $\tau=1-3$ Myr. We apply the Markov Chain Monte Carlo (MCMC) method for sampling the parameter space in the fitting process.

\subsection{Spatially Resolved SED Fitting} \label{sec:resolved_sedfit}

With the generated spectrophotometric data cube, we perform spatially resolved SED fitting using \piXedfit\ to map the stellar population properties and emission lines in the bright clump region. This analysis is restricted to the area covered by the NIRSpec IFU, where spectrophotometric SEDs are available. To improve the signal-to-noise ratios (S/N) of the resolved SEDs, we first apply pixel binning using \piXedfit, requiring a minimum S/N of 4 in all filters redward of F250M. Although no explicit S/N thresholds are imposed on the spectra, the binning process still provides a reasonable boost.

We then fit the spectrophotometric SEDs of the spatial bins with \piXedfit, which simultaneously models the spectra and photometry following the procedure described in \citet{Abdurrouf2021}, with an improvement that now includes Gaussian fitting of emission lines. In brief, the stellar continuum (with spectral regions around possible emission lines masked) and the photometry are fitted with models using a Bayesian approach with MCMC sampling. To fit the emission lines, first, the best-fit stellar continuum is subtracted from the observed spectrum to isolate the emission lines, after which each line is individually fitted with a Gaussian profile.

The SED modeling setup and assumptions are the same as those described in Section~\ref{sec:pixedfit_fitting}. For the SFH, we adopt a delayed-$\tau$ model, as the clumps have extended size and could harbor a stellar population with a range of ages.

\subsection{Lens Modeling} \label{sec:lens_modeling}

The detailed lens modeling and source-plane reconstruction will be presented in a companion paper (Resseguier et al., in prep.). Here we provide a brief description of the methodologies and constraints relevant to this study. The macro-lens model of the galaxy cluster MACS J0308+2645 is constructed using a parametric approach combining dual Pseudo-Isothermal Elliptical Mass Distributions \citep[dPIE][]{Eliasdottir2007_dPIE} to represent the main dark matter halo and cluster member galaxies. While a previous lens model by \cite{Acebron2018} relied on only three multiple-image families identified in HST data, the superior depth and spatial resolution of our new JWST NIRCam imaging enabled the identification of 10 additional multiple-image systems. This brought the total constraints to 13 distinct image families consisting of 33 individual multiple images. The macro-lens model achieves the spatial-position recovery of the observed multiple images with an RMS of $0\farcs8$. 

These constraints were used to optimize the parameter inferences via MCMC sampling implemented in \textsc{Lenstool} \citep{Jullo2007_Lenstool}. Crucially, the model is anchored by two background families with secure spectroscopic redshifts. The photometric redshifts from \textsc{EAZY} \citep{Brammer2008} were applied as priors for the remaining families.

The macro-lensing potential alone, however, cannot account for the unique morphology of the Cosmic Spear's tail. The intense stretching and duplication of the compact star clusters are driven by a low-mass cluster member galaxy situated immediately southwest of the arc that acts as a local perturber. To achieve the sub-arcsecond precision necessary to map the Cosmic Spear, this perturber is decoupled from the scaling relation and optimized independently as an individual dPIE component, with its coordinates fixed at its observed light center. To fine-tune the local mass parameters and resolve degenerate lensing configurations near the critical line, pixel-level forward modeling was executed using \textsc{Lenstronomy} \citep{BIRRER2018189}. While the global macro-lens model recovers the positions of the broader cluster's multiple images with an RMS of $0\farcs8$, this localized forward modeling of the Cosmic Spear system yields an effective RMS of $0\farcs0287$ (approximately the size of a single pixel). This sub-pixel precision ensures the accuracy of magnification estimates for the star clusters analyzed in this work. Alternative lens modeling setups were tested assuming different mirror symmetry pairings; however, these alternative scenarios were rejected as they required unphysical mass profiles for the perturber and generated significant residuals in the image-plane forward modeling.

Magnification factors and lensing quantities are extracted directly from the fine-tuned lens model, at the image plane positions corresponding to the sources optimized using the forward modeling method. Magnification values are calculated across the entire forward modeling MCMC chain, and the uncertainties are derived from the $16^{\text{th}}$ and $84^{\text{th}}$ percentiles of the resulting distributions. We compound them with systematics, conservatively taken to be 10\% for magnification values, based on lens model comparisons in past works. The resulting critical curves at $z=6.2$ and constant magnification contours are overlaid on the imaging data in Figure~\ref{fig:macs0308-zD1_img}.

\section{Results} \label{sec:results}

\subsection{Physical Properties of the Star Clusters} \label{sec:properties_clusters}

\begin{figure*}[ht]
\centering
\includegraphics[width=0.42\textwidth]{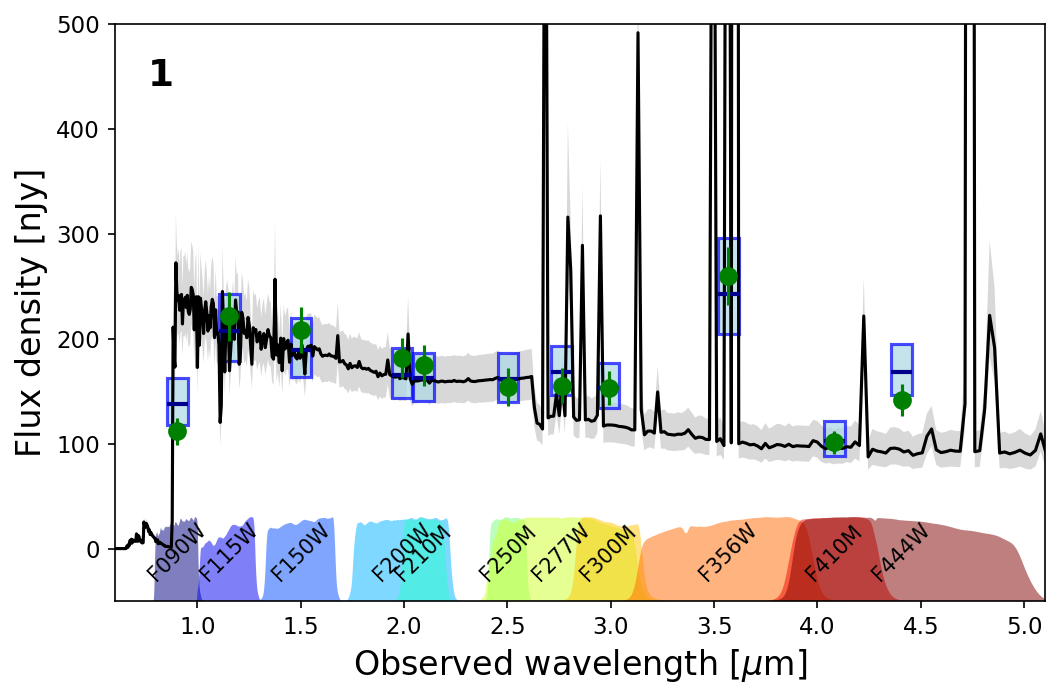}
\includegraphics[width=0.42\textwidth]{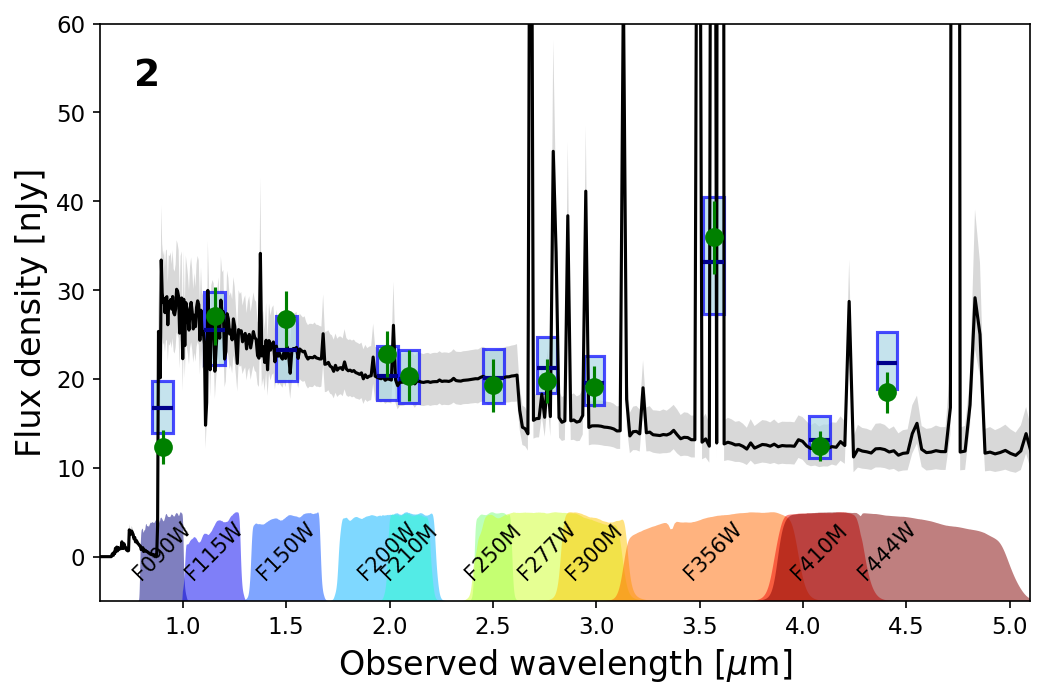}
\includegraphics[width=0.42\textwidth]{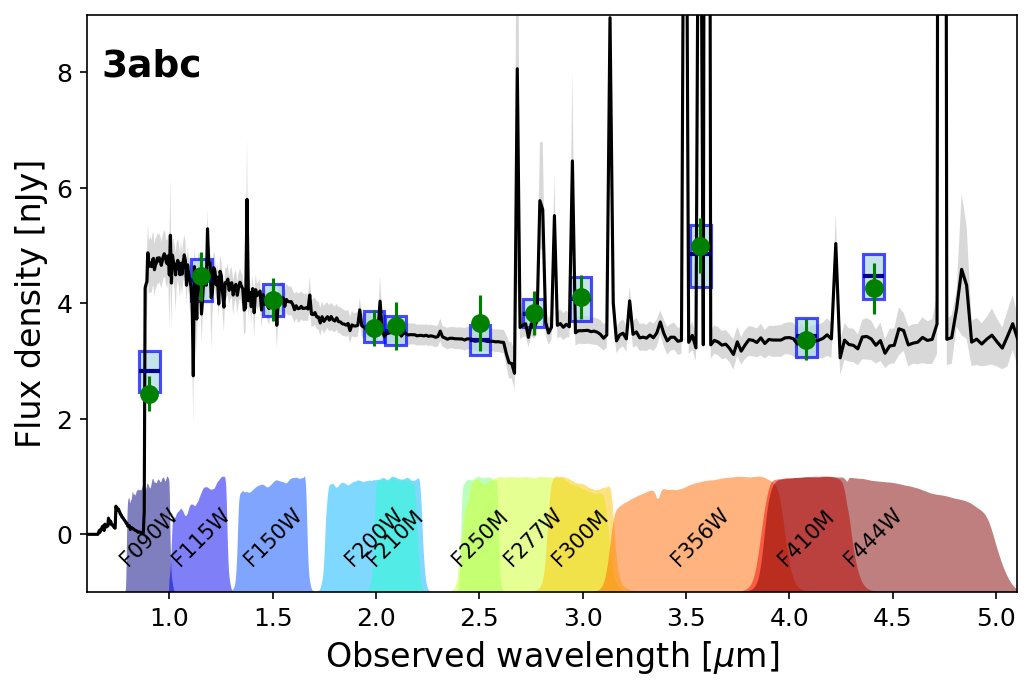}
\includegraphics[width=0.42\textwidth]{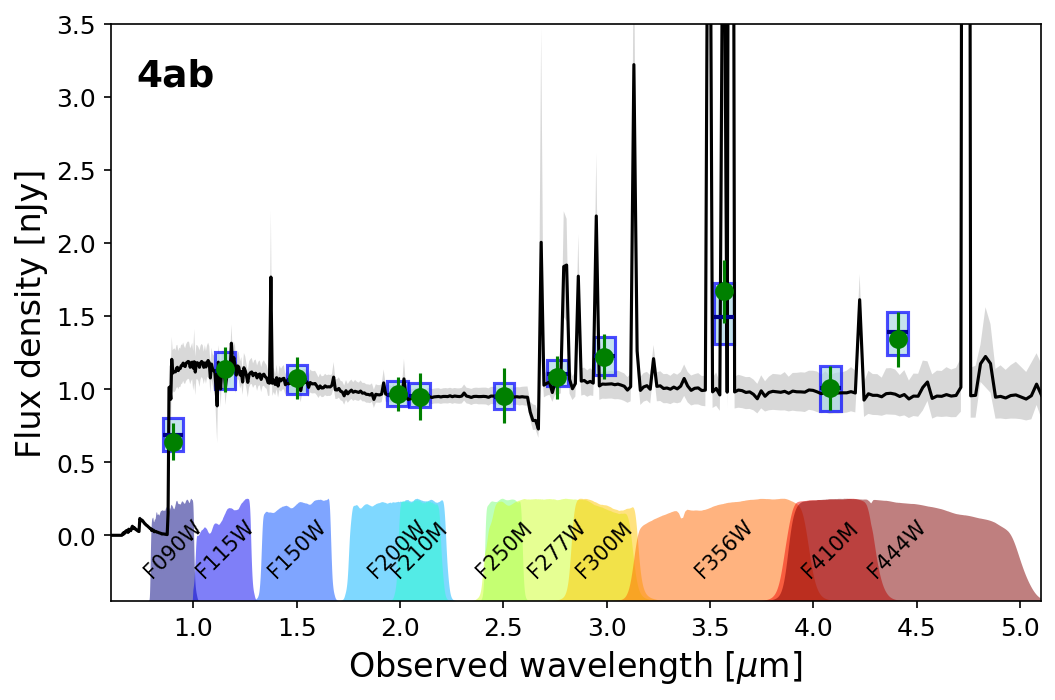}
\includegraphics[width=0.42\textwidth]{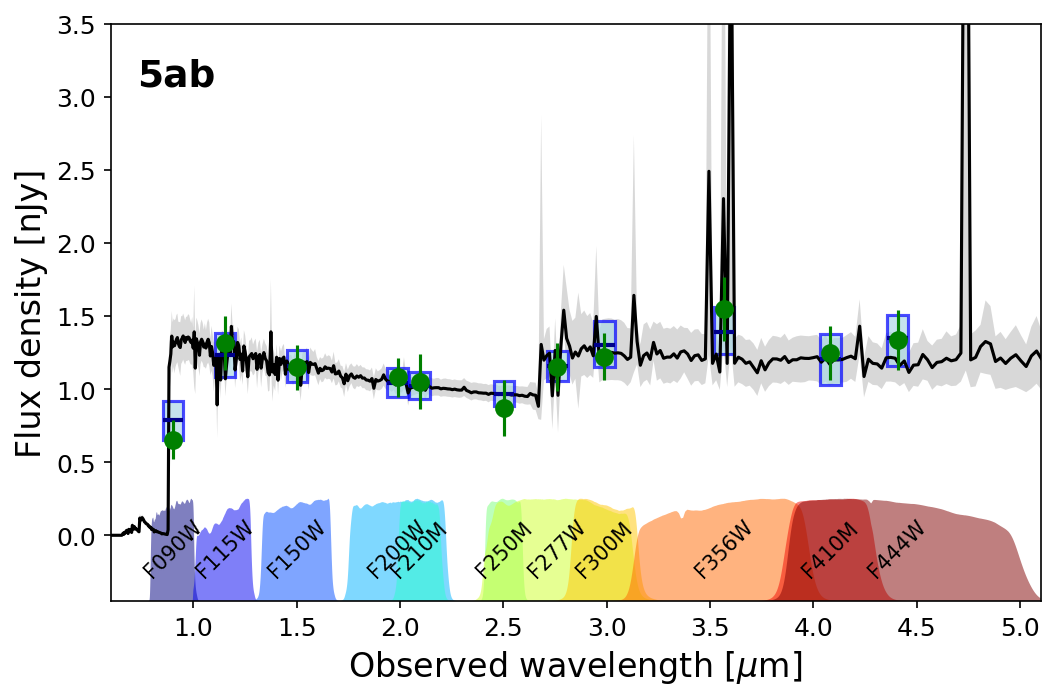}
\includegraphics[width=0.42\textwidth]{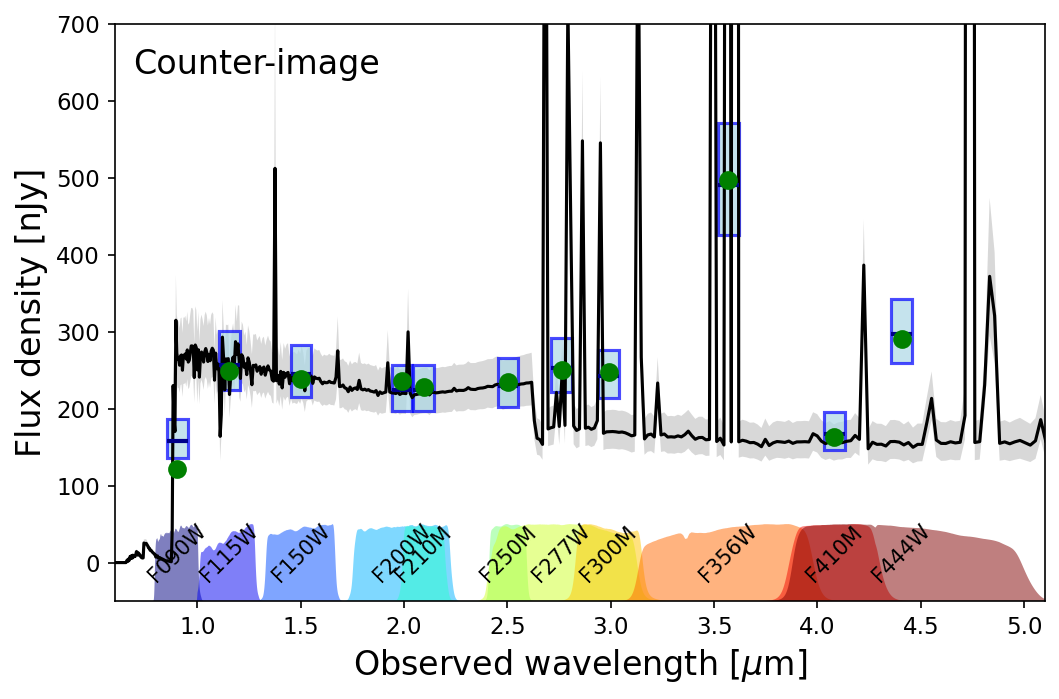}
\caption{Results of the SED fitting for the clumps, star clusters (using stacked photometry), and the counter-image, performed with \bagpipes. All the SEDs presented here are corrected for the lensing magnification. The SEDs of the star clusters (3, 4, and 5) are obtained from stacking the delensed photometry of their multiple images. In each panel, the observed photometry is shown as green circles. The best-fit model is represented by the blue squares (photometry) and the black line (spectrum).}
\label{fig:plot_comb_sedfits_bagpipes}
\end{figure*}

\begin{deluxetable*}{lcccccccc}
\tablecaption{\label{tab:props_stacked_clumps}Measured Properties of the Clumps, Star Clusters, and Counter-image
}
\tablewidth{\columnwidth}
\tablehead{
\colhead{ID} &
\colhead{$R_{\rm eff,obs}$} &
\colhead{$R_{\rm eff,int}$} &
\colhead{$M_{*,\rm{int}}$} &
\colhead{$\Sigma_{*,\rm{int}}$} &
\colhead{Age} &
\colhead{$\Pi$} &
\colhead{$\mu_{\rm tot}$} &
\colhead{$\mu_{\rm tan}$}
\vspace{-0.07in}\\
\colhead{} &
\colhead{(pix)} &
\colhead{(pc)} &
\colhead{($10^{6}\,$\Msun)} &
\colhead{($10^{2}\,$\sigmaMsunpc)} &
\colhead{(Myr)} &
\colhead{} &
\colhead{} &
\colhead{}
\vspace{-0.07in}\\
\colhead{(1)} &
\colhead{(2)} &
\colhead{(3)} &
\colhead{(4)} &
\colhead{(5)} &
\colhead{(6)} &
\colhead{(7)} &
\colhead{(8)} &
\colhead{(9)}
}
\startdata
1
& $7 \pm 0$
& $117 \pm 18$
& ${270}_{-97}^{+157}$
& ${31}_{-17}^{+36}$
& ${82}_{-60}^{+103}$
& ${7}_{-6}^{+18}$
& $10 \pm 1$
& $10 \pm 1$ \\
2
& $5 \pm 0$
& $73 \pm 8$
& ${35}_{-14}^{+22}$
& ${10}_{-5}^{+10}$
& ${83}_{-55}^{+107}$
& ${5}_{-4}^{+12}$
& $11 \pm 1$
& $11 \pm 1$ \\
3abc
& $<1$
& $<8$
& ${8}_{-3}^{+4}$
& $>214$
& ${7}_{-1}^{+2}$
& $>6$
& $17 \pm 2$
& $18 \pm 2$ \\
4ab
& $<1$
& $<3$
& ${1}_{-1}^{+2}$
& $>292$
& ${6}_{-2}^{+2}$
& $>10$
& $44 \pm 5$
& $43 \pm 4$ \\
5ab
& $<1$
& $<5$
& ${5}_{-3}^{+5}$
& $>336$
& ${11}_{-4}^{+13}$
& $>15$
& $36 \pm 4$
& $36 \pm 4$ \\
CI
& $7 \pm 1$
& $483 \pm 97$
& ${443}_{-154}^{+232}$
& ${3}_{-2}^{+4}$
& ${115}_{-83}^{+123}$
& ${2}_{-1}^{+4}$
& $3.0 \pm 0.3$
& $2.5 \pm 0.3$
\enddata
\tablecomments{The photometry of star clusters 3, 4, and 5 is represented by the stacked photometry of their multiple images. 
(1) Clump ID (i.e.,~label). CI refers to the counter image.
(2) Observed and PSF-deconvolved effective radius measured using \galfit\ with a 2D Gaussian model. $R_{\rm eff,obs}$ is calculated as half of the FWHM (full width at half maximum) of the best-fit Gaussian model. The modelling is performed on the F150W image, except for CI, which is based on the F200W image because it has stronger detection of the diffuse tail component. The images have spatial sampling of $0\farcs03$ pixel$^{-1}$. 
(3) Intrinsic effective radius, corrected for lensing tangential magnification ($\mu_{\rm tan}$). 
(4) Intrinsic stellar mass, corrected for lensing total magnification ($\mu_{\rm tot}$).
(5) Intrinsic stellar mass surface density, corrected for lensing.
(6) Stellar mass-weighted age. 
(7) Dynamical age defined as $\Pi \equiv \rm age / t_{\rm cross}$ with $t_{\rm cross}=10 \sqrt{R_{\rm eff}^{3}/(G,M_{*})}$.
(8) Total lensing magnification.
(9) Tangential lensing magnification.
}
\end{deluxetable*}

Star cluster candidates 3, 4, and 5 are very compact, with observed PSF-deconvolved effective radii (obtained from \galfit\ fitting on the F150W image) of $<1$ pixel. Correcting for tangential lensing magnification ($\mu_{\rm tan}$), this corresponds to upper limits in the intrinsic (i.e.,~delensed) size of $8$, $3$, and $5$~pc for the clusters 3, 4, and 5, respectively. These radius estimates are anchored to the individual multiple images of each cluster that exhibits the highest lensing magnification, namely 3b, 4a, and 5a. The corresponding tangential magnifications for these images are $18 \pm 2$, $43 \pm 4$, and $36 \pm 4$, respectively, with total lensing magnifications ($\mu_{\rm tot}$) of $17 \pm 2$, $44 \pm 5$, and $36 \pm 4$, respectively.

The intrinsic stellar masses of clusters 3, 4, and 5 are $8_{-3}^{+4} \times 10^{6}$, $1_{-1}^{+2} \times 10^{6}$, and $5_{-3}^{+5} \times 10^{6}\,M_{\odot}$, respectively. These values imply highly dense structures, with lower limits on the intrinsic stellar mass surface density of $\sim 214\times 10^{2}$, $292\times 10^{2}$, and $336 \times 10^{2}\,M_{\odot}\,\rm pc^{-2}$, respectively. These physical parameters are summarized in Table~\ref{tab:props_stacked_clumps}.

We fit the delensed, stacked photometry of the clusters (i.e., 3abc, 4ab, and 5bc) using \bagpipes\ with a single-burst SFH model. This simple model reproduces the observed photometry well ($\chi_{\nu}^{2}\lesssim 0.5$), as shown in Figure~\ref{fig:plot_comb_sedfits_bagpipes}. The flux excesses in F356W (driven by H$\beta$+[OIII]) and F444W (driven by H$\alpha$) are well-reproduced by the models, indicating very young stellar populations. This is consistent with the spectroscopic data presented in Figure~\ref{fig:plot_specphoto_fits}, which display strong H$\beta$, [OIII]$\lambda\,5007$, and H$\alpha$ emission lines. The inferred cluster ages are $7_{-1}^{+2}$, $6_{-2}^{+2}$, and $11_{-4}^{+13}$~Myr for clusters 3, 4, and 5, respectively. Despite these young ages, their compact sizes yield dynamical ages ($\Pi \equiv \rm age / t_{\rm cross}$) greater than unity: $6_{-3}^{+6}$, $10_{-6}^{+14}$, and $15_{-14}^{+476}$, respectively. This further supports their classification as gravitationally bound star clusters. Here, the crossing time is defined as $t_{\rm cross}=10 \sqrt{R_{\rm eff}^{3}/(GM_{*})}$, where $G$ is the gravitational constant \citep{Gieles2011}.

SED fitting with an exponentially declining SFH model with a fixed, short timescale of $\tau=1$~Myr yields stellar masses and ages that are consistent within uncertainties with the single-burst model. To test the sensitivity to dust modeling, we also implement an alternative dust attenuation prescription using the \citet{Salim2018} curve for both SFH models. The results are broadly consistent, yielding stellar masses and ages within the errors of those derived using the standard \citet{Calzetti2000} dust law, and consistently indicating low dust attenuation ($A_{V}\lesssim 0.3$~mag). A summary of the parameters derived from these alternative SED fitting configurations is provided in Table~\ref{tab:sedfit_varies_sc}. 

In addition to analyzing the stacked photometry, we perform SED fitting on the individual lensed images. The individual SED fits are presented in Figure~\ref{fig:plot_sed_bagpipes_indiv_clumps} (Appendix~\ref{sec:aper_photo_clumps_arc}), and the inferred stellar population parameters across the various model assumptions are summarized in Table~\ref{tab:sedfit_varies_sc} (Appendix~\ref{sec:sedfits_varies_clumps}). As an independent check, we also fit both the stacked and individual photometry using \piXedfit, applying a short exponentially declining SFH with $\tau$ ranging from $1-3$~Myr. These fits yield stellar masses and ages that are broadly consistent within the uncertainties with our primary \bagpipes\ results. Finally, an evaluation of the individual mirror images demonstrates that their derived properties are mutually consistent within their respective uncertainties, and compatible with the results obtained from the stacked photometry. This agreement independently validates the geometric configurations predicted by our lens model.

\begin{figure*}[ht]
\centering
\includegraphics[width=0.95\textwidth]{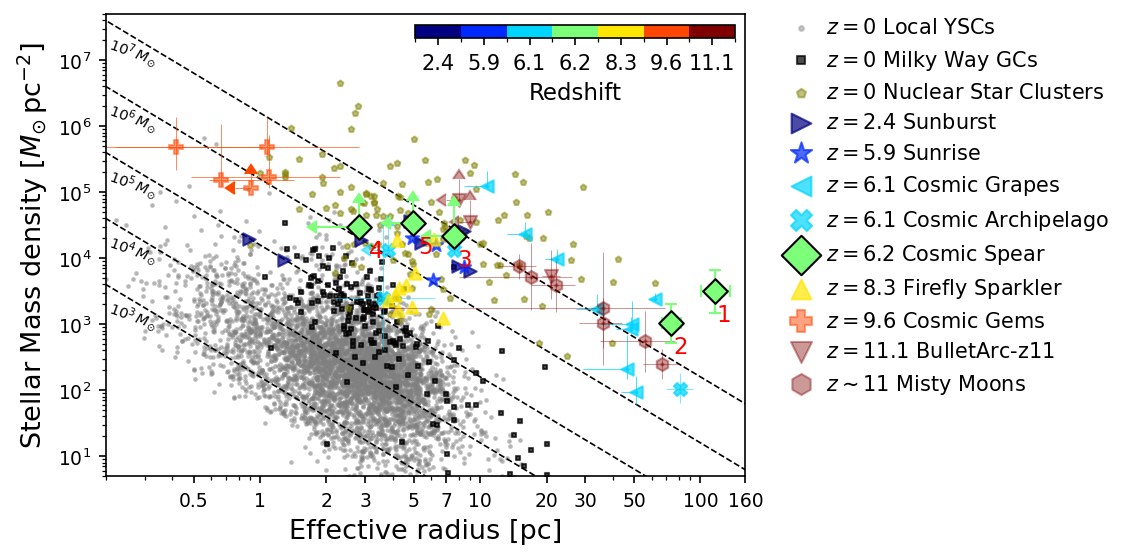}
\caption{Size (i.e.,~effective radius) and stellar mass surface density ($\Sigma_{*}$) of the clumps and star clusters in the Cosmic Spear compared with star clusters observed in galaxies at high-redshift and nearby universe: BulletArc-z11 \citep{2025Bradac}, Misty Moon \citep{Nakane2025}, Cosmic Gems \citep{Adamo2024}, Firefly Sparkler \citep{Mowla2024}, Cosmic Archipelago (combined clumps analyzed by \citealt{Messa2025a} and \citealt{Bolamperti2026}), Cosmic Grapes \citep{2025Fujimoto}, Sunrise \citep{Vanzella2023}, Sunburst \citep{Rivera-Thorsen2019}, local young star clusters \citep[YSCs;][]{Brown2021}, globular clusters in the Milky Way \citep{2018Baumgardt}, and nuclear star clusters \citep{Neumayer2020, Erwin2012, Georgiev2009}.}
\label{fig:plot_effrad_vs_mass_sd_new}
\end{figure*}

In Figure~\ref{fig:plot_effrad_vs_mass_sd_new}, we plot our star clusters on the size-density plane and put them in context with a diverse sample of star clusters from the local universe to high redshift that have been observed so far. The comparison samples include local young star clusters (YSCs) taken from \citet{Brown2021}, ancient Milky Way globular clusters from \citet{2018Baumgardt}, nuclear star clusters (NSC) from a catalog by \citet{Neumayer2020} (sourced from compiled measurements by \citealt{Erwin2012} and \citealt{Georgiev2009}), and other lensed star clusters observed at high redshifts out to $z\sim 11$, including the Sunburst \citep{Rivera-Thorsen2019}, Sunrise arc \citep{Vanzella2023}, Cosmic Grapes \citep{2025Fujimoto}, Cosmic Archipelago (combined four clumps analyzed by \citealt{Messa2025a} and \citealt{Bolamperti2026}), Firefly Sparkler \citep{Mowla2024}, Cosmic Gems \citep{Adamo2024,Bradley2025,Vanzella2025}, BulletArc-z11 \citep{2025Bradac}, and Misty Moons \citep{Nakane2025}. The Sunrise Arc has been recently spectroscopically confirmed at a redshift of z=5.9 \citep{Pascale2025}, while the Cosmic Gems is confirmed at z=9.6 \citep{Messa2025}.

The star clusters in the Cosmic Spear occupy a similar region of the size-density plane as some of the star clusters observed in other high-redshift lensed galaxies (Sunburst, Sunrise, Firefly Sparkler, Cosmic Archipelago, and BulletArc-z11). They are characterized by very compact sizes, with effective radii of only a few parsecs, and high stellar mass surface densities, ranging from $10^4$ to $10^5 \, M_{\odot} \, \mathrm{pc}^{-2}$. When compared to local populations, their surface densities are significantly higher than those of typical YSCs and globular clusters, but comparable to nuclear star clusters. It is important to notice that these clusters are intrinsically denser since we get only an upper limit to the real sizes. For similar masses, the dashed lines show that the Cosmic Spear (and Firefly Sparkler) star clusters would move toward the top left corner, where the Cosmic Gems clusters are. In relation to other high-redshift objects, our clusters fall within the general population of compact, high-density systems, sharing similar properties with other clusters observed at $z \sim 6$ (e.g., the Sunrise arc) and the Sunburst at $z = 2.4$.

\subsection{Physical Properties of the Star-forming Clumps} \label{sec:properties_clumps}

\begin{figure*}[ht]
\centering
\includegraphics[width=0.9\textwidth]{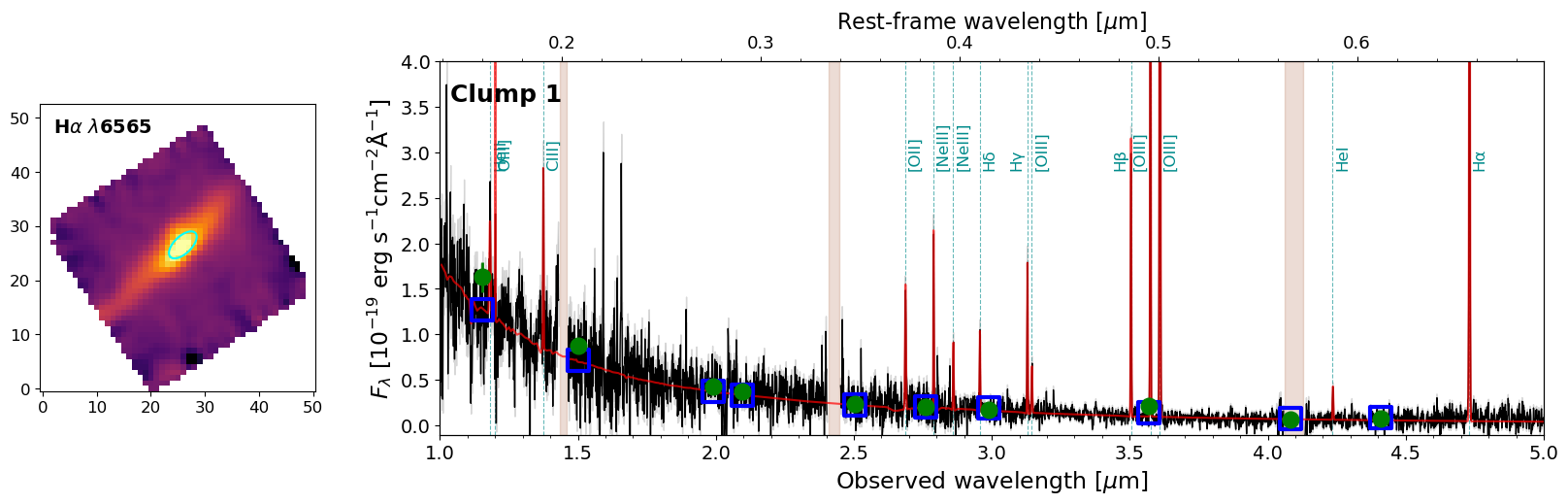}
\includegraphics[width=0.9\textwidth]{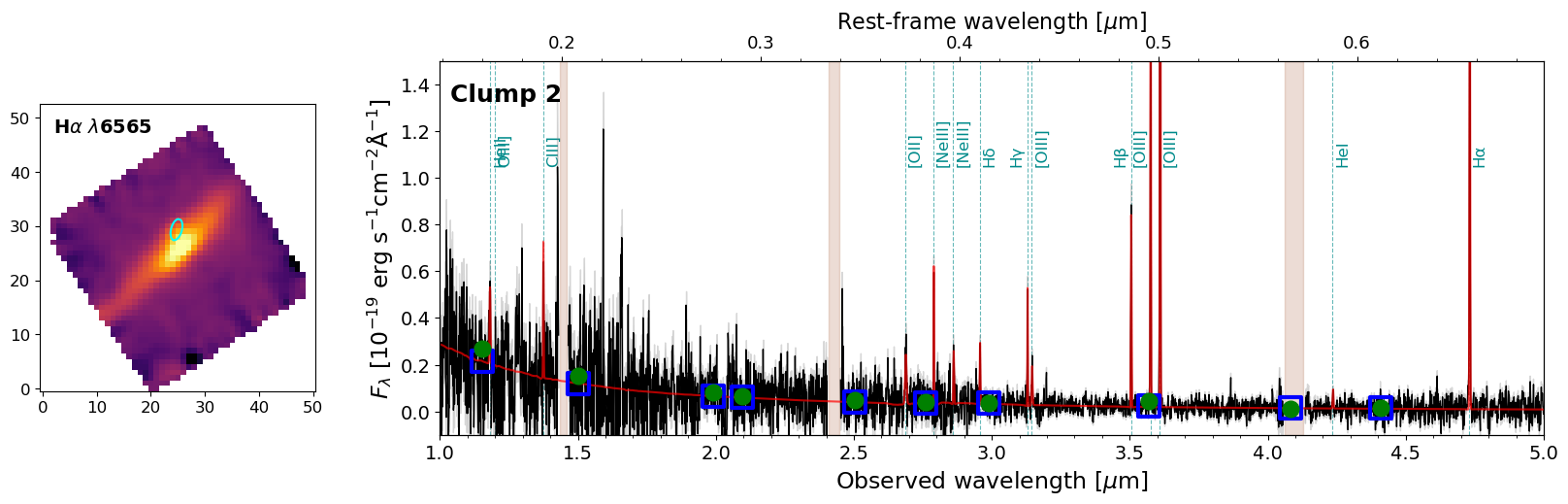}
\includegraphics[width=0.9\textwidth]{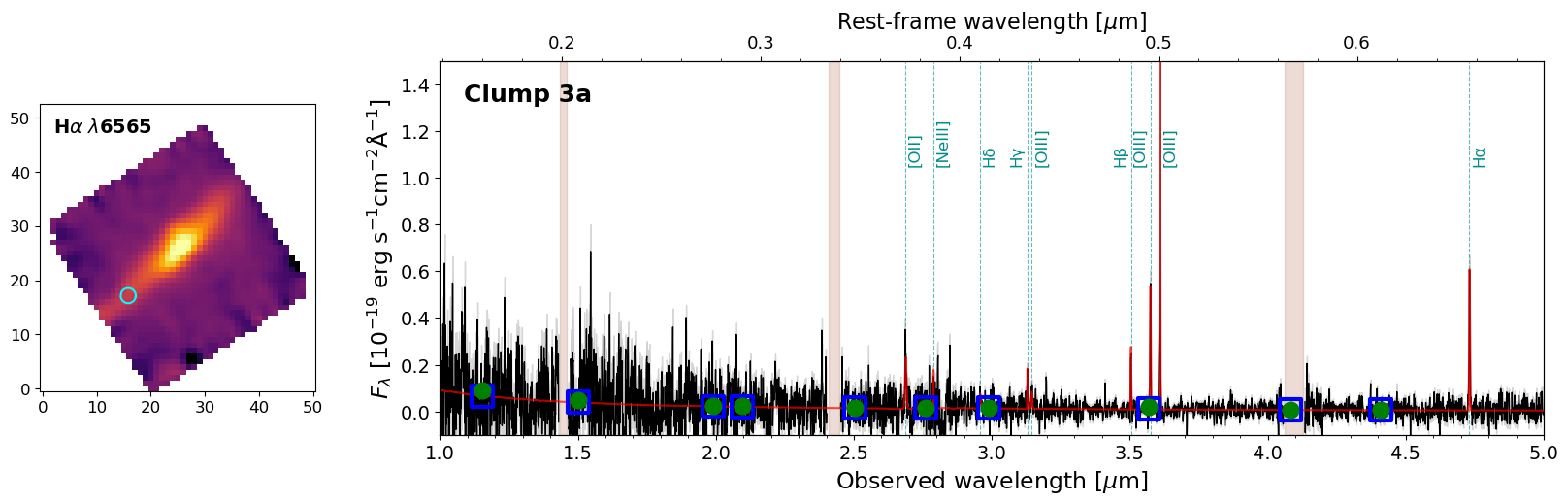}
\includegraphics[width=0.9\textwidth]{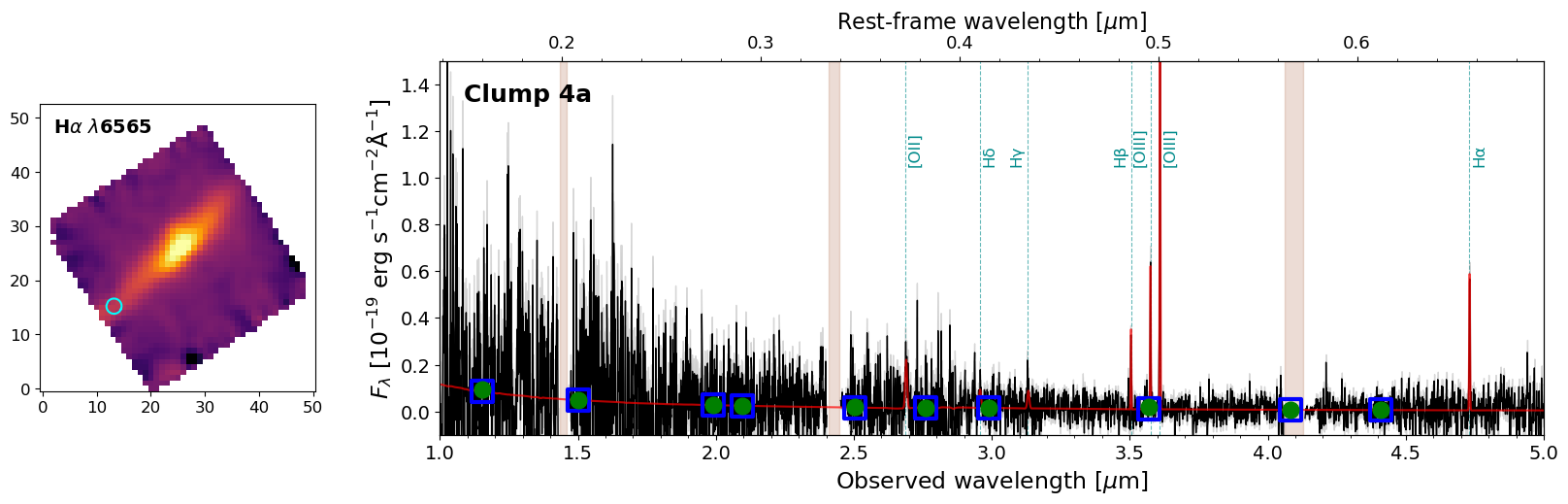}
\caption{Spectrophotometric SED fits of clumps 1 and 2, and star clusters 3a and 4a, obtained using \piXedfit. \textit{Left panel}: \Halpha\ map obtained by integrating the NIRSpec IFU data cube $\pm 15$~\AA\ around the line's wavelength. \textit{Right panel}: spectrophotometric SED fits. The black and red spectra represent the observed and best-fit model spectra, respectively, while the green circles and blue open squares show the corresponding observed and best-fit model photometry. The vertical dashed cyan lines indicate the detected emission lines: \HeIIw, \CIIIdw, \OIIdw, \NeIIIw, \NeIIIwb, \Hdelta, \Hgamma, \OIIIwa, \Hbeta, \OIIIwc, \OIIIw, HeI\,$\lambda$5877, and \Halpha. The emission lines are fitted with a single Gaussian profile. The vertical shaded regions indicate the gaps between the detector chips. The cyan ellipses and circles denote the apertures used for calculating the integrated spectrophotometric SEDs. The spectra presented here are empirical as they are measured from the reduced data cube and not corrected for lensing magnification.}
\label{fig:plot_specphoto_fits}
\end{figure*}

Clump 1 is a very bright lensed star-forming structure with an apparent F115W (i.e.,~rest-frame UV) magnitude of $\sim 23.2$, making it the brightest lensed $z>6$ system observed to date. It is magnified by a factor of $10 \pm 1$ by the foreground galaxy cluster. From SED fitting of its photometry with \bagpipes, adopting a non-parametric SFH model with the continuity prior \citep{Leja2019} and the \citet{Calzetti2000} dust attenuation law, we infer a stellar mass of $\sim 270_{-97}^{+157} \times 10^{6}\,M_{\odot}$ and a star formation rate\footnote{averaged over the last 10 Myr} of $12_{-5}^{+7}\,M_{\odot},\rm{yr}^{-1}$ (both corrected for lensing magnification). This corresponds to a specific SFR (sSFR) of $5_{-3}^{+7}\times 10^{-8}$ yr$^{-1}$, implying a short mass-doubling timescale of $22_{-13}^{+36}$ Myr. The SED fit is shown in Figure~\ref{fig:plot_comb_sedfits_bagpipes}. The best-fit model predicts strong H$\beta+$\OIIIww\ emission lines, consistent with the spectroscopic data (see Figure~\ref{fig:plot_specphoto_fits}) and indicative of young, massive stellar populations within the clump.

\begin{figure*}[ht]
\centering
\includegraphics[width=0.325\linewidth]{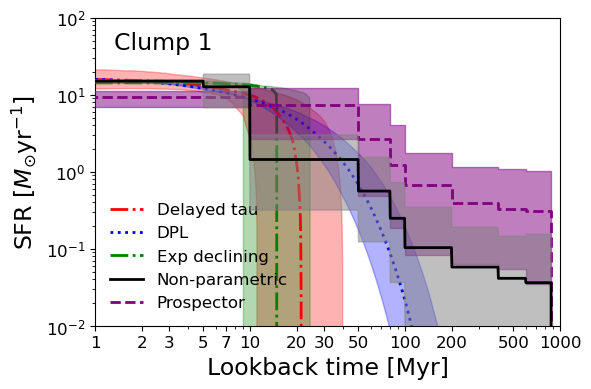}
\includegraphics[width=0.325\linewidth]{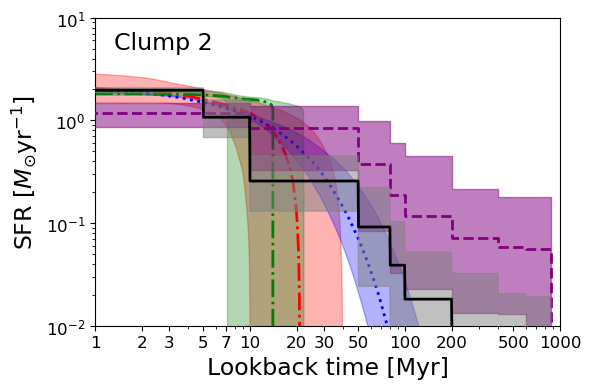}
\includegraphics[width=0.325\linewidth]{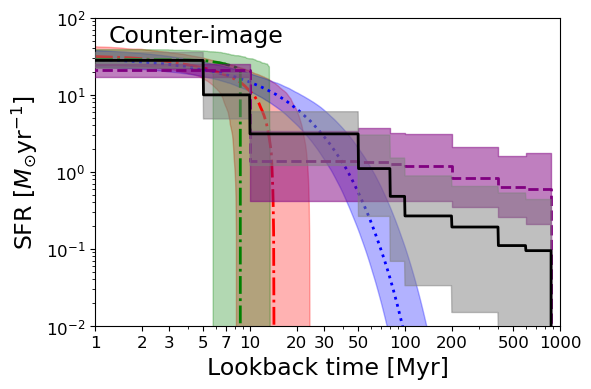}
\caption{Reconstructed SFH of the clumps 1, 2, and counter-image, obtained from fitting photometry with \bagpipes\ using multiple SFH models: non-parametric SFH with continuity prior \citep{Leja2019}, delayed tau, double-power law, and exponentially declining. In each panel, different colors represent different SFH models applied in the SED fitting, and the purple line and shaded region show the SFH inferred by \prospector\ for comparison. The SFHs are corrected for lensing magnification.}
\label{fig:plot_comp_sfh_intgsedfit_clump1n2}
\end{figure*}

The inferred non-parametric SFH (Figure~\ref{fig:plot_comp_sfh_intgsedfit_clump1n2}) suggests extended star formation with a major burst occurring in the past few Myr: half of the current stellar mass formed within the last $10_{-5}^{+28}$ Myr, indicating that the stellar population is dominated by young stars. Fits with simpler SFH models (delayed-$\tau$ or exponentially declining) yield similarly good fits with low reduced chi-square values ($\chi_{\nu}^{2}\lesssim 1$). However, these models fail to capture the possible contribution of older stars, whose light is overwhelmed by bright young stars, a phenomenon known as ``outshining'' \citep[e.g.,][]{Sorba2018,Abdurrouf2018,Gimenez-Arteaga2024,Narayanan2024,Harvey2025}. As a result, the inferred SFHs are very short, with mass-weighted ages of $\lesssim 10$ Myr. 

In contrast, the non-parametric SFH recovers the presence of older populations, yielding a mass-weighted age of $82_{-60}^{+103}$ Myr. A more flexible parametric SFH model in the form of a double power-law function can also capture slightly older stars, resulting in a mass-weighted age of $16_{-5}^{+7}$ Myr. Further details are provided in Appendix~\ref{sec:sedfits_varies_clumps} and Table~\ref{tab:sedfit_varies_clumps1n2}. Given its extended size, with an intrinsic effective radius of $117 \pm 18$ pc (corrected for tangential magnification of $\mu_{\rm tan}=10 \pm 1$), it is more plausible that its stellar population comprises stars spanning a range of ages, rather than being uniformly coeval as in star clusters. SED fitting with \prospector\ infers an SFH shape that is broadly consistent with the \bagpipes\ results, though skewed slightly more toward older stellar populations. 

The clump 1 has a stellar mass density ($\Sigma_{*}$) of $31_{-17}^{+36}\times 10^{2}\,M_{\odot}\,\text{pc}^{-2}$ and
SFR density ($\Sigma_{\rm SFR}$) of $44 \pm 7\,M_{\odot}\,\rm{yr}^{-1}\,\rm{kpc}^{-2}$. The SFR density is well above the canonical threshold of $0.1\,M_{\odot}\,\rm{yr}^{-1}\,\rm{kpc}^{-2}$ for a system capable of driving an outflow \citep[e.g.,][]{2002Heckman,2021Prusinski} and producing abundant ionizing photons (X. Xu et al. in prep). 

We also fit the integrated photometry of clump 2 with \bagpipes, adopting a non-parametric SFH model. This choice is motivated by its relatively extended size, with an effective radius of $73 \pm 8$ pc (corrected for lensing with $\mu_{\rm tan}=11 \pm 1$). We also test several alternative SFH models (Figure~\ref{fig:plot_comp_sfh_intgsedfit_clump1n2}). Similar to clump 1, the simple models capture only the young stellar populations and miss the contribution of older stars, resulting in a very young mass-weighted age of $\lesssim 10$ Myr. In contrast, the non-parametric model recovers star formation episodes extending further back in time, yielding an older mass-weighted age of $83_{-55}^{+107}$ Myr, quite similar to clump 1. Overall, the recovered SFH of clump 2 closely resembles that of clump 1, with half of its current stellar mass formed in the last $23_{-16}^{+22}$ Myr, indicating that the two clumps host similar stellar populations. The SFH inferred by \prospector\ shows a broadly consistent picture, though skewed slightly more toward older stellar populations. 

The spectrophotometric SED fits of the central region of clumps 1 and 2 are shown in the first and second rows in Figure~\ref{fig:plot_specphoto_fits}. In each SED plot, the best-fit SED model obtained from fitting using \piXedfit\ is shown with the red line (spectrum) and blue squares (photometry). The left panel shows the aperture with which the spectrophotometric SED was integrated (see Section~\ref{sec:clump_photometry} for the aperture's geometry).  Overall, we detect 13 emission lines in clumps 1 and 2: \HeIIw, \CIIIdw, \OIIdw, \NeIIIw, \NeIIIwb, \Hdelta, \Hgamma, \OIIIwa, \Hbeta, \OIIIwc, \OIIIw, HeI\,$\lambda$5877, and \Halpha, indicated by the dashed cyan lines.
A comprehensive spatially resolved study of the interstellar medium (ISM) conditions inferred from these emission lines will be performed, spatially resolved study of the interstellar medium (ISM) conditions, inferred from these emission lines, will be presented in N. Roy et al. (in prep.).

\begin{figure*}[ht]
\centering
\includegraphics[width=0.6\textwidth]{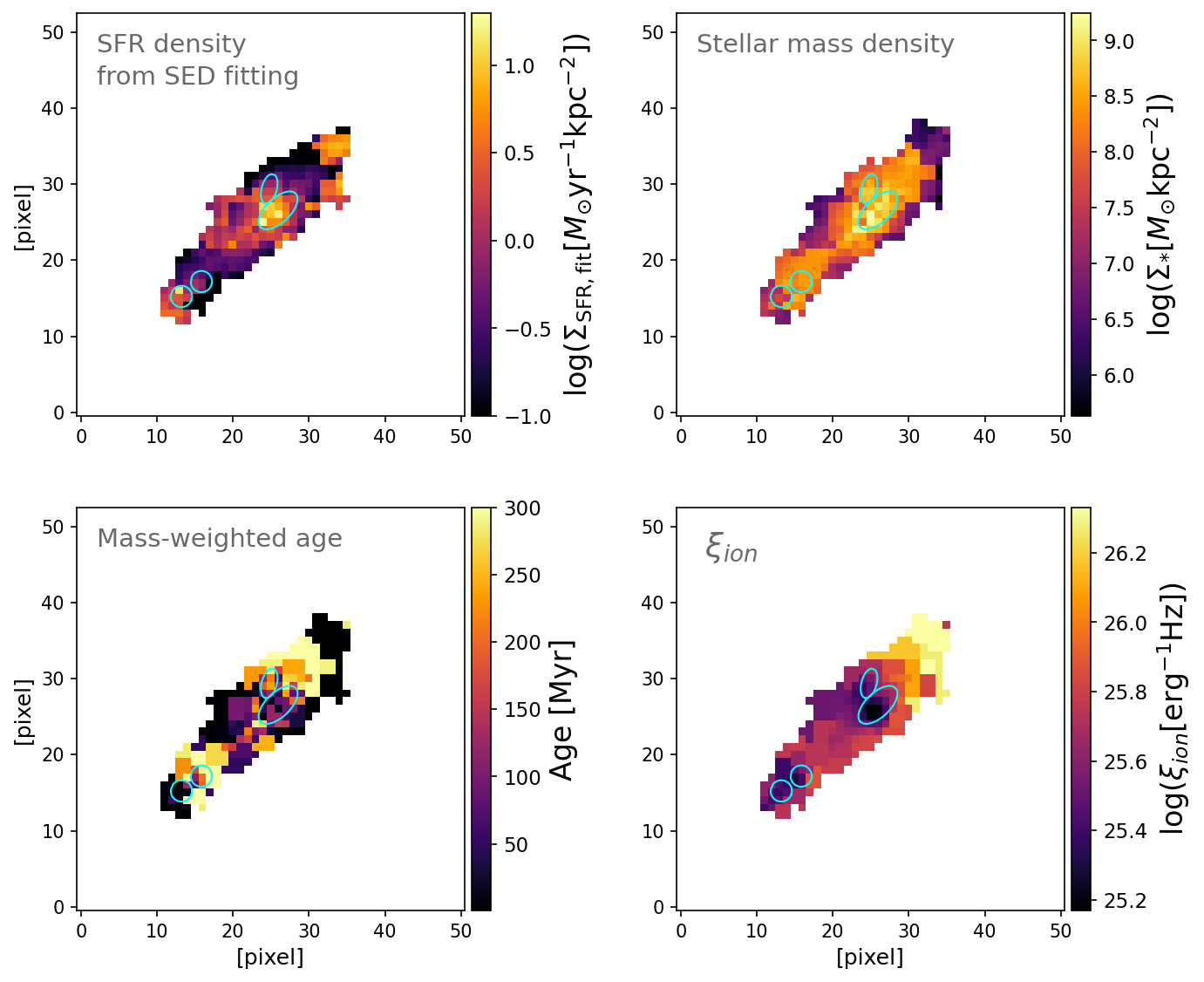}
\caption{Physical property maps of the star-forming clumps (1 and 2) and part of the arc that is covered within the NIRSpec IFU field-of-view (FoV), including the star cluster images of 3a and 4a. The apertures of the clumps 1, 2, 3a, and 4a are shown by the cyan ellipses and circles. The stellar population properties are derived directly from the resolved spectrophotometric SED fitting with \piXedfit, while the ionizing photon production efficiency ($\xi_{\rm ion}$; bottom right panel) is derived from the \Halpha\ and UV luminosity (see Section~\ref{sec:xi_ion}). The cyan apertures are identical to those used to measure the photometry of clumps 1 and 2, as well as cluster images 3a and 4a (see Figure~\ref{fig:plot_arc_sed_apertures} in Appendix~\ref{sec:aper_photo_clumps_arc}).}
\label{fig:maps_resolved_properties}
\end{figure*}

To further characterize star-forming clumps 1 and 2, we performed spatially resolved spectrophotometric SED fitting using \piXedfit\ (described in Section~\ref{sec:resolved_sedfit}). The resulting maps of their stellar population properties, including SFR density ($\Sigma_{\rm SFR}$), stellar mass density ($\Sigma_{*}$), and mass-weighted age, are presented in Figure~\ref{fig:maps_resolved_properties}. The analysis reveals that both the stellar mass and SFR are centrally concentrated. The stellar mass distribution peaks centrally at $\sim 2\times 10^{9}\,M_{\odot}\,\rm{kpc}^{-2}$. The SFR map shows a similarly compact, centrally peaked morphology, reaching a maximum $\Sigma_{\rm SFR}$ of $\sim 20\,M_{\odot}\,\rm{yr}^{-1}\,\rm{kpc}^{-2}$, indicating an intensely star-forming core. These surface densities are not corrected for gravitational lensing, as this effect is expected to preserve surface densities.

The fitting process also measures nebular emission line fluxes, which will be analyzed in detail in a forthcoming paper (N. Roy et al., in prep.). From the resolved H$\alpha/$H$\beta$ map, we infer very low dust attenuation throughout the clump regions, consistent with a dust-free environment as has been indicated by the non-detection of dust continuum around the clumps \citep{Fudamoto2024}. The only exception is the central region of clump 1, which exhibits slight dust attenuation reaching $A_{V}\sim 0.15$ mag. 

The mass-weighted age map reveals spatial variations across the clump regions. Such details are difficult to capture when fitting the integrated SED due to the outshining effect of young stellar populations within the clumps. The central pixel, which corresponds to the highest SFR density, hosts a young stellar population with a mass-weighted age of $\sim 4$ Myr. Moving outward to a radial distance of 1 kpc, the stellar age increases, yielding a median age of $\sim 100$ Myr within this aperture.

\subsection{Global Properties of the Counter-image} \label{sec:host_counter_properties}

Leveraging the gravitational lens model, we identify a counter-image (hereafter called CI) of the Cosmic Spear at R.A.$=$03:08:57.2 and decl.$=$26:45:48.2. Our lens model estimates modest magnifications ($\mu_{\rm tot} = 3.0 \pm 0.3$ and $\mu_{\rm tan} = 2.5 \pm 0.3$) for this CI. Despite its low magnification, CI has an apparent F115W magnitude of 24.6, bright enough to be detected substantially at S/N$>20$ (with an average of 48) across filters redward of F090W. We measure its photometric redshift using \textsc{EAZY} \citep{Brammer2008} and \bagpipes, finding consistent redshifts within uncertainties: $6.12 \pm 0.17$ and ${6.28}_{-0.13}^{+0.14}$, respectively (see Appendix~\ref{sec:photo_z_counter_image}), which align with the Cosmic Spear's redshift. Its SED shape is similar to those of clumps 1 and 2, showing significant flux excess in F356W and F444W due to H$\beta$+[OIII] and H$\alpha$, respectively. These results provide supporting evidence that the object is a counter-image of the Cosmic Spear.  

Because the CI is only moderately magnified and lacks significant shear distortion, its morphology and integrated photometry can serve as a robust representation of the entire galaxy system. We therefore use it to infer the galaxy's global physical properties. Deriving intrinsic global (i.e.,~integrated) properties, such as stellar mass and SFR, from the strongly lensed Cosmic Spear itself is difficult due to the significantly varying magnification across the arc, compounded by the critical curve intersecting the arc (which creates mirror symmetry). While the properties of the individual star clusters and clumps have been successfully derived, determining the intrinsic properties of the underlying diffuse arc is difficult owing to the differential magnification across the region. 

The CI appears with a clumpy morphology where three sub-components are clearly distinguishable (labeled C1, C2, and C3 in the top-left inset of Figure~\ref{fig:macs0308-zD1_img}). Interestingly, while the CI cleanly mirrors the major star-forming clumps (1 and 2), the compact parsec-scale knots seen in the main Spear tail are absent in the CI, confirming that the tail features are not global structures but are individual young stellar associations caught in a zone of extreme local magnification ($\mu \sim 15$--$43$) due to critical line crossings.

\begin{figure}[ht]
\includegraphics[width=1.0\linewidth]{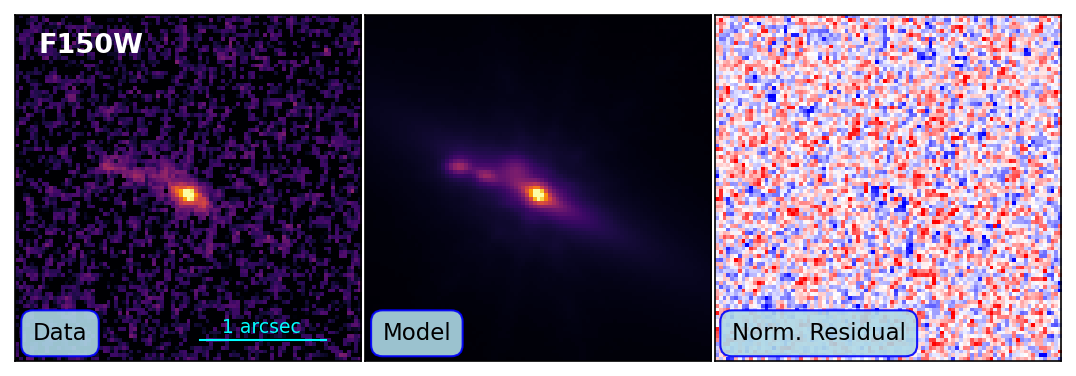}
\caption{Morphological modeling of the counter-image using \galfit. The fitting is performed on the F150W image. The main clump component (C1) has a delensed effective radius of $483 \pm 97$ pc (corrected for lensing tangential magnification of $\mu_{\rm tan}=2.5\pm 0.3$).}
\label{fig:host_galfit_fits}
\end{figure}

To measure the intrinsic size of CI, we model the F150W image of the CI using \galfit\ with a single S\'ersic profile for the brightest clump (C1) and a Gaussian for the other two small clumps (C2 and C3). The result of this fit is shown in Figure~\ref{fig:host_galfit_fits}. The estimated delensed effective radius of CI is $483 \pm 97$ pc. This size is significantly larger than the individual star-forming clumps analyzed previously, consistent with the CI representing the entire host galaxy.

To derive the integrated properties of CI, we first extract its integrated SED through image processing with \piXedfit, which includes PSF matching. This integrated SED is obtained by summing the flux of all pixels within the galaxy's segmentation map (i.e., the region of interest, RoI). We then fit the integrated SED using \bagpipes\ to infer the global stellar population properties. We fix the redshift to $z=6.2$, matching that of the lensed arc system. To account for a potentially complex stellar population, our fiducial fit applies a non-parametric SFH. As shown in the bottom-right panel of Figure~\ref{fig:plot_comb_sedfits_bagpipes}, this model provides an excellent fit to the data. The inferred SFH (right panel in Figure~\ref{fig:plot_comp_sfh_intgsedfit_clump1n2}) is extended over several hundred million years, yielding a mass-weighted age of $115_{-83}^{+123}$ Myr and a delensed total stellar mass of $4.4_{-1.5}^{+2.3}\times 10^{8}\,M_{\odot}$, slightly higher than the total stellar mass of clumps 1, 2, and star clusters.

For comparison, we also fit simpler SFH models in the forms of delayed-tau and exponentially declining, which consistently yield very young ages of $\sim 5$ Myr (see Table~\ref{tab:sedfit_varies_counterimg}). This emphasizes the ``outshining'' effect. Finally, the SED fitting infers a low dust attenuation, with $A_{V} = 0.17 \pm 0.07$ mag, consistent with that of Clumps 1 and 2 (see Table~\ref{tab:sedfit_varies_clumps1n2}). These results show a young, rapidly assembling galaxy with low dust content in the early universe (around the end of the reionization epoch). 

\section{Discussions} \label{sec:discuss}

\subsection{Clustered Formation Efficiency} \label{sec:cfe}

Having estimated the global properties of the host galaxy through analysis of the CI (see Section~\ref{sec:host_counter_properties}), we compute the Clustered Formation Efficiency (CdFE), defined as the ratio of stellar mass formed in clustered aggregates (i.e.,~star clusters and clumps) to the total stellar mass formed in the host galaxy over the same period. While the precise cluster formation efficiency (CFE) is difficult to constrain for our lensed galaxy system due to potentially unresolved individual star clusters within clumps 1 and 2, the CdFE serves as a robust proxy for the overall efficiency of clustered stellar aggregate formation.

We restrict this estimation to a lookback timescale of $20$ Myr, a window within which star clusters 3, 4, and 5 safely formed. Using the star formation histories of the individual clumps (1 and 2) and the integrated host galaxy (CI), we calculate the respective stellar masses formed within this timescale (see Figure~\ref{fig:plot_comp_sfh_intgsedfit_clump1n2}). The stellar mass formed in the host galaxy during this period is estimated to be $2.2_{-0.7}^{+1.1} \times 10^{8}\,M_{\odot}$. In comparison, the total stellar mass formed in clustered aggregates is $1.9_{-0.8}^{+1.2} \times 10^{8}\,M_{\odot}$, which comprises $1.5_{-0.7}^{+1.0} \times 10^{8}\,M_{\odot}$ from Clump 1, $1.8_{-0.6}^{+1.0} \times 10^{7}\,M_{\odot}$ from Clump 2, and an additional $1.4_{-0.6}^{+1.1} \times 10^{7}\,M_{\odot}$ from the combined mass of the star clusters. This yields a high clustered formation efficiency of $\text{CdFE} = 86_{-11}^{+8}\%$, indicating that star formation in this high-redshift galaxy is overwhelmingly dominated by, and concentrated within, massive stellar aggregates.

Additionally, we estimate that the stellar mass currently locked in the clustered aggregates accounts for $72_{-2}^{+3} \%$ of the galaxy's total stellar mass. For comparison, if we do not use the CI and instead use the observed (i.e.,~no correction for lensing) total stellar mass of all components in the Cosmic Spear, including the lensed images of the three star clusters, two clumps (1 and 2), and the diffuse arc of the host (i.e.,~tail), we obtain $56_{-2}^{+3}\%$ mass locked in the clustered aggregates. From fitting the integrated SED of the diffuse arc, we obtain $M_{*}=3_{-1}^{+2}\times 10^{9}\,M_{\odot}$ (not corrected for lensing magnification; see Appendix~\ref{sec:aper_photo_clumps_arc}).

\subsection{The Nature and Evolutionary Fate of the Star Clusters in the Cosmic Spear} \label{sec:formation_protogc}

The physical properties of the Cosmic Spear's star clusters, specifically their compact sizes ($R_{\rm{eff}} \lesssim 8$ pc), high stellar mass surface densities ($\Sigma_{*} \gtrsim 2 \times 10^{4}\,M_{\odot}\,\rm{pc}^{-2}$), and dynamical ages $\Pi >> 1$, are consistent with gravitationally bound systems massive enough to possibly survive long-term evolution. The formation of such dense stellar systems is thought to be a more prevalent mode of star formation in the high-pressure, turbulent, and gas-rich environments characteristic of early galaxies \citep[e.g.,][]{Krumholz2019}. This has been confirmed by the results of high-resolution cosmological simulations of star formation in galaxies at redshifts $z>6$ \citep{Garcia2025}. While these properties provide evidence that we are witnessing the formation of bound clusters, their substantially high masses ($\log(M_{*}/M_{\odot}) \sim 6-7$) suggest their evolutionary fates may be bifurcated, potentially evolving into either ancient GCs or the seeds of a nuclear star cluster (NSC). 

The formation epoch of the star clusters in the Cosmic Spear provides a critical constraint for theoretical models of globular cluster assembly. A recent comparison of six different GC formation models by \citet{Valenzuela2025} reveals that while most models predict the peak formation of surviving GCs to occur at $z \sim 1-3$, there is significant disagreement regarding the contribution of high-redshift ($z > 6$) star formation to the present-day GC population. Our detection of massive, bound clusters at $z=6.2$ is consistent with ``optimistic'' models that incorporate rapid mass growth or specific formation pathways in dwarf galaxies, such as those by \citet{Valenzuela2024} and \citet{Chen2024}. These models predict a significant population of surviving GCs forming at $z=6-10$ and even up to $z \sim 12$. In contrast, models that produce younger overall populations \citep[e.g.,~][]{DeLucia2024,Reina-Campos2022} predict fewer surviving GCs originating from this epoch. The existence of the Cosmic Spear clusters adds to the growing census of gravitationally bound structures detected during the Epoch of Reionization. As argued by \citet{Valenzuela2025}, the detection of such systems strongly supports scenarios where a substantial fraction of the oldest GCs in Milky Way-mass galaxies were assembled during the first billion years of cosmic history.

The total stellar mass locked in the Cosmic Spear's clusters represents a significant fraction of the host galaxy's mass (see Section~\ref{sec:cfe}). In the Milky Way, the total mass of metal-poor GCs constitutes about 2\% of the stellar halo mass. On the other hand, massive GCs in a low-mass dwarf galaxy like Fornax account for 30-50\% of the total stellar mass \citep{Larsen2012}. The high fraction of clustered mass observed in the Cosmic Spear is therefore substantial and falls well within the expected range for star cluster formation during the primary phase of galaxy assembly. This indicates that clustered star formation was a key contributor to the stellar mass buildup of this galaxy during the epoch of reionization \citep{Ricotti2002, Katz2013}. With cluster and clump ages of $\sim 6-100$ Myr at $z=6.20$ (Table 1), we can place the Cosmic Spear's cluster formation epochs at $z \sim 6.24$--$6.8$. This timeline coincides with the conclusion of cosmic reionization, implying that these young, massive aggregates belonged to the stellar populations that actively sustained and completed reionization at $z \gtrsim 6$.

However, the substantially high masses of these clusters warrant a discussion on their dynamical survival. While their high densities ($>$ Milky Way GCs; see Fig.~\ref{fig:plot_effrad_vs_mass_sd_new}) can be explained by formation in extreme high-$z$ conditions \citep[extreme star-forming conditions in the early universe, such as more compact galaxies, harder ionizing radiation fields, and higher electron densities; e.g.,~][]{2024Morishita,2024Roberts-Borsani,Abdurrouf2024}, their location in the size-density plane places them in a regime where dynamical friction becomes highly efficient. The timescale for a cluster to spiral into the galaxy center due to dynamical friction scales inversely with mass ($t_{\rm{df}} \propto M_{*}^{-1}$). For clusters with $M_{*} \gtrsim 10^6 M_{\odot}$ residing in the inner regions of a host galaxy, this migration can occur within a few Gyrs \citep[e.g.,][]{Tremaine1975, Capuzzo-Dolcetta2008}. This suggests that the most massive star clusters in the Cosmic Spear may migrate inward and coalesce to form a Nuclear Star Cluster (NSC) or contribute to the bulge.

Conversely, two mechanisms may allow these massive clusters to survive orbital decay and evolve into the GCs we observe today. First, impulsive scattering events or tidal interactions during hierarchical merging could eject these clusters from the inner disk into the galaxy's halo, where the dynamical friction timescale is significantly longer \citep{Kruijssen2012}. Second, internal stellar and dynamical evolution play a critical dual role. This evolution over a Hubble time is driven by internal processes like stellar-evolutionary mass loss, N-body relaxation, and stellar black hole dynamics \citep[e.g.,][]{2010Gieles,2020Antonini}. The adiabatic expansion driven by the mass loss not only increases the cluster radius, bringing the densities roughly in line with local massive GCs (a factor of $\sim 8$ decrease for 50\% mass loss; \citealt{Gieles2011}), but also reduces the total cluster mass. This mass loss increases the dynamical friction timescale ($t_{\rm{df}}$), effectively ``stalling'' the inspiral and allowing the cluster to survive as an independent system in the galaxy's potential.

\subsection{Ionizing Photon Production Efficiency} \label{sec:xi_ion}

We investigate the ionizing properties of the star-forming clumps in the Cosmic Spear by calculating the ionizing photon production efficiency, $\xi_{\rm{ion}}$. This quantity is defined as the production rate of hydrogen-ionizing photons, $\dot{N}_{\rm{ion}}$, per unit of UV continuum luminosity, $L_{\rm{UV}}$ (i.e., $\xi_{\rm{ion}} = \dot{N}_{\rm{ion}} / L_{\rm{UV}}$). The $\dot{N}_{\rm{ion}}$ is inferred from the dust-corrected H$\alpha$ luminosity using the prescription from \citet{Leitherer1995}: $L({\rm H}\alpha)[\text{erg s}^{-1}] = 1.36 \times 10^{-12} \dot{N}_{\rm{ion}}[\text{s}^{-1}]$. This conversion was derived assuming case B recombination and that no ionizing photons escape the system ($f_{\rm{esc}} = 0$). If a fraction of ionizing photons escapes the region before they can produce recombination events, the observed H$\alpha$ flux will underestimate the true intrinsic production rate. Consequently, the H$\alpha$-derived $\xi_{\rm{ion}}$ should be interpreted as a lower limit. We calculate UV luminosity from the F115W flux, corresponding to a rest-frame wavelength of $\sim 1500$ \AA.

We first estimate this quantity for individual spatial bins in our spectrophotometric data cube (see Section~\ref{sec:generate_datacubes}). The resulting $\xi_{\rm{ion}}$ map, presented in the bottom right panel of Figure~\ref{fig:maps_resolved_properties}, reveals values ranging from $\log(\xi_{\rm{ion}}[\rm{erg}^{-1}\,\rm{Hz}]) = 25.2$ to $26.3$ with a median of $25.7$. We note that the high values ($>26$) on the northwestern side of the clump are less reliable due to low S/N ($\lesssim 2$). Across the map, the fractional uncertainty on $\xi_{\rm{ion}}$ ranges between $3\%$ and $57\%$. Additionally, we calculate this quantity for the specific clumps (1 and 2) and star clusters (3a and 4a) falling within the IFU FoV, utilizing their spectrophotometric SEDs (see Figure~\ref{fig:plot_specphoto_fits}). We obtain $\log(\xi_{\rm{ion}}[\rm{erg}^{-1}\,\rm{Hz}])$ values of $25.4 \pm 0.1$, $25.5 \pm 0.1$, $25.5 \pm 0.1$, and $25.4 \pm 0.1$ for clumps 1, 2, 3a, and 4a, respectively. These results are consistent with the large body of literature that measured this quantity for galaxies at $z>4$, which finds values in the range of $\sim 24.5 - 26.0$ \citep[e.g.,][]{Robertson2022,Vanzella2023,Llerena2025}. For instance, the intensely star-forming regions within the Sunrise Arc at $z=5.9$ show a high efficiency of $\log(\xi_{\rm{ion}}) \sim 25.7$ \citep{Vanzella2023}.

The $\xi_{\rm{ion}}$ map reveals that the core of Clump 1 exhibits slightly lower values than the surrounding region, despite having intense star formation (as indicated by the SFR map; see Section~\ref{sec:properties_clumps}). This discrepancy may indicate a higher escape fraction of Lyman continuum (LyC) photons in the central region, rather than an intrinsically inefficient production of ionizing photons. We estimate an escape fraction excess of $0.4\pm 0.1$ relative to the outer surrounding region, derived by anchoring the expected \Halpha\ luminosity to the $\text{H}\alpha\text{--UV}$ luminosity correlation of spatial bins in the outer disk.

This interpretation is physically well-motivated; the high star-formation rate in the clump is expected to drive powerful radiative and mechanical feedback, creating porous pathways in the surrounding ISM. Such channels would allow LyC photons to escape unimpeded into the intergalactic medium. \citet{Fudamoto2024} reported evidence for strong outflows based on multiple velocity components and a clear velocity shift in the [CII] $158\mu$m emission line. A thorough analysis of the spatially resolved UV-to-optical emission-line maps to characterize these outflows and ISM conditions will be presented in forthcoming papers (X. Xu et al., in prep.; Roy et al., in prep.).

\section{Conclusions} \label{sec:conclusions}

We conducted a spatially resolved analysis of stellar population properties in a lensed galaxy dubbed the Cosmic Spear at $z=6.2$ using a combination of JWST/NIRCam imaging in 11 filters and NIRSpec high-resolution IFU spectroscopy with three gratings (G140H/F100LP, G235H/F170LP, and G395H/F290LP), spanning the rest-frame UV to optical. Through morphological modeling and spatially resolved SED fitting of the galaxy's components, we studied resolved stellar populations of its distinct components on scales of a few parsecs. Our main findings are as follows:

\begin{enumerate}
    \item The galaxy's structure is resolved into two main components: a pair of large, bright star-forming clumps (1 and 2) that constitute the bulk of the stellar mass, and an arc containing three distinct, compact star clusters (3, 4, and 5) that are multiply imaged along the arc.

    \item The three star clusters are extremely compact and dense, with intrinsic effective radii of $R_{\rm{eff}} \lesssim 8$ pc and stellar mass surface densities of $\Sigma_{*} \gtrsim 2\times 10^{4}\,M_{\odot}\,\rm{pc}^{-2}$. With stellar masses of $M_{*} \sim 1-8 \times 10^{6}\,M_{\odot}$, very young ages of $\lesssim 11$ Myr, and dynamical ages $\Pi >> 1$, indicating they are gravitationally bound systems. While their properties identify them as candidate proto-globular clusters, their extreme masses suggest a bifurcated evolutionary fate: they may survive as GCs through adiabatic expansion and mass loss, or spiral inward via dynamical friction to form the seed of a Nuclear Star Cluster.

    \item The spatially resolved analysis of the main star-forming clumps reveals a very compact and intensely star-forming core with the central stellar density of $\sim 2\times 10^{9}\,M_{\odot}\,\rm{kpc}^{-2}$ and central SFR density of $\sim 20\,M_{\odot}\,\rm{yr}^{-1}\,\rm{kpc}^{-2}$.

    \item Analysis of a modestly magnified, and less distorted counter-image reveals the estimated global properties of the host galaxy, yielding a delensed stellar mass of $M_{*}=4.4_{-1.5}^{+2.3} \times 10^{8}\,M_{\odot}$ and a mass-weighted age of $115_{-83}^{+123}$ Myr. The total mass in the clustered aggregates (the two clumps and three star clusters) accounts for $\sim 72\%$ of the host's total stellar mass, with a global clustered formation efficiency estimated to be $\sim 86\%$.

    \item We derive spatially resolved ionizing photon production efficiencies of $\log(\xi_{\rm{ion}}[\rm{erg}^{-1}\rm{Hz}]) \sim 25.2 - 26.3$ across the brightest-clump region, consistent with typical values at $z>4$. Notably, the core of Clump 1 exhibits reduced $\xi_{\rm{ion}}$ despite intense star formation. We interpret this as an indication of a locally elevated Lyman-continuum escape fraction facilitated by feedback-driven channels in the ISM. We estimate an escape fraction excess of $0.4 \pm 0.1$ relative to the outer surrounding region. 
    
\end{enumerate}

Ultimately, this study demonstrates the great synergy between JWST and gravitational lensing of galaxy clusters to resolve parsec-scale structures within galaxies in the early Universe. These findings open a new observational window for detailed statistical studies of young star clusters, providing crucial constraints on the formation scenarios of the ancient GCs and NSCs we observe in local galaxies today.

\begin{acknowledgments}

We thank the anonymous referee for the valuable comments that helped to improve our manuscript. This work is based on observations made with the NASA/ESA/CSA James Webb Space Telescope. The data were obtained from the Mikulski Archive for Space Telescopes at the Space Telescope Science Institute, which is operated by the Association of Universities for Research in Astronomy, Inc., under NASA contract NAS 5-03127 for JWST. These observations are associated with program GO 6882 and GO 5293. The JWST and HST data presented in this article were obtained from the Mikulski Archive for Space Telescopes (MAST) at the Space Telescope Science Institute. The specific observations analyzed can be accessed via \dataset[doi: 10.17909/pxsv-0758]{https://doi.org/10.17909/pxsv-0758} and \dataset[doi: 10.17909/k9b4-sb85]{https://doi.org/10.17909/k9b4-sb85}. The authors acknowledge the VENUS team for developing their observing program with a zero-exclusive-access period. The authors acknowledge the use of the Canadian Advanced Network for Astronomy Research (CANFAR) Science Platform operated by the Canadian Astronomy Data Center (CADC) and the Digital Research Alliance of Canada (DRAC), with support from the National Research Council of Canada (NRC), the Canadian Space Agency (CSA), CANARIE, and the Canadian Foundation for Innovation (CFI). Abdurro'uf acknowledges support from the Sullivan Prize Fellowship from the Department of Astronomy at IU. PD warmly acknowledges support from an NSERC discovery grant (RGPIN-2025-06182). MB acknowledges support from the ERC Grant FIRSTLIGHT and Slovenian national research agency ARIS through grants N1-0238 and P1-0188. YF acknowledges support from JSPS KAKENHI Grant Numbers JP22K21349 and JP23K13149. RAW acknowledges support from NASA JWST Interdisciplinary Scientist grants NAG5-12460, NNX14AN10G and 80NSSC18K0200 from GSFC. RA acknowledges support of Grant PID2023-147386NB-I00 funded by MICIU/AEI/10.13039/501100011033 and by ERDF/EU, and the Severo Ochoa award to the IAA-CSIC CEX2021-001131-S. EV and MM acknowledge financial support through grants INAF GO Grant 2022 ``The revolution is around the corner: JWST will probe globular cluster precursors and Population III stellar clusters at cosmic dawn'', INAF GO Grant 2024 ``Mapping Star Cluster Feedback in a Galaxy 450 Myr after the Big Bang'' and by the European Union – NextGenerationEU within PRIN 2022 project n.20229YBSAN - Globular clusters in cosmological simulations and lensed fields: from their birth to the present epoch. AZ acknowledges support by the Israel Science Foundation Grant No. 864/23.
We acknowledge the support of the Canadian Space Agency (CSA) [25JWGO4A18].

\end{acknowledgments}

\begin{contribution}

All authors contributed to the work that resulted in this paper.


\end{contribution}

%
\facilities{JWST, HST}

\software{\bagpipes\ \citep{Carnall2018,Carnall2019}, 
        \prospector\ \citep{Johnson2021},
        \piXedfit\ \citep{Abdurrouf2021,Abdurrouf2022},
        \textsc{EAZY} \citep{Brammer2008},
        \texttt{photutils} \citep{larry_bradley_2024_13989456}
    }


\appendix

\section{Aperture Photometry and individual SED fits of the clumps and the diffuse arc}
\label{sec:aper_photo_clumps_arc}

\begin{figure*}[ht]
\centering
\includegraphics[width=0.9\linewidth]{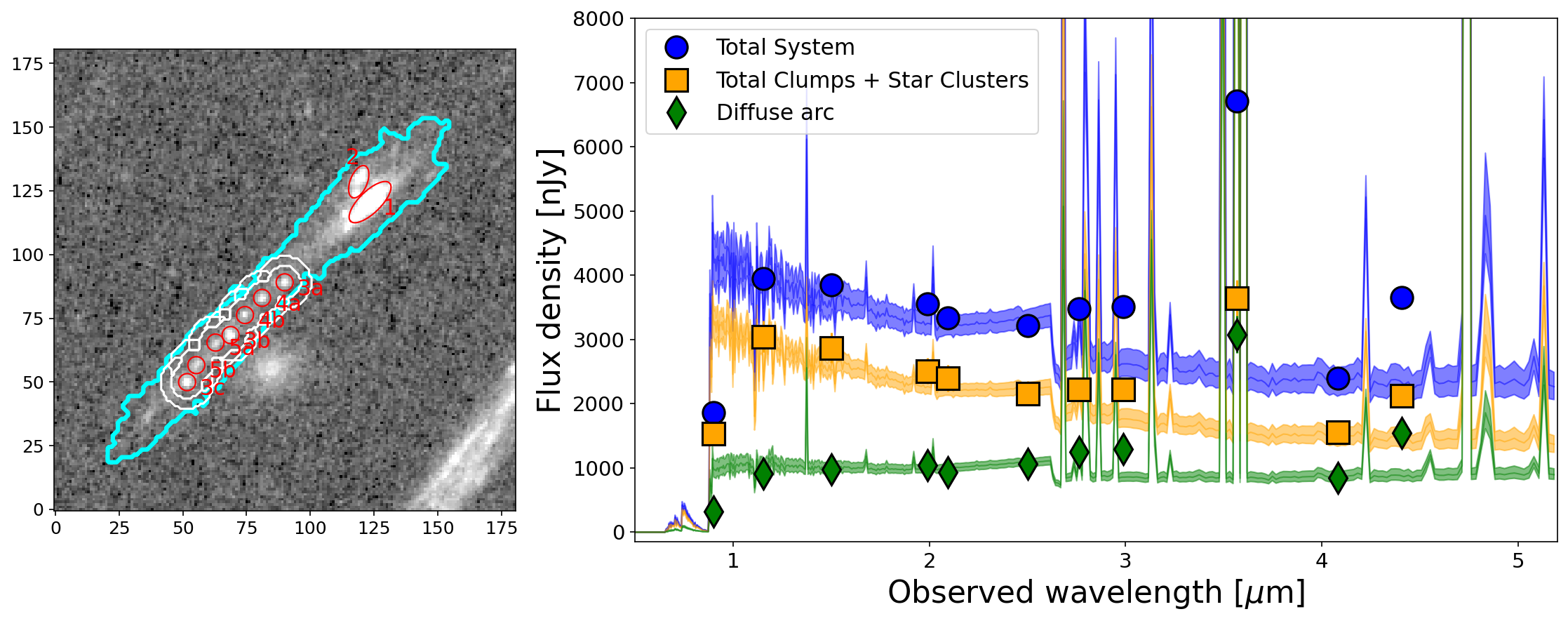}
\caption{Aperture photometry of the clumps (1 and 2), star cluster images, and the surrounding diffuse arc. The left panel shows the apertures used for measuring the photometry (red circles and ellipses) and the background flux from the host's diffuse arc. The right panel shows the derived SEDs and their best-fit model spectra for the diffuse arc (green), the total SED of the clumps and star cluster images (orange), and the total SED of the whole system (blue).}
\label{fig:plot_arc_sed_apertures}
\end{figure*}

As detailed in Section~\ref{sec:clump_photometry}, we measure the photometry of the star clusters using an aperture photometry technique. The aperture geometries are described in Section~\ref{sec:clump_photometry} and visualized here in Figure~\ref{fig:plot_arc_sed_apertures}. The left panel shows the apertures (shown in red) used in measuring the photometry of the clumps (1 and 2) and star cluster images (3a, 3b, 3c, 4a, 4b, 5a, and 5b). The apertures for the star clusters are circular with a radius of $0\farcs1$, while those for the clumps 1 and 2 are elliptical with a semi-major radius of $0\farcs32$ and $0\farcs2$, respectively, and geometry adopting the \galfit\ fits. The white apertures surrounding the clusters are used to estimate the diffuse background flux from the host galaxy. The right panel shows the measured SEDs of the diffuse arc (green), total clumps and star clusters (orange), and the whole system (blue).

Fitting the diffuse-arc SED with a non-parametric SFH model yields an observed stellar mass of $3.4_{-1.3}^{+1.9}\times 10^{9}\,M_{\odot}$, a SFR of $49_{-13}^{+26}\,M_{\odot}\rm{yr}^{-1}$, and a mass-weighted age of $168_{-104}^{+145}$ Myr. The stellar mass and SFR are not corrected for lensing because of the complex spatial variation of the magnification across the arc. 

\begin{figure*}[ht]
\centering
\includegraphics[width=0.31\textwidth]{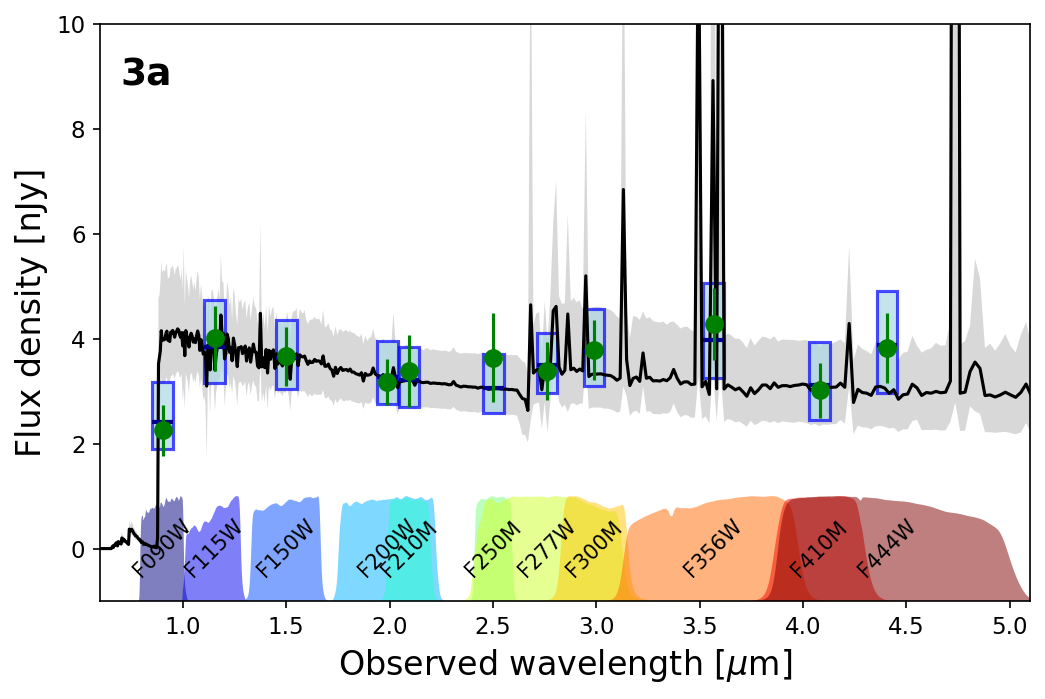}
\includegraphics[width=0.31\textwidth]{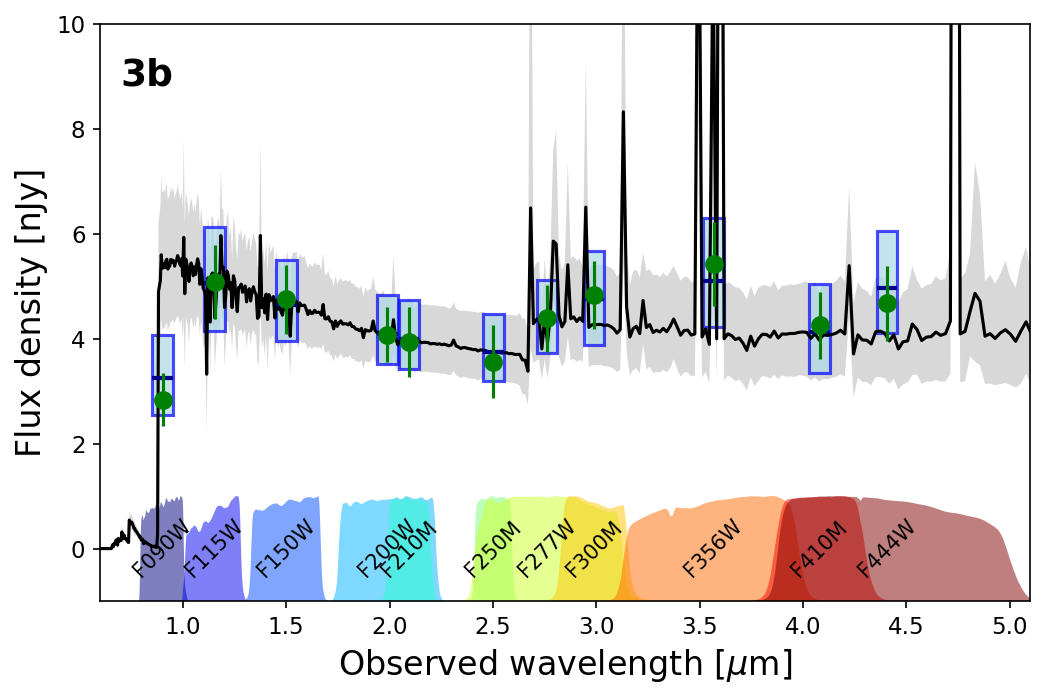}
\includegraphics[width=0.31\textwidth]{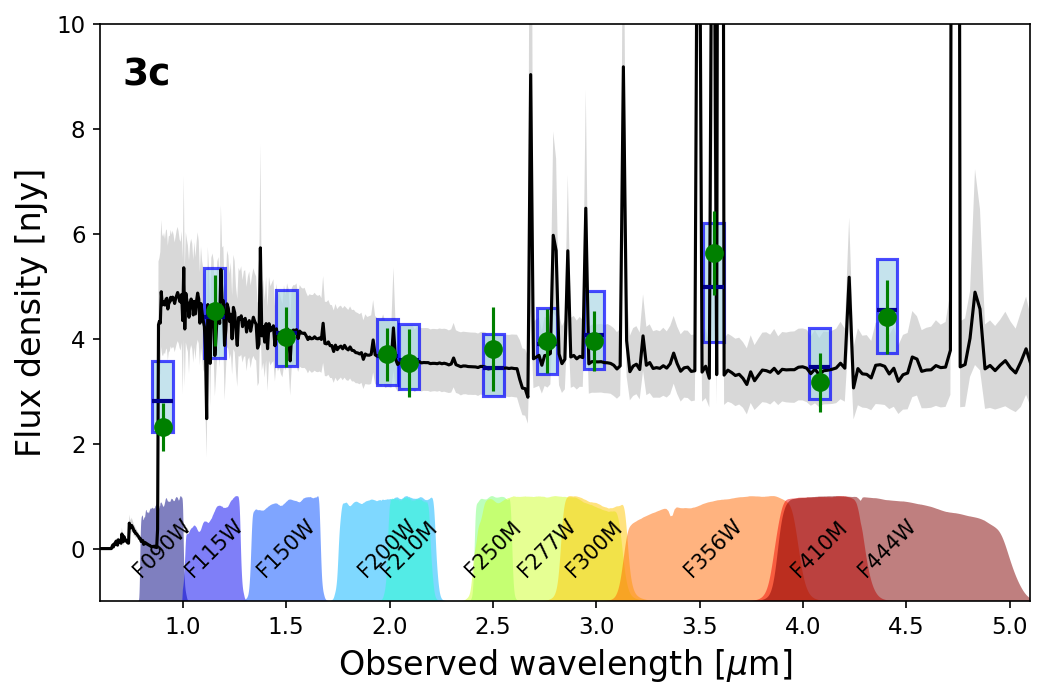}
\includegraphics[width=0.31\textwidth]{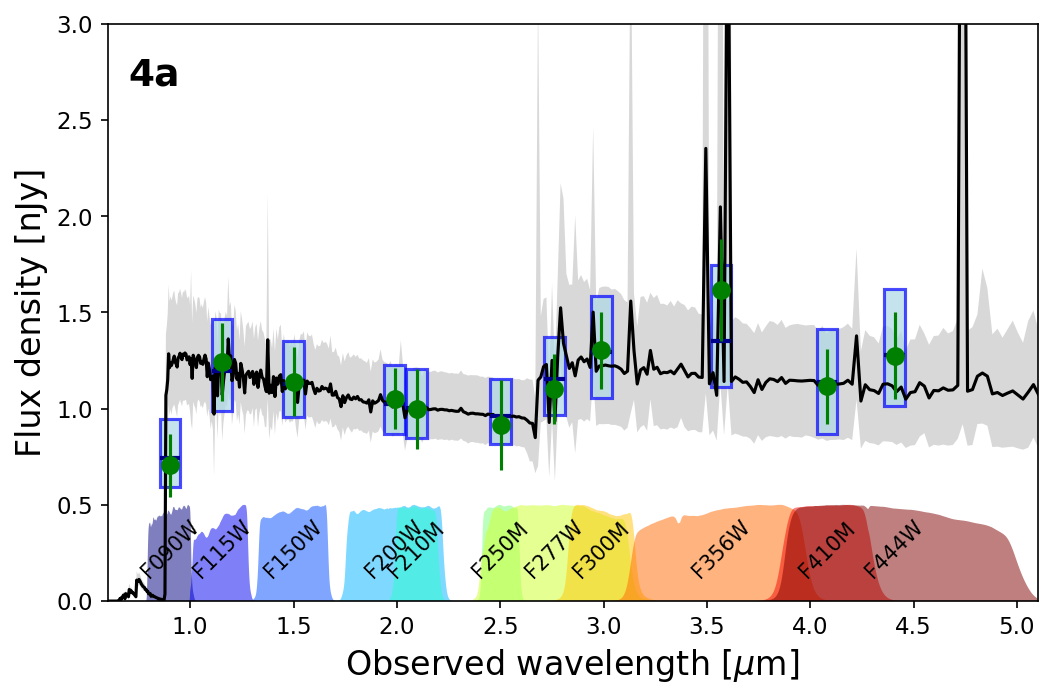}
\includegraphics[width=0.31\textwidth]{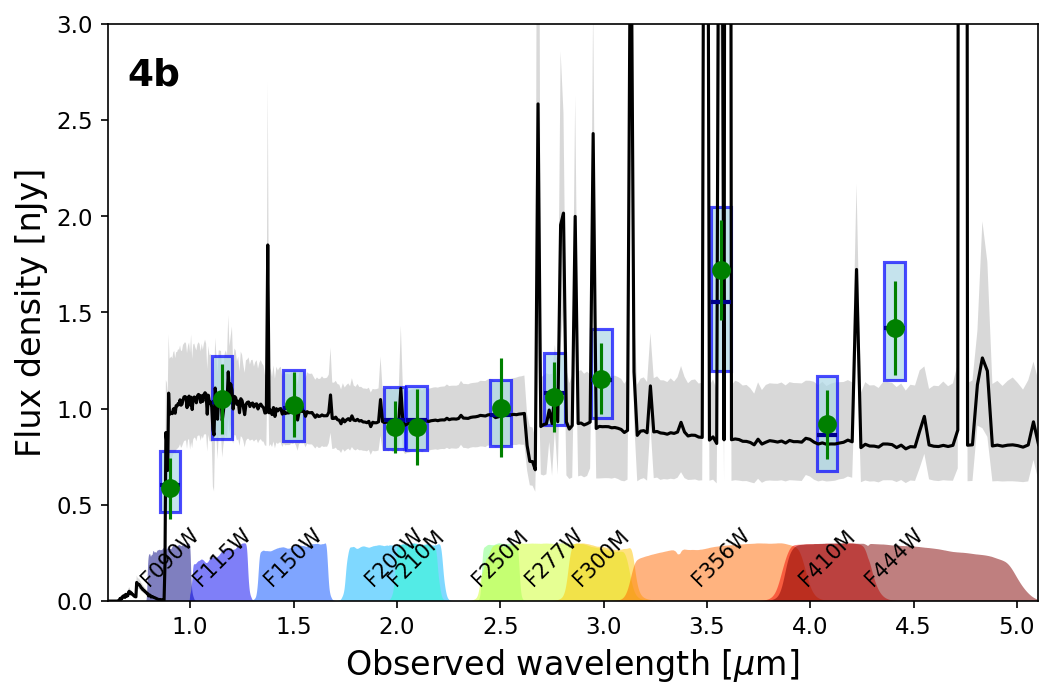}
\includegraphics[width=0.31\textwidth]{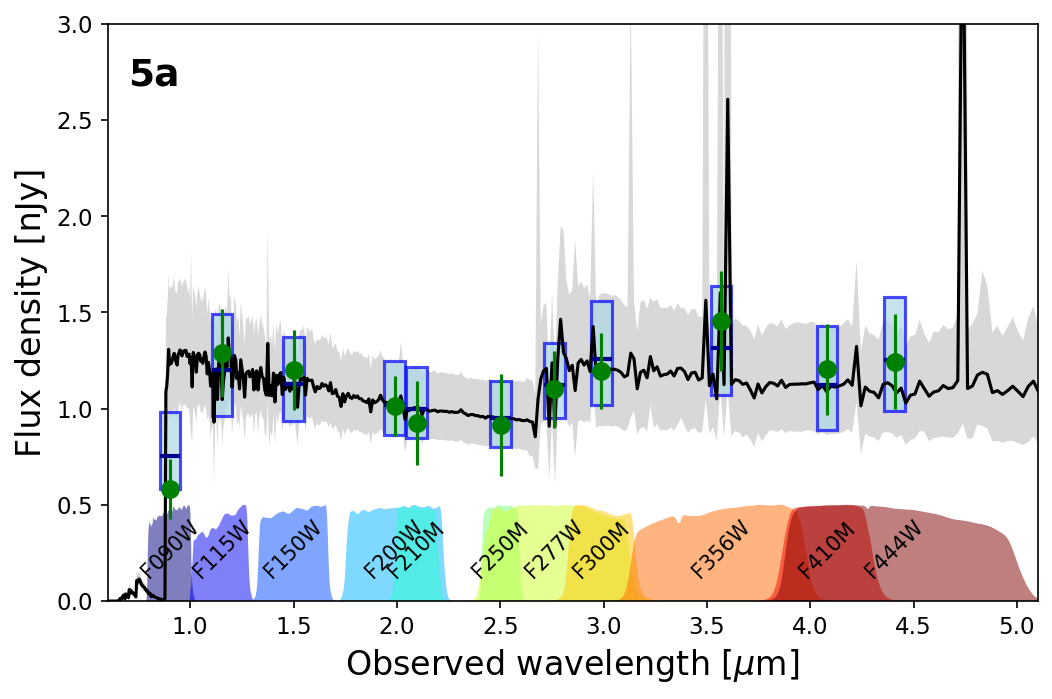}
\includegraphics[width=0.31\textwidth]{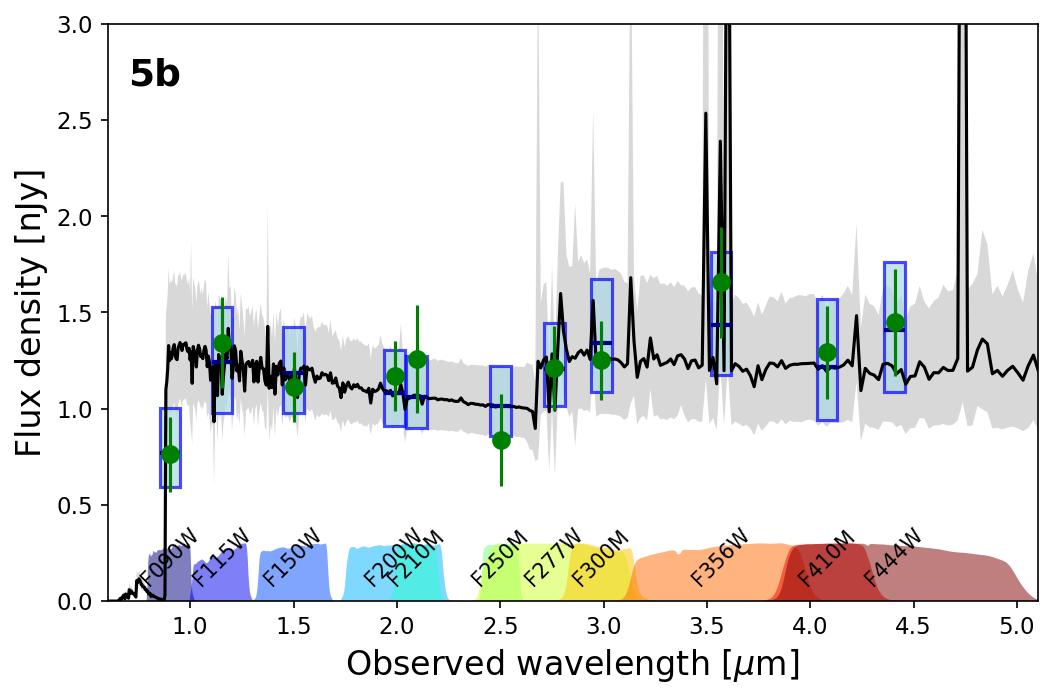}
\caption{SED fits of individual lensed images of the star clusters, obtained using \bagpipes\ with a single-burst SFH assumption. All the SEDs are corrected for lensing magnifications. The symbols are the same as those in Figure~\ref{fig:plot_comb_sedfits_bagpipes}.}
\label{fig:plot_sed_bagpipes_indiv_clumps}
\end{figure*}

The SED fits for individual lensed images of the star clusters are shown in Figure~\ref{fig:plot_sed_bagpipes_indiv_clumps}, with measured properties listed in Table~\ref{tab:sedfit_varies_sc}. The photometry of the clumps, star cluster images, diffuse host's arc, and the counter-image are summarized in Table~\ref{tab:clump_photometry}. 

\section{Stacking the Photometry of Star Clusters} \label{sec:photometry_stacking}

\begin{figure*}[ht]
\centering
\includegraphics[width=0.325\linewidth]{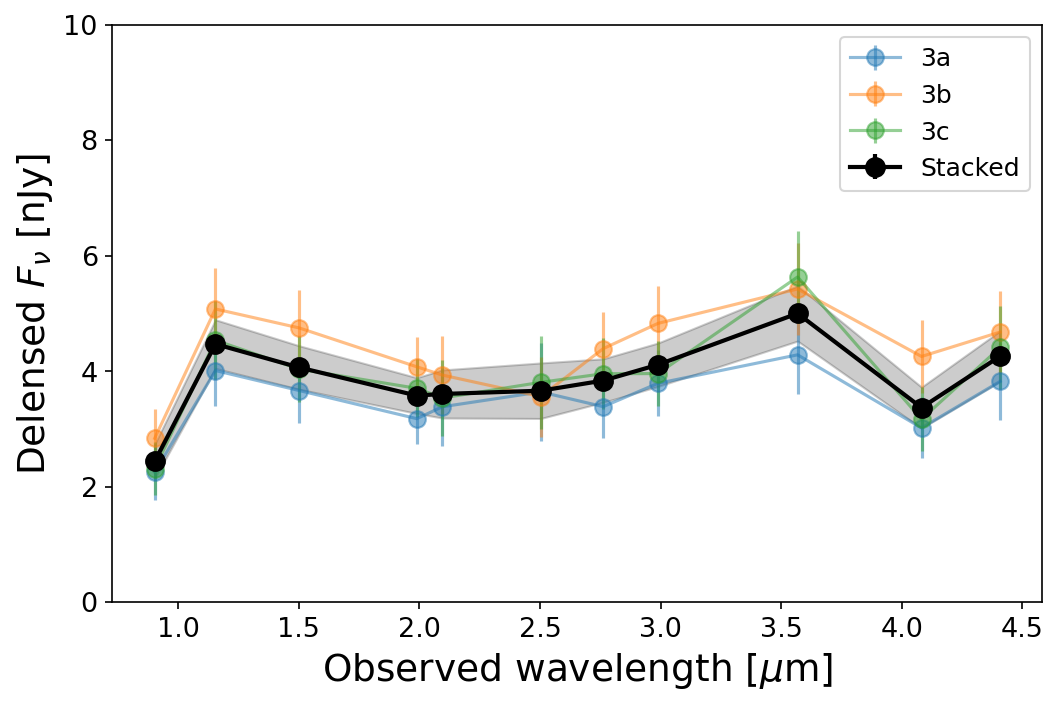}
\includegraphics[width=0.325\linewidth]{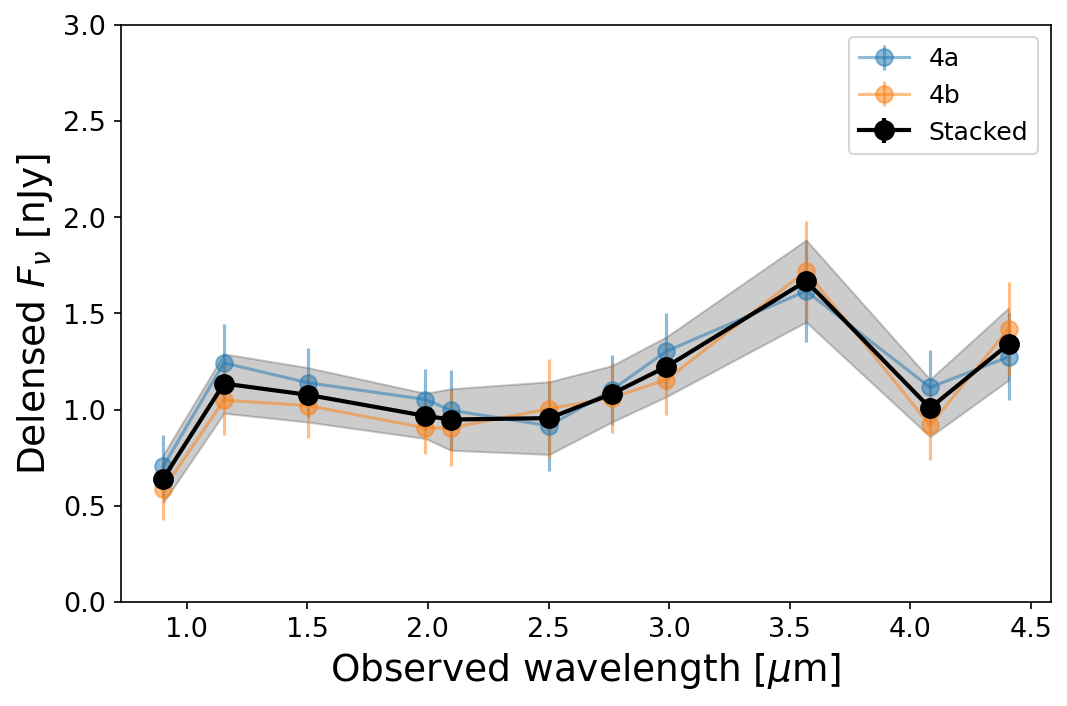}
\includegraphics[width=0.325\linewidth]{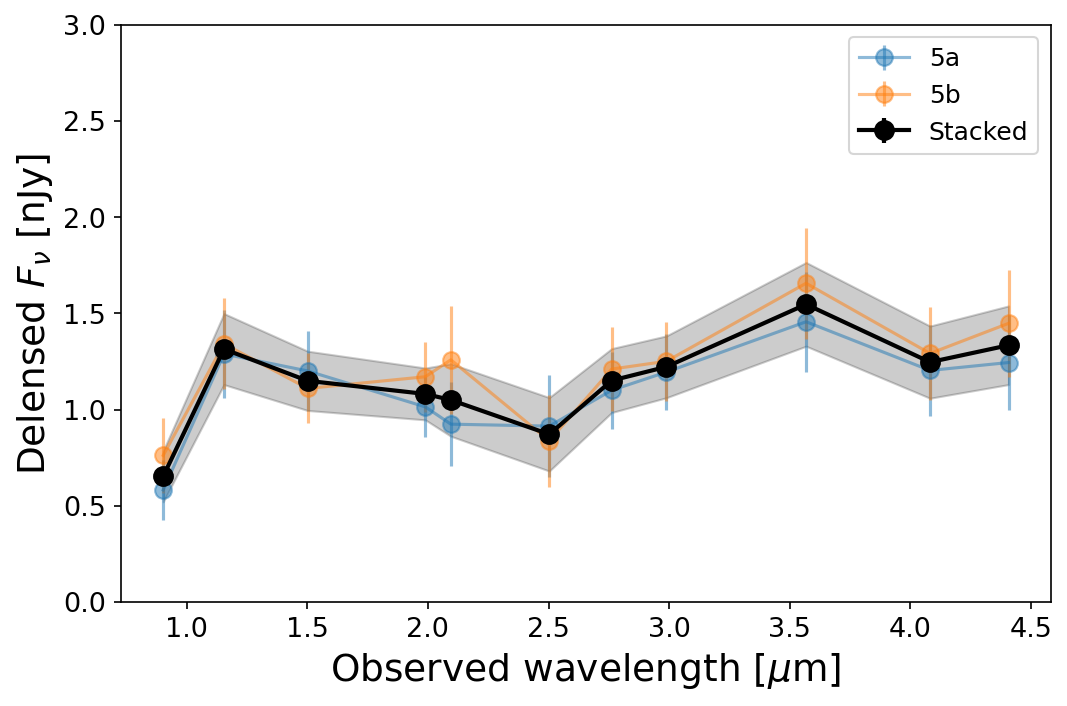}
\caption{Stacking SEDs of star cluster images obtained by inverse-variance weighted averaging. In each panel, the stacked SED is shown in black, while the SEDs of individual mirror images are shown in other colors. The SEDs of individual cluster images are corrected for lensing magnification before the stacking.}
\label{fig:stacking_photonetry}
\end{figure*}

To obtain representative photometry for each star cluster, we stack the photometry of their individual lensed images through inverse-variance-weighted averaging. The stacking is performed after correcting each multiple-image SED for the lensing magnification. To properly account for the lensing effect, the magnification uncertainties are propagated into the final flux uncertainties. 
Figure~\ref{fig:stacking_photonetry} presents the resulting stacked SEDs (shown in black) alongside those of the individual cluster images (shown in colors). As can be seen in the figure, the SEDs of the mirror images of each star cluster are consistent with one another within their uncertainties, which provides supporting evidence that they are multiple images of the same objects. 

\section{Dependence of Physical Parameters on SED Modeling Assumptions} 
\label{sec:sedfits_varies_clumps}

Throughout this paper, we perform SED fitting using multiple codes to explore various assumptions and priors, focusing particularly on SFH and dust attenuation models. With \bagpipes, the age estimates for clumps 1 and 2 depend systematically on the assumed SFH model: simple parametric forms yield significantly younger ages than non-parametric models. This trend holds for both dust attenuation models (\citealt{Calzetti2000} and \citealt{Salim2018}). As discussed in Section~\ref{sec:properties_clumps}, this discrepancy is likely driven by the outshining effect from young stellar populations in the system. 

In comparison, the \piXedfit\ fits, which apply simple parametric SFH models (delayed-$\tau$ and double power-law), yield moderate ages of $\sim 30$--$50$~Myr and slightly higher stellar masses for the clumps than the non-parametric \bagpipes\ fits. Similarly, \prospector\ fits yield slightly older ages and larger stellar masses than the non-parametric \bagpipes\ runs. Notably, all fitting configurations consistently produce low $A_{\rm V}$ attenuation values.

For the star cluster images, the \bagpipes\ fits employing single-burst and short exponentially declining ($\tau=1$~Myr) SFH models yield consistent ages, stellar masses, and $A_{\rm V}$ values within their uncertainties, regardless of the choice of dust attenuation model. Furthermore, \piXedfit\ fits using a short exponentially declining SFH model with $\tau$ restricted to $1$--$3$~Myr provide ages, stellar masses, and $A_{\rm V}$ values that are broadly consistent within the uncertainties with those derived from \bagpipes. 

We present a comprehensive summary of the fitting results for the star-forming clumps (1 and 2), the individual star clusters and their stacked photometry, and the counter image in Tables~\ref{tab:sedfit_varies_clumps1n2}, \ref{tab:sedfit_varies_sc}, and \ref{tab:sedfit_varies_counterimg}, respectively. 

\begin{table*}[h!]
\centering
\caption{\label{tab:sedfit_varies_clumps1n2}Combined SED fitting results for clumps 1 and 2} 
\begin{tabular}{lccccclccccc}
\toprule
ID & Age & $M_{*,\rm{obs}}$ & $M_{*,\rm{int}}$ & $A_V$ & $\chi_{\nu}^{2}$ & ID & Age & $M_{*,\rm{obs}}$ & $M_{*,\rm{int}}$ & $A_V$ & $\chi_{\nu}^{2}$ \\
 & (Myr) & $10^7 M_{\odot}$ & $10^7 M_{\odot}$ & (mag) &  &  & (Myr) & $10^7 M_{\odot}$ & $10^7 M_{\odot}$ & (mag) &  \\
\midrule
\multicolumn{6}{c}{\textbf{BAGPIPES, non-parametric, Salim attenuation}} & \multicolumn{6}{c}{\textbf{BAGPIPES, Delayed tau, Salim attenuation}} \\
\midrule
1 & $90_{-57}^{+96}$ & $321_{-87}^{+97}$ & $31_{-10}^{+14}$ & $0.03_{-0.02}^{+0.03}$ & $0.3$ &
1 & $9_{-4}^{+7}$ & $199_{-56}^{+101}$ & $19_{-7}^{+13}$ & $0.01_{-0.01}^{+0.02}$ & $0.4$ \\
2 & $112_{-71}^{+113}$ & $44_{-12}^{+17}$ & $4_{-1}^{+2}$ & $0.02_{-0.01}^{+0.02}$ & $0.3$ &
2 & $7_{-3}^{+7}$ & $23_{-7}^{+13}$ & $2_{-1}^{+2}$ & $0.02_{-0.01}^{+0.03}$ & $0.3$ \\
\midrule
\multicolumn{6}{c}{\textbf{BAGPIPES, Exp declining, Salim attenuation}} & \multicolumn{6}{c}{\textbf{BAGPIPES, Double power-law, Salim attenuation}} \\
\midrule
1 & $7_{-3}^{+5}$ & $200_{-56}^{+81}$ & $19_{-7}^{+11}$ & $0.02_{-0.01}^{+0.03}$ & $0.3$ &
1 & $16_{-5}^{+8}$ & $253_{-51}^{+71}$ & $24_{-7}^{+10}$ & $0.02_{-0.01}^{+0.02}$ & $0.4$ \\
2 & $7_{-3}^{+5}$ & $25_{-9}^{+12}$ & $2_{-1}^{+1}$ & $0.02_{-0.01}^{+0.03}$ & $0.3$ &
2 & $17_{-6}^{+9}$ & $34_{-8}^{+11}$ & $3_{-1}^{+1}$ & $0.02_{-0.01}^{+0.03}$ & $0.3$ \\
\midrule
\multicolumn{6}{c}{\textbf{BAGPIPES, non-parametric, Calzetti attenuation}} & \multicolumn{6}{c}{\textbf{BAGPIPES, Delayed tau, Calzetti attenuation}} \\
\midrule
1 & $82_{-60}^{+103}$ & $283_{-84}^{+120}$ & $27_{-10}^{+16}$ & $0.03_{-0.02}^{+0.03}$ & $0.3$ &
1 & $8_{-3}^{+7}$ & $192_{-55}^{+96}$ & $18_{-6}^{+12}$ & $0.03_{-0.02}^{+0.04}$ & $0.3$ \\
2 & $83_{-55}^{+107}$ & $40_{-13}^{+18}$ & $4_{-1}^{+2}$ & $0.04_{-0.03}^{+0.04}$ & $0.2$ &
2 & $7_{-3}^{+6}$ & $25_{-7}^{+13}$ & $2_{-1}^{+1}$ & $0.03_{-0.02}^{+0.05}$ & $0.2$ \\ 
\midrule
\multicolumn{6}{c}{\textbf{BAGPIPES, Exp declining, Calzetti attenuation}} & \multicolumn{6}{c}{\textbf{BAGPIPES, Double power-law, Calzetti attenuation}} \\
\midrule
1 & $7_{-3}^{+5}$ & $203_{-59}^{+87}$ & $19_{-7}^{+11}$ & $0.03_{-0.02}^{+0.04}$ & $0.3$ &
1 & $16_{-5}^{+7}$ & $244_{-47}^{+70}$ & $23_{-6}^{+10}$ & $0.03_{-0.02}^{+0.03}$ & $0.4$ \\
2 & $7_{-3}^{+4}$ & $25_{-9}^{+10}$ & $2_{-1}^{+1}$ & $0.04_{-0.03}^{+0.05}$ & $0.3$ &
2 & $14_{-4}^{+9}$ & $31_{-6}^{+12}$ & $3_{-1}^{+1}$ & $0.03_{-0.02}^{+0.04}$ & $0.3$ \\
\midrule
\multicolumn{6}{c}{\textbf{\textsc{piXedfit}, Delayed tau, Calzetti attenuation}} & \multicolumn{6}{c}{\textbf{\textsc{piXedfit}, Double power-law, Calzetti attenuation}} \\
\midrule
1 & $47_{-26}^{+38}$ & $617_{-279}^{+374}$ & $59_{-29}^{+46}$ & $0.03_{-0.02}^{+0.05}$ & $1.4$ & 
1 & $38_{-20}^{+31}$ & $706_{-333}^{+414}$ & $67_{-35}^{+51}$ & $0.03_{-0.02}^{+0.04}$ & $1.2$ \\
2 & $38_{-23}^{+40}$ & $73_{-35}^{+52}$ & $7_{-3}^{+6}$ & $0.04_{-0.03}^{+0.06}$ & $1.4$ &
2 & $28_{-17}^{+32}$ & $68_{-35}^{+51}$ & $6_{-3}^{+6}$ & $0.03_{-0.02}^{+0.04}$ & $1.3$ \\
\midrule
\multicolumn{6}{c}{\textbf{\textsc{Prospector}, non-parametric, Calzetti attenuation}} & \multicolumn{6}{c}{} \\
\midrule
1 & $185_{-89}^{+63}$ & $1000_{-535}^{+812}$ & $95_{-55}^{+96}$ & $0.05_{-0.03}^{+0.06}$ & $1.0$  \\ 
2 & $225_{-82}^{+54}$ & $157_{-90}^{+142}$ & $14_{-9}^{+16}$ & $0.05_{-0.04}^{+0.06}$ & $0.7$ \\
\bottomrule
\end{tabular}
\parbox{\linewidth}{
\tablecomments{Stellar mass values are corrected for lensing magnification.}}
\end{table*}

\begin{table*}[h!]
\centering
\caption{\label{tab:sedfit_varies_sc}Combined SED fitting results for star clusters} 
\begin{tabular}{lccccclccccc}
\toprule
ID & Age & $M_{*,\rm{obs}}$ & $M_{*,\rm{int}}$ & $A_V$ & $\mu_{\rm tot}$ & ID & Age & $M_{*,\rm{obs}}$ & $M_{*,\rm{int}}$ & $A_V$ & $\mu_{\rm tot}$ \\
 & (Myr) & $10^6 M_{\odot}$ & $10^6 M_{\odot}$ & (mag) &  &  & (Myr) & $10^6 M_{\odot}$ & $10^6 M_{\odot}$ & (mag) &  \\
\midrule
\multicolumn{6}{c}{\textbf{BAGPIPES, $\tau=1$\,Myr, Salim attenuation}} & \multicolumn{6}{c}{\textbf{BAGPIPES, Burst, Salim attenuation}} \\
\midrule
3abc & $7_{-2}^{+2}$ &  \nodata & $7_{-3}^{+4}$ & $0.06_{-0.04}^{+0.09}$ & \nodata & 
3abc & $7_{-2}^{+1}$ &  \nodata & $7_{-2}^{+3}$ & $0.06_{-0.04}^{+0.10}$ & \nodata \\
4ab & $6_{-2}^{+2}$ &  \nodata & $2_{-1}^{+1}$ & $0.10_{-0.06}^{+0.10}$ & \nodata &
4ab & $6_{-2}^{+2}$ &  \nodata & $1_{-1}^{+1}$ & $0.09_{-0.05}^{+0.11}$ & \nodata \\
5ab & $10_{-3}^{+9}$ &  \nodata & $4_{-2}^{+4}$ & $0.09_{-0.06}^{+0.09}$ & \nodata &
5ab & $10_{-3}^{+10}$ &  \nodata & $4_{-2}^{+4}$ & $0.10_{-0.07}^{+0.11}$ & \nodata \\
3a & $8_{-3}^{+8}$ & $138_{-85}^{+184}$ & $9_{-6}^{+14}$ & $0.09_{-0.06}^{+0.10}$ & $16 \pm 2$ & 
3a & $8_{-3}^{+8}$ & $131_{-68}^{+223}$ & $8_{-5}^{+17}$ & $0.09_{-0.06}^{+0.11}$ & $16 \pm 2$ \\
3b & $11_{-4}^{+13}$ & $273_{-128}^{+331}$ & $16_{-8}^{+23}$ & $0.07_{-0.05}^{+0.08}$ & $17 \pm 2$ &
3b & $9_{-3}^{+2}$ & $209_{-86}^{+79}$ & $12_{-6}^{+6}$ & $0.08_{-0.05}^{+0.10}$ & $17 \pm 2$ \\
3c & $6_{-2}^{+2}$ & $87_{-47}^{+68}$ & $5_{-3}^{+5}$ & $0.09_{-0.06}^{+0.13}$ & $16 \pm 2$ & 
3c & $7_{-1}^{+2}$ & $118_{-41}^{+55}$ & $7_{-3}^{+5}$ & $0.09_{-0.06}^{+0.11}$ & $16 \pm 2$ \\
4a & $10_{-4}^{+14}$ & $160_{-102}^{+253}$ & $4_{-2}^{+7}$ & $0.10_{-0.06}^{+0.13}$ & $44 \pm 5$ &
4a & $9_{-4}^{+13}$ & $157_{-98}^{+238}$ & $4_{-2}^{+6}$ & $0.10_{-0.06}^{+0.12}$ & $44 \pm 5$ \\
4b & $5_{-2}^{+3}$ & $46_{-19}^{+54}$ & $1_{-0}^{+2}$ & $0.14_{-0.07}^{+0.13}$ & $43 \pm 4$ & 
4b & $4_{-2}^{+2}$ & $43_{-18}^{+54}$ & $1_{-0}^{+2}$ & $0.14_{-0.07}^{+0.12}$ & $43 \pm 4$ \\
5a & $12_{-5}^{+11}$ & $184_{-116}^{+134}$ & $5_{-3}^{+5}$ & $0.09_{-0.05}^{+0.09}$ & $36 \pm 4$ &
5a & $10_{-4}^{+11}$ & $150_{-97}^{+145}$ & $4_{-3}^{+5}$ & $0.09_{-0.06}^{+0.12}$ & $36 \pm 4$ \\
5b & $10_{-4}^{+8}$ & $158_{-105}^{+142}$ & $4_{-3}^{+5}$ & $0.13_{-0.08}^{+0.13}$ & $35 \pm 4$ & 
5b & $10_{-4}^{+10}$ & $160_{-88}^{+161}$ & $5_{-3}^{+6}$ & $0.14_{-0.08}^{+0.10}$ & $35 \pm 4$ \\
\midrule
\multicolumn{6}{c}{\textbf{BAGPIPES, $\tau=1$\,Myr, Calzetti attenuation}} & \multicolumn{6}{c}{\textbf{BAGPIPES, Burst, Calzetti attenuation}} \\
\midrule
3abc & $7_{-2}^{+2}$ &  \nodata & $7_{-3}^{+5}$ & $0.13_{-0.08}^{+0.17}$ &  \nodata &
3abc & $7_{-1}^{+2}$ &  \nodata & $8_{-3}^{+4}$ & $0.15_{-0.10}^{+0.18}$ &  \nodata \\
4ab & $6_{-2}^{+3}$ &  \nodata & $2_{-1}^{+2}$ & $0.23_{-0.11}^{+0.14}$ &  \nodata &
4ab & $6_{-2}^{+2}$ &  \nodata & $1_{-1}^{+2}$ & $0.18_{-0.09}^{+0.16}$ &  \nodata \\
5ab & $11_{-3}^{+11}$ &  \nodata & $5_{-2}^{+4}$ & $0.19_{-0.11}^{+0.16}$ &  \nodata &
5ab & $11_{-4}^{+13}$ &  \nodata & $5_{-3}^{+5}$ & $0.21_{-0.12}^{+0.14}$ &  \nodata \\
3a & $9_{-3}^{+9}$ & $175_{-110}^{+217}$ & $11_{-7}^{+16}$ & $0.19_{-0.13}^{+0.18}$ & $16 \pm 2$ & 
3a & $8_{-3}^{+9}$ & $152_{-101}^{+225}$ & $10_{-7}^{+17}$ & $0.17_{-0.11}^{+0.16}$ & $16 \pm 2$ \\
3b & $12_{-4}^{+12}$ & $342_{-165}^{+359}$ & $20_{-10}^{+25}$ & $0.18_{-0.12}^{+0.17}$ & $17 \pm 2$ &
3b & $8_{-2}^{+2}$ & $213_{-91}^{+120}$ & $12_{-6}^{+9}$ & $0.17_{-0.10}^{+0.15}$ & $17 \pm 2$ \\
3c & $5_{-2}^{+3}$ & $88_{-45}^{+100}$ & $5_{-3}^{+7}$ & $0.17_{-0.11}^{+0.14}$ & $16 \pm 2$ & 
3c & $7_{-1}^{+2}$ & $130_{-42}^{+72}$ & $8_{-3}^{+6}$ & $0.20_{-0.13}^{+0.15}$ & $16 \pm 2$ \\
4a & $11_{-5}^{+16}$ & $229_{-156}^{+269}$ & $5_{-4}^{+7}$ & $0.20_{-0.12}^{+0.15}$ & $44 \pm 5$ & 
4a & $10_{-4}^{+15}$ & $206_{-138}^{+273}$ & $5_{-3}^{+7}$ & $0.20_{-0.13}^{+0.17}$ & $44 \pm 5$ \\
4b & $4_{-2}^{+2}$ & $47_{-16}^{+47}$ & $1_{-0}^{+1}$ & $0.27_{-0.12}^{+0.12}$ & $43 \pm 4$ & 
4b & $4_{-1}^{+3}$ & $41_{-15}^{+52}$ & $1_{-0}^{+1}$ & $0.24_{-0.12}^{+0.15}$ & $43 \pm 4$ \\
5a & $12_{-5}^{+13}$ & $182_{-118}^{+190}$ & $5_{-3}^{+6}$ & $0.20_{-0.13}^{+0.14}$ & $36 \pm 4$ & 
5a & $11_{-4}^{+11}$ & $182_{-113}^{+187}$ & $5_{-3}^{+6}$ & $0.20_{-0.12}^{+0.18}$ & $36 \pm 4$ \\
5b & $10_{-4}^{+11}$ & $185_{-117}^{+164}$ & $5_{-3}^{+6}$ & $0.23_{-0.14}^{+0.15}$ & $35 \pm 4$ &
5b & $10_{-4}^{+13}$ & $175_{-116}^{+203}$ & $5_{-3}^{+7}$ & $0.24_{-0.15}^{+0.15}$ & $35 \pm 4$ \\
\midrule
\multicolumn{6}{c}{\textbf{\textsc{piXedfit}, $\tau=1-3$\,Myr, Calzetti attenuation}} \\
\midrule
3abc & $6_{-2}^{+4}$ &  \nodata & $3_{-1}^{+4}$ & $0.06_{-0.04}^{+0.06}$ &  \nodata \\
4ab & $6_{-2}^{+5}$ &  \nodata & $1_{-0}^{+1}$ & $0.08_{-0.05}^{+0.05}$ &  \nodata \\
5ab & $15_{-8}^{+12}$ &  \nodata & $4_{-3}^{+4}$ & $0.06_{-0.04}^{+0.06}$ &  \nodata \\
3a & $6_{-2}^{+6}$ & $51_{-13}^{+98}$ & $3_{-1}^{+7}$ & $0.07_{-0.05}^{+0.06}$ & $16 \pm 2$ \\
3b & $11_{-4}^{+11}$ & $171_{-101}^{+301}$ & $10_{-6}^{+20}$ & $0.06_{-0.04}^{+0.07}$ & $17 \pm 2$ \\
3c & $5_{-1}^{+4}$ & $49_{-8}^{+57}$ & $3_{-1}^{+4}$ & $0.08_{-0.05}^{+0.06}$ & $16 \pm 2$ \\
4a & $12_{-6}^{+19}$ & $126_{-83}^{+273}$ & $3_{-2}^{+7}$ & $0.06_{-0.05}^{+0.06}$ & $44 \pm 5$ \\
4b & $5_{-2}^{+4}$ & $37_{-7}^{+47}$ & $1_{-0}^{+1}$ & $0.11_{-0.06}^{+0.04}$ & $43 \pm 4$ \\
5a & $13_{-6}^{+16}$ & $112_{-74}^{+200}$ & $3_{-2}^{+7}$ & $0.07_{-0.05}^{+0.06}$ & $36 \pm 4$ \\
5b & $16_{-9}^{+13}$ & $173_{-127}^{+176}$ & $5_{-4}^{+6}$ & $0.08_{-0.05}^{+0.05}$ & $35 \pm 4$ \\
\bottomrule
\end{tabular}
\parbox{\linewidth}{
\tablecomments{Stellar mass values are not corrected for lensing magnification.}}
\end{table*}
\begin{table*}[h!]
\centering
\caption{\label{tab:sedfit_varies_counterimg}Combined SED Fitting Results for the Counter-image}
\begin{tabular*}{\textwidth}{@{\extracolsep{\fill}} llcccccc}
\toprule
Code & SFH model & Dust curve & Age & $M_{*,\rm{obs}}$ & $M_{*,\rm{int}}$ & $A_V$ & $\chi_{\nu}^{2}$ \\
 & & & (Myr) & $10^7 M_{\odot}$ & $10^7 M_{\odot}$ & (mag) &  \\
\midrule
\bagpipes\ & non-parametric & Calzetti & $115_{-83}^{+123}$ & $135_{-38}^{+50}$ & $44_{-15}^{+23}$ & $0.17_{-0.07}^{+0.07}$ & $0.1$ \\
\bagpipes\ & non-parametric & Salim & $152_{-94}^{+120}$ & $145_{-38}^{+60}$ & $48_{-16}^{+27}$ & $0.07_{-0.03}^{+0.05}$ & $0.1$ \\
\bagpipes\ & Delayed tau & Calzetti & $5_{-2}^{+3}$ & $75_{-22}^{+30}$ & $24_{-9}^{+14}$ & $0.19_{-0.08}^{+0.07}$ & $0.1$ \\
\bagpipes\ & Delayed tau & Salim & $6_{-3}^{+4}$ & $71_{-21}^{+33}$ & $23_{-8}^{+15}$ & $0.09_{-0.04}^{+0.06}$ & $0.1$ \\
\bagpipes\ & Double power-law & Calzetti & $12_{-3}^{+5}$ & $112_{-21}^{+25}$ & $37_{-10}^{+13}$ & $0.20_{-0.07}^{+0.06}$ & $0.1$ \\
\bagpipes\ & Double power-law & Salim & $13_{-3}^{+6}$ & $99_{-17}^{+28}$ & $33_{-8}^{+14}$ & $0.09_{-0.04}^{+0.06}$ & $0.1$ \\
\bagpipes\ & Exp declining & Calzetti & $4_{-1}^{+2}$ & $74_{-22}^{+29}$ & $24_{-9}^{+13}$ & $0.20_{-0.06}^{+0.08}$ & $0.1$ \\
\bagpipes\ & Exp declining & Salim & $5_{-2}^{+3}$ & $67_{-18}^{+29}$ & $22_{-7}^{+13}$ & $0.09_{-0.04}^{+0.06}$ & $0.1$ \\
\textsc{piXedfit} & Delayed tau & Calzetti & $3_{-1}^{+3}$ & $66_{-7}^{+17}$ & $22_{-4}^{+9}$ & $0.11_{-0.06}^{+0.04}$ & $1.0$ \\
\textsc{piXedfit} & Double power-law & Calzetti & $4_{-2}^{+4}$ & $71_{-7}^{+18}$ & $23_{-4}^{+9}$ & $0.11_{-0.06}^{+0.03}$ & $1.2$ \\
\textsc{Prospector} & non-parametric & Calzetti & $290_{-57}^{+50}$ & $306_{-108}^{+209}$ & $63_{-26}^{+55}$ & $0.07_{-0.04}^{+0.05}$ & $0.4$ \\
\bottomrule
\end{tabular*}
\parbox{\linewidth}{
\tablecomments{Stellar mass values are corrected for lensing magnification.}}
\end{table*}

\section{Photometric Redshift of the Counter-image}
\label{sec:photo_z_counter_image}

\begin{figure}[ht]
\centering
\includegraphics[width=0.6\linewidth]{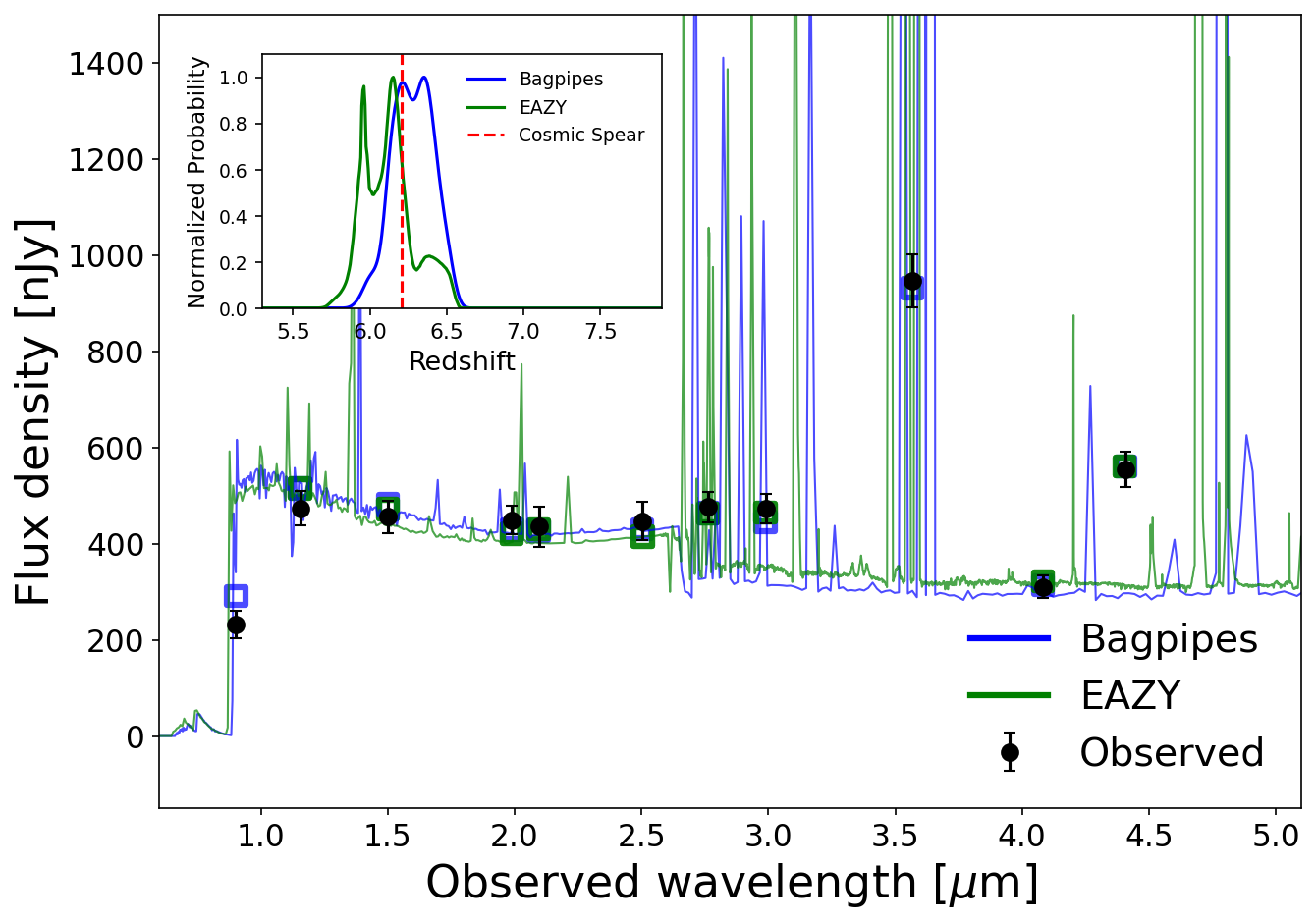}
\caption{Photometric redshift estimation for the CI using \textsc{EAZY} and \bagpipes. Black circles denote the observed photometry. The best-fit models (spectra and photometry) from \textsc{EAZY} and \bagpipes\ are shown in green and blue, respectively. The derived photometric redshifts are $6.12 \pm 0.17$ (\textsc{EAZY}) and ${6.28}_{-0.13}^{+0.14}$ (\bagpipes). The inset panel shows the redshift posterior probability distributions. The vertical dashed red line indicates the spectroscopic redshift of the Cosmic Spear.}
\label{fig:plot_photoz_CI}
\end{figure}

To derive the photometric redshift of the CI, we perform SED fitting using \textsc{EAZY} \citep{Brammer2008} and \bagpipes. The \textsc{EAZY} analysis employs the blue SED templates of \citet{Larson2023}, which incorporate BPASS stellar population models and nebular emission models with \textsc{CLOUDY} photoionization code. For the \bagpipes\ fitting, we adopt a setup similar to that used for the clumps in the Cosmic Spear, with the exception that the redshift is allowed to vary within the range of $0 < z < 15$. To reduce the number of free parameters and focus on constraining the redshift, we assume a delayed tau SFH model. The results are presented in Figure~\ref{fig:plot_photoz_CI}. The photometric redshifts derived with \textsc{EAZY} and \bagpipes\ are consistent within the uncertainties, yielding $6.12 \pm 0.17$ and ${6.28}_{-0.13}^{+0.14}$, respectively. The redshift probability distributions from both codes tend to be bimodal, with one peak overlapping the spectroscopic redshift of the Cosmic Spear. This redshift consistency provides supporting evidence that the CI is a counter-image of the Cosmic Spear.  

\begin{deluxetable*}{lccccccccccc}
\tablecaption{\label{tab:clump_photometry}Observed Photometry of the Clumps, Star Clusters, Host diffuse arc, and Counter-image
}
\tablewidth{\columnwidth}
\tablehead{
\colhead{ID} &
\colhead{F090W} &
\colhead{F115W} &
\colhead{F150W} &
\colhead{F200W} &
\colhead{F210M} &
\colhead{F250M} &
\colhead{F277W} &
\colhead{F300M} &
\colhead{F356W} &
\colhead{F410M} &
\colhead{F444W} \\
\colhead{} &
\colhead{(nJy)} &
\colhead{(nJy)} &
\colhead{(nJy)} &
\colhead{(nJy)} &
\colhead{(nJy)} &
\colhead{(nJy)} &
\colhead{(nJy)} &
\colhead{(nJy)} &
\colhead{(nJy)} &
\colhead{(nJy)} &
\colhead{(nJy)}
}
\startdata
1
& $1173 \pm 134$
& $2324 \pm 246$
& $2188 \pm 231$
& $1910 \pm 199$
& $1832 \pm 203$
& $1617 \pm 190$
& $1629 \pm 176$
& $1605 \pm 172$
& $2725 \pm 287$
& $1063 \pm 117$
& $1485 \pm 161$ \\
2
& $138 \pm 21$
& $303 \pm 36$
& $299 \pm 35$
& $255 \pm 29$
& $228 \pm 32$
& $216 \pm 33$
& $222 \pm 27$
& $214 \pm 26$
& $403 \pm 46$
& $139 \pm 19$
& $207 \pm 26$ \\
3a
& $36 \pm 8$
& $64 \pm 10$
& $58 \pm 9$
& $51 \pm 7$
& $54 \pm 11$
& $58 \pm 14$
& $54 \pm 9$
& $60 \pm 9$
& $68 \pm 11$
& $48 \pm 8$
& $61 \pm 11$ \\
3b
& $49 \pm 9$
& $88 \pm 12$
& $83 \pm 11$
& $71 \pm 9$
& $68 \pm 12$
& $62 \pm 12$
& $76 \pm 11$
& $84 \pm 11$
& $94 \pm 14$
& $74 \pm 11$
& $81 \pm 12$ \\
3c
& $37 \pm 7$
& $73 \pm 11$
& $65 \pm 9$
& $60 \pm 8$
& $57 \pm 10$
& $61 \pm 13$
& $64 \pm 10$
& $64 \pm 9$
& $91 \pm 13$
& $51 \pm 9$
& $71 \pm 11$ \\
4a
& $31 \pm 7$
& $55 \pm 9$
& $50 \pm 8$
& $46 \pm 7$
& $44 \pm 9$
& $40 \pm 10$
& $48 \pm 8$
& $57 \pm 9$
& $71 \pm 12$
& $49 \pm 9$
& $56 \pm 10$ \\
4b
& $25 \pm 7$
& $45 \pm 8$
& $44 \pm 7$
& $39 \pm 6$
& $39 \pm 9$
& $43 \pm 11$
& $46 \pm 8$
& $50 \pm 8$
& $74 \pm 11$
& $40 \pm 8$
& $61 \pm 10$ \\
5a
& $21 \pm 6$
& $46 \pm 8$
& $43 \pm 7$
& $36 \pm 6$
& $33 \pm 8$
& $33 \pm 10$
& $39 \pm 7$
& $43 \pm 7$
& $52 \pm 9$
& $43 \pm 8$
& $45 \pm 9$ \\
5b
& $27 \pm 7$
& $47 \pm 8$
& $39 \pm 6$
& $41 \pm 6$
& $45 \pm 10$
& $30 \pm 8$
& $43 \pm 8$
& $44 \pm 7$
& $59 \pm 10$
& $46 \pm 9$
& $51 \pm 10$ \\
Host arc
& $322 \pm 140$
& $907 \pm 251$
& $981 \pm 236$
& $1046 \pm 203$
& $934 \pm 210$
& $1061 \pm 199$
& $1254 \pm 180$
& $1293 \pm 176$
& $3077 \pm 293$
& $840 \pm 122$
& $1539 \pm 167$ \\
CI
& $371 \pm 45$
& $758 \pm 57$
& $729 \pm 53$
& $718 \pm 48$
& $697 \pm 67$
& $715 \pm 63$
& $761 \pm 52$
& $755 \pm 49$
& $1514 \pm 89$
& $496 \pm 37$
& $886 \pm 59$
\enddata
\tablecomments{The fluxes are not corrected for lensing magnification.}
\end{deluxetable*}

\begin{deluxetable*}{lccccccccccc}
\tablecaption{\label{tab:clump_stacked_photometry}Delensed Stacked Photometry of the Star Clusters
}
\tablewidth{\columnwidth}
\tablehead{
\colhead{ID} &
\colhead{F090W} &
\colhead{F115W} &
\colhead{F150W} &
\colhead{F200W} &
\colhead{F210M} &
\colhead{F250M} &
\colhead{F277W} &
\colhead{F300M} &
\colhead{F356W} &
\colhead{F410M} &
\colhead{F444W} \\
\colhead{} &
\colhead{(nJy)} &
\colhead{(nJy)} &
\colhead{(nJy)} &
\colhead{(nJy)} &
\colhead{(nJy)} &
\colhead{(nJy)} &
\colhead{(nJy)} &
\colhead{(nJy)} &
\colhead{(nJy)} &
\colhead{(nJy)} &
\colhead{(nJy)}
}
\startdata
3abc
& $2.4 \pm 0.3$
& $4.5 \pm 0.4$
& $4.1 \pm 0.4$
& $3.6 \pm 0.3$
& $3.6 \pm 0.4$
& $3.7 \pm 0.5$
& $3.8 \pm 0.4$
& $4.1 \pm 0.4$
& $5.0 \pm 0.5$
& $3.4 \pm 0.4$
& $4.3 \pm 0.4$ \\
4ab
& $0.6 \pm 0.1$
& $1.1 \pm 0.2$
& $1.1 \pm 0.1$
& $1.0 \pm 0.1$
& $0.9 \pm 0.2$
& $1.0 \pm 0.2$
& $1.1 \pm 0.1$
& $1.2 \pm 0.2$
& $1.7 \pm 0.2$
& $1.0 \pm 0.1$
& $1.3 \pm 0.2$ \\
5ab
& $0.7 \pm 0.1$
& $1.3 \pm 0.2$
& $1.1 \pm 0.2$
& $1.1 \pm 0.1$
& $1.0 \pm 0.2$
& $0.9 \pm 0.2$
& $1.2 \pm 0.2$
& $1.2 \pm 0.2$
& $1.5 \pm 0.2$
& $1.2 \pm 0.2$
& $1.3 \pm 0.2$
\enddata
\tablecomments{The fluxes are corrected for lensing magnification.}
\end{deluxetable*}


\bibliography{z6arc}{}
\bibliographystyle{aasjournalv7}



\end{document}